\documentclass[11pt,letterpaper]{article}
\pdfoutput=1

\usepackage{jcappub}

\usepackage{graphicx}
\usepackage{dcolumn}
\usepackage{bm}
\usepackage{natbib}
\usepackage{amsmath}
\usepackage{amssymb}
\usepackage{epstopdf}
\usepackage{hhline}
\usepackage{color}
\usepackage[T1]{fontenc}
\usepackage{wasysym}
\usepackage{url}
\newcommand{\be}{\begin{equation}}
\newcommand{\ee}{\end{equation}}
\newcommand{\bea}{\begin{eqnarray}}
\newcommand{\eea}{\end{eqnarray}}

\newcommand{\s}{{\rm ~s}}

\newcommand{\mo}{\mathcal{O}}
\newcommand{\ml}{\mathcal{L}}
\newcommand{\nn}{\nonumber}

\newcommand{\bra}[1]{ \langle {#1} | }
\newcommand{\ket}[1]{ | {#1} \rangle }

\definecolor{darkgreen}{rgb}{0,0.5,0}

\begin{document}

\title{Capture and Decay of Electroweak WIMPonium}

\author[a]{Pouya Asadi,}

\affiliation[a]{New High Energy Theory Center, Rutgers University, Piscataway, NJ 08854}

\author[a]{Matthew Baumgart,}

\author[b]{Patrick J. Fitzpatrick,}

\author[b]{Emmett Krupczak}

\author[b]{and Tracy R. Slatyer}

\affiliation[b]{Center for Theoretical Physics, Massachusetts Institute of Technology, Cambridge, MA 02139, USA}

\emailAdd{asadi@physics.rutgers.edu}
\emailAdd{baumgart@physics.rutgers.edu}
\emailAdd{ekrupcza@mit.edu}
\emailAdd{fitzppat@mit.edu}
\emailAdd{tslatyer@mit.edu}

\date{today}

\abstract{The spectrum of Weakly-Interacting-Massive-Particle (WIMP) dark matter generically possesses bound states when the WIMP mass becomes sufficiently large relative to the mass of the electroweak gauge bosons. The presence of these bound states enhances the annihilation rate via resonances in the Sommerfeld enhancement, but they can also be produced directly with the emission of a low-energy photon. In this work we compute the rate for SU(2) triplet dark matter (the wino) to bind into WIMPonium -- which is possible via single-photon emission for wino masses above 5 TeV for relative velocity $v < O(10^{-2})$  -- and study the subsequent decays of these bound states. We present results with applications beyond the wino case, e.g.~for dark matter inhabiting a nonabelian dark sector; these include analytic capture and transition rates for general dark sectors in the limit of vanishing force carrier mass, efficient numerical routines for calculating positive and negative-energy eigenstates of a Hamiltonian containing interactions with both massive and massless force carriers, and a study of the scaling of bound state formation in the short-range Hulth\'{e}n potential. In the specific case of the wino, we find that the rate for bound state formation is suppressed relative to direct annihilation, and so provides only a small correction to the overall annihilation rate. The soft photons radiated by the capture process and by bound state transitions could permit measurement of the dark matter's quantum numbers; for wino-like dark matter, such photons are rare, but might be observable by a future ground-based gamma-ray telescope combining large effective area and a low energy threshold.

Due to a correction to the relative sign between the two diagrams in figure 1, a new version of the original paper is provided, since this sign propagates through the paper and gives rise to changes in some equations and figures. This affects both capture rate into the bound states and transition rates between different bound states. The topline qualitative results of the paper remain unchanged.
}

%\pacs{95.35.+d,98.80.Es}

\preprint{MIT-CTP/4845}
\maketitle

%%%%%%%%%%%%%%%%
\section{Introduction}
\label{sec:intro}
%%%%%%%%%%%%%%%%

Cold Dark Matter (DM) remains a compelling, economical explanation for a variety of phenomena at scales from the galactic (velocity rotation curves) to the cosmological (peaks in the anisotropy power spectrum of the cosmic microwave background).  Although the particle content of the Standard Model (SM) does not contain such a ``magic bullet,'' it is straightforward to add new degrees of freedom with the necessary properties: the so-called Weakly Interacting Massive Particles (WIMPs) \cite{Cirelli:2005uq,Jungman:1995df}. The coldness of Cold DM implies we are immersed in a sea of slowly-moving particles, and giving the DM couplings of similar strength to the SM (perturbative but not ultra-weak), the correct DM relic abundance is naturally obtained for masses of $\mo$(1 TeV) \cite{Dimopoulos:1990gf}. 

Thus, we are moved to consider the physics of heavy, slow particles, with simulations suggesting a mean velocity $\langle v \rangle \sim 10^{-3}$ \cite{2009ApJ...704.1704B}.  We can therefore work in the nonrelativistic limit, setting up an effective field theory for the DM in analogy with NRQCD and NRQED \cite{Beneke:2012tg,Baumgart:2014vma,Bauer:2014ula,Ovanesyan:2014fwa}.  In this limit, the interactions of the DM with long-range force carriers (e.g.~electroweak bosons, dark-sector photons) are properly treated as a nonperturbative, nonlocal, but instantaneous potential.  This leads to the well-known phenomenon of Sommerfeld enhancement in DM annihilations \cite{Hisano:2004ds,Hisano:2006nn,ArkaniHamed:2008qn,Cirelli:2007xd, Cassel:2009wt,Slatyer:2009vg,Beneke:2014gja}.  The potential deforms the two-particle DM wavefunction near the origin, leading to large deviations from a calculation treating the initial state as a plane wave.  Schematically, for annihilation of DM in an $s$-wave state, the annihilation rate goes as 
\be
\sigma v \,=\, \Gamma |\psi(0)|^2,
\label{eq:schema}
\ee
where $\Gamma$ is the perturbatively-calculated, short-distance annihilation rate, and $\psi(0)$ is the wavefunction of the two-particle DM-DM state at the origin.  In the limit that the potential turns off, $|\psi(0)|$ = 1, and we recover the perturbative result.  

The wavefunction in eq.~\ref{eq:schema} is for a positive-energy scattering state.  However, the spectrum of the long-range potential may also include negative-energy bound states. When the binding energy for one of these states approaches zero, it induces a large resonant enhancement to the scattering-state wavefunction at the origin $\psi(0)$, and hence to the Sommerfeld enhancement \cite{Hisano:2003ec,Hisano:2004ds}.

The presence of bound states in the spectrum can have effects beyond an enhanced Sommerfeld factor. In particular, capture of DM particles into these bound states gives rise to an alternative annihilation channel for the DM, analogous to formation and annihilation of positronium, which in some circumstances may dominate over the Sommerfeld-enhanced direct annihilation. Transitions into and between bound states can also produce particles at energies parametrically suppressed relative to  the DM mass. There has been considerable interest in the literature in such WIMPonium states and their properties \cite{MarchRussell:2008tu,Shepherd:2009sa,Braaten:2013tza,Laha:2013gva,Wise:2014ola,Wise:2014jva,Petraki:2014uza,vonHarling:2014kha,Petraki:2015hla,Tsai:2015ugz,An:2015pva,An:2016gad,An:2016kie,Bi:2016gca,Kouvaris:2016ltf}; however, most of the work on indirect signatures to date has focused on models where the DM couples to only a single mediator (a dark photon or scalar), and where the mass of the mediator is sufficiently light that the resulting potential can be approximated by the Coulomb potential.

In this work, we extend these considerations to the electroweak potential, where these simplifying assumptions do not apply: the DM is generally part of a multiplet of states of similar masses, and these states may couple to both massive and massless gauge bosons. DM transforming under SU(2)$_L \times$U(1)$_Y$ is known to receive large Sommerfeld corrections for masses above $\sim$1 TeV, with the first resonance -- signaling the presence of a bound state -- occurring for the SU(2) triplet, or wino, at a DM mass $\sim$ 2.5 TeV.  Interestingly, it is for similar wino masses ($\sim 3$ TeV) that the present-day abundance of DM is naturally obtained, i.e. the wino is a thermal relic.  Unfortunately, as several groups have independently shown \cite{Cohen:2013ama,Fan:2013faa,Bauer:2014ula,Ovanesyan:2014fwa,Baumgart:2014saa,Baumgart:2015bpa}, thermal wino DM is now in severe tension with constraints on gamma-ray lines from the HESS experiment~\cite{Abramowski:2013ax}.  Nonetheless, we will consider here the phenomenology of heavy wino bound states, with the following motivations:
\begin{itemize}

\item The results of any indirect detection experiment come with large astrophysical uncertainties due to the poorly-constrained DM halo density profile.  Thus, we should continue to explore new phenomena that could allow for additional constraints.

\item Even if the wino is not a thermal relic, nature could still realize a high-scale MSSM as a means of resolving most of the hierarchy problem along with providing grand-unification.  Current and future Cherenkov telescopes like CTA and HAWC will set limits on DM masses up to 100 TeV and 1000 TeV, respectively, albeit with sensitivity less than the rates predicted for electroweak DM \cite{Doro:2012xx,Abeysekara:2014ffg,Harding:2015bua}.  We should explore the physics of electroweak WIMPs in this regime, even if the mechanism for providing their relic density is unspecified.

\item Dark-sector models have provided a WIMP DM candidate unshackled by the specific couplings of the SM.  It is worth considering scenarios where the hidden-sector gauge group is more complex than the dark U(1) of simple dark photon models (e.g. \cite{ArkaniHamed:2008qn,Chen:2009ab}), and the DM can be part of a nontrivial multiplet.  In such scenarios, the dark gauge group may feature large hierarchies between force carrier masses, just as we see in SU(2)$_L \times$U(1)$_Y$.  Our wino calculations are therefore a toy model for studying bound state physics in the presence of nearly-degenerate matter fields that may experience both long- and short-range forces, where the particle radiated in the formation of the bound state may be different from the force carrier primarily responsible for the potential.   Lastly, the nonabelian potential contains richer structures, including the ability of force carriers to emit radiation and the possibility of multiple attractive and repulsive channels.
\end{itemize}

\begin{figure}[h]
\begin{center}
\vspace{0.5cm}
\includegraphics[scale=0.5]{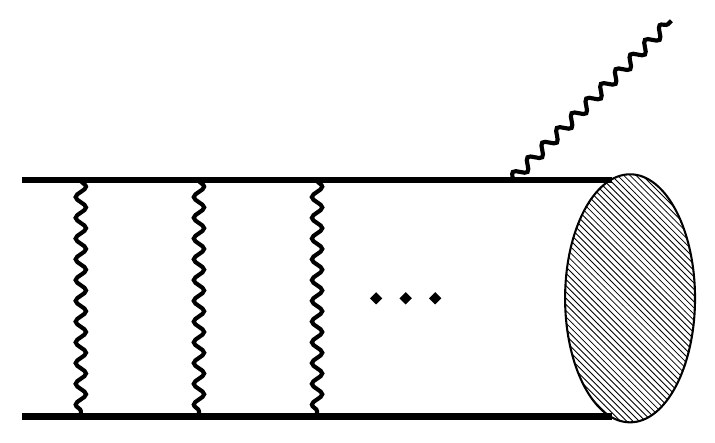}
\includegraphics[scale=0.5]{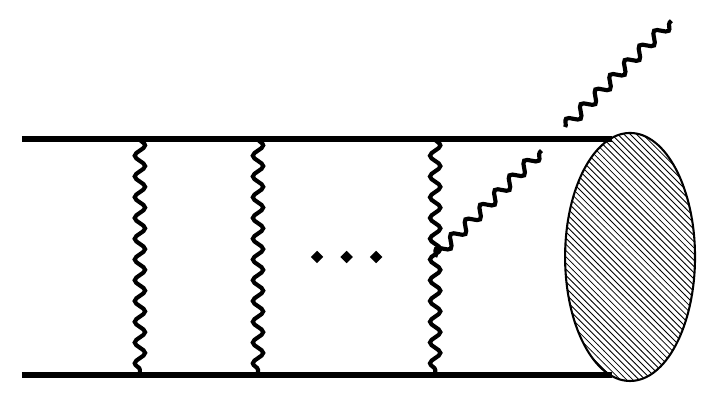}
\caption{WIMPs exchange a ladder of weak gauge bosons, which gives rise to a non-local potential in the nonrelativistic limit.  Finally, the dipole emission of a single photon can convert the initial, positive-energy scattering state to a negative-energy bound state, WIMPonium.  ({\bf Right:}) Since the potential contains charged force carriers, $W^\pm$, they can also emit radiation to capture into the bound state.}
\label{fig:capt}
\end{center}
\end{figure}

\begin{figure}[h]
\begin{center}
\vspace{0.5cm}
\includegraphics[scale=0.5]{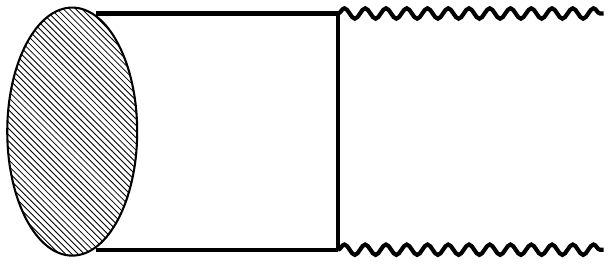}
\hspace{0.5cm}
\includegraphics[scale=0.5]{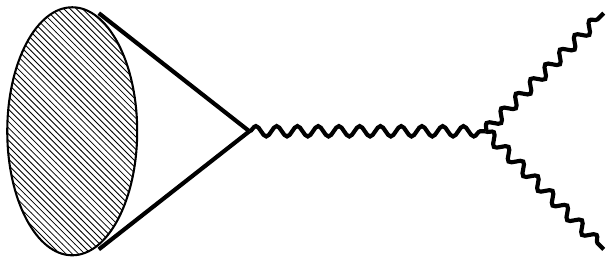}
\caption{Since we consider WIMPonium constituents that are charged under the electroweak gauge group, its lifetime is set by weak-scale physics.  Excited states typically transition to deeper bound states, but the deeper bound states will annihilate to SM particles.  We note that if the WIMPonium is in a $^1S_0$ state, $s$-channel annihilation through a gauge boson is forbidden.}
\label{fig:ann}
\end{center}
\end{figure}

In section \ref{sec:spectrum}, we discuss winos in the nonrelativistic limit, the potential that governs their evolution, and its spectrum of bound and continuum states.  In section \ref{sec:decay}, we develop the necessary formalism to calculate the rate of bound state formation by radiative capture (cf.~figure \ref{fig:capt}) in the case of wino DM, 
as well as the rates for bound states to transition among themselves and annihilate to SM particles (cf.~figure \ref{fig:ann}).  
In section \ref{sec:anr}, we apply the results of section \ref{sec:decay}, present our numerical results, and discuss observational possibilities, before presenting our conclusions in section \ref{sec:conclusion}.  

Finally, in appendices \ref{app:init}-\ref{app:bs} we detail our numerical procedures for computing wavefunctions. In appendices \ref{app:coulomb}-\ref{app:hulthen} we discuss two illuminating toy problems: (1) the nonabelian analogue of positronium, which the wino case approaches in the limit of very high DM mass, and (2) the bound states of the Hulth\'{e}n potential, which provides an analytically tractable approximation to the Yukawa potential and hence allows us to study the effects of reducing the range of the potential in a simple system. In appendix \ref{app:anndictionary} we discuss how to translate existing results in the literature to our formalism for WIMPonium annihilation, and in appendix \ref{app:integrals} we derive and present several useful integrals.

%%%%%%%%%%%%%%%%
\section{Winos in the nonrelativistic limit}
\label{sec:spectrum}
%%%%%%%%%%%%%%%%

The specific WIMP whose capture and annihilation we compute is an SU(2)$_L$ triplet Majorana fermion, denoted $\chi^a$, with mass $M_\chi$:\footnote{In previous versions of this article, a different convention was used for the relative sign between the two terms in the definition of covariant derivatives. Both these sign conventions are commonly used in the literature.}  
\be
\ml = i \chi^{a\, \dag} \left( \bar{\sigma}^\mu \partial_\mu \,-i\,g\, \bar{\sigma}^\mu W_\mu^b \, T^b_{ac} \right)\chi^c -\frac{1}{2} M_\chi (\chi^a \chi^a + {\rm h.c.})
\label{2.1}
\ee
We refer to it as the wino even though it is the only field beyond the SM we include.  One can think of it as either a minimal extension of the SM to provide DM or as the lightest supersymmetric particle (LSP) of an otherwise decoupled SUSY sector.  Although we are interested in the multi-TeV regime, it is necessary to include the effects of electroweak symmetry breaking in the $W$ and $Z$ masses and to work in the wino mass eigenstate basis, with the neutralino $\chi^0 = \chi^3$, and the chargino $\chi^\pm = \frac{1}{\sqrt{2}}(\chi^1 \mp i \, \chi^2)$.  There is a small, but important, mass splitting between the charged and neutral states, arising from radiative corrections from SM fields: 
\be
\delta M \equiv M_{\chi^\pm} - M_{\chi^0} \,=\, 165 \, {\rm MeV},
\label{2.2}
\ee
and we will take $M_{\chi^0} \equiv M_\chi$ \cite{Ibe:2012sx,Cohen:2013ama}.  

%%%
\subsection{General considerations and symmetries}
\label{subsec:gencon}
%%%

In the nonrelativistic limit of electroweak WIMPs, the interactions of the fermions with gauge bosons whose momenta have ``potential'' scaling, $(E, {\bf p}) \sim (M_\chi v^2,\, M_\chi v)$, can be integrated out to give a nonlocal potential.   Furthermore, for all of our processes of interest -- Sommerfeld-enhanced annihilation, capture into bound states, transitions between bound states, and annihilation of bound states to SM fields -- it is more useful to work with two-particle states, rather than single-particle quantum fields.  If the state has positive energy, it will be a plane-wave-normalized, two-particle state.\footnote{There is a subtlety in this normalization for states consisting of identical fermions, which must be appropriately antisymmetrized, as we will discuss below.} If it is a negative-energy bound state, then it will have the standard single-particle normalization (i.e. integrating over the norm-squared of the position-space wavefunction gives 1).  We will detail a formalism below that can handle both cases.

Whether the state is positive or negative-energy, the potential due to gauge boson exchange experienced by a two-particle state with even-$L+S$ is:
\bea
V_{L+S \, {\rm even}}(r) = \left( \begin{array}{cc}
0  & -\sqrt{2}\alpha_W \frac{e^{-m_W r}}{r}   \\
-\sqrt{2}\alpha_W  \frac{e^{-m_W r}}{r}  & \;\;  2\delta M -\frac{\alpha}{r} - \alpha_W c_W^2 \frac{ e^{-m_Z r}}{r}
\end{array} \right).
\label{eq:potltwo}
\eea
Here $L$ and $S$ denote the total orbital and spin angular momentum quantum numbers for the two-particle state, respectively (we will generally use upper-case letters to denote the quantum numbers of an arbitrary two-particle state, while using lower-case $nlm$ to label the quantum numbers of the bound states). For a detailed derivation of this potential and the construction of two-body quantum-mechanical states starting from the fully relativistic quantum field theory, see \cite{Hisano:2004ds,Baumgart:2014saa}.  This potential enters the Hamiltonian via,
\be
i \partial_t \Psi = H^0 \, \Psi = \left[ -\frac{\nabla_X^2}{4M_\chi} -\frac{\nabla_r^2}{M_\chi} + V(r) \right] \Psi,
\label{eq:hamil}
\ee
where $X$ is the center of mass coordinate and $\Psi$ is a two-component wavefunction,
\bea
\Psi = \left( \begin{array}{c}
\psi_N \, (\equiv \chi^0 \chi^0) \\
\psi_C \, (\equiv \chi^+ \chi^-)
\end{array} \right).
\label{eq:psidecomp}
\eea
The nonzero off-diagonal terms in $V_{L+S \, {\rm even}}(r)$ mix the charged and neutral components, so we must evolve them simultaneously.

As noted by \cite{Cirelli:2007xd, Beneke:2014gja}, in this basis even the lowest-order nonrelativistic potential is dependent on the spin and angular momentum of the two-particle states. The potential of eq.~\ref{eq:potltwo} applies to spin-singlet states with even $L$ and spin-triplet states with odd $L$. For spin-singlet states with odd $L$ or spin-triplet states with even $L$, so $L+S$ is odd, the wavefunction is symmetric and there can be no two-particle state consisting of the identical neutral fermions $\chi^0 \chi^0$; consequently, the potential is non-zero only for the charged two-particle state $\chi^+ \chi^-$,

\bea
V_{L+S \, {\rm odd}}(r) = \left( \begin{array}{cc}
0  & 0   \\
0 &  2\delta M -\frac{\alpha}{r} - \alpha_W c_W^2 \frac{ e^{-m_Z r}}{r}
\end{array} \right).
\label{eq:potlone}
\eea

We will not consider in this work the on-shell emission of $W$ or $Z$ bosons.  Since the parametric size of the binding energy $E_n \lesssim \mathcal{O}(\alpha_W^2 M_\chi)$, this process is kinematically forbidden for DM lighter than $\sim 100$ TeV.\footnote{We can estimate this a bit more precisely.  In this high mass limit, electroweak symmetry is approximately restored.  Thus, we just need the binding energy for a Coulomb potential with coupling $\alpha_W$, $E_n = -\frac{\alpha_W^2 M_\chi}{4n^2}$.  For our dominant single-photon capture at high masses to either $s$-wave or $p$-wave bound states, $n=2$, and sufficient energy to produce an on-shell $Z$ requires $M_\chi$ = 1284 TeV.  This is higher than the mass regime we study in detail, which goes up to 300 TeV.} Off-shell production of $W$ and $Z$ bosons which subsequently decay is allowed, but will be strongly suppressed relative to processes involving the emission of a photon, by a factor $\sim \frac{\alpha_W}{\pi} \left( \frac{E_n}{m_W} \right)^4$.  Accordingly, we only consider the $Q=0$ sector of two-particle states (i.e. the total electric charge of the state is zero).

Electric dipole transitions with single-photon emission do not flip the spin of the two-particle state, but change its angular momentum by $\Delta L = \pm 1$. Since the initial two-particle state, far from the point of interaction, will consist of neutral identical fermions, it must have even $L+S$ (the $s$-wave piece is purely spin-singlet; the $p$-wave piece is purely spin-triplet, etc). The two-particle state resulting from a single photon emission will then have odd $L+S$, and so must be purely $\chi^+ \chi^-$.  

Computing the capture rate, $\sigma v \, (\chi^0 \chi^0 \rightarrow {\rm WIMPonium + \gamma})$, will be very similar to the standard quantum-mechanical calculation of radiative transitions between hydrogenic bound states.  Instead of our initial state being negative-energy with a compact wavefunction, it will be a positive-energy solution to the Schr\"{o}dinger equation, eq.~\ref{eq:hamil}, with potential given by eq.~\ref{eq:potltwo}, and energy $M_\chi v_\text{rel}^2/4$ in the center-of-momentum (CM) frame.  Additionally, we will have to account for the fact that the potential itself is charged.  Although our Hamiltonian requires numerical analysis due to the Yukawa terms, one can calculate analytically the pure QED process for $e^+ e^-$ to bind into positronium after electric dipole emission \cite{1996JPhB...29.2135A,Pospelov:2008jd}.  In appendix \ref{app:coulomb}, we present exact analytic results for the SU(2) analog of positronium, with potentials corresponding to those in eqs.~\ref{eq:potltwo} and  \ref{eq:potlone} in the limit $\delta M,\, m_W,\, m_Z \rightarrow 0$.

For the WIMP bound states, we will need to find the negative-energy solutions with the single-component potential in eq.~\ref{eq:potlone}. Bound states supported by the potential of eq.~\ref{eq:potltwo} do exist, but cannot be accessed from our initial state by single-photon emission; nonetheless, we will discuss their properties. We can obtain parametric intuition for the effect of short-range potentials by studying the Hulth\'{e}n potential, a close cousin of the Yukawa.  We collect detailed results on this potential in appendix \ref{app:hulthen}.  

Note that our convention for zero energy is set by two $\chi^0$ particles far apart at rest; the $2\delta M$ term in eq.~\ref{eq:potlone} can therefore set the energies of some of the $\chi^+\chi^-$ bound states to be positive, although they would have negative energy in the alternate convention where zero is set by the constituents' rest masses at infinity.  We will briefly discuss the behavior of these ``positive-energy'' bound-states, although we do not expect them to be important for generic parameters.

%%%%%%%%%%%%%%%%%%%%
\subsection{The bound state spectrum in the high-mass limit}
\label{subsec:himass}
%%%%%%%%%%%%%%%%%%%%

Let us consider the spectrum of bound states present in the case where the SU(2)$_L$ symmetry is unbroken, the force carriers are massless, and there is no mass splitting between the charginos and neutralinos. The potential matrices simplify to:

\begin{align}
V(r) & = \frac{\alpha_W}{r} \left( \begin{array}{cc}
0  & -\sqrt{2}   \\
-\sqrt{2}  &   -1
\end{array} \right), \quad L + S \; \text{even}, \qquad
V(r) & = \frac{\alpha_W}{r} \left( \begin{array}{cc}
0  & 0   \\
0 &  -1
\end{array} \right), \quad L + S \; \text{odd}.
\label{eq:highmasslimit}
\end{align}
In this limit, the Hamiltonian can be diagonalized and the solutions to the Schr\"{o}dinger equation can be immediately written down in terms of the eigenstates of the Coulomb potential. For the case of odd $L+S$ this is trivial. For the case of even $L+S$, the matrix potential has eigenvalues $- \lambda_i \frac{\alpha_W}{r}$ where $\lambda_1 = 2$, $\lambda_2=-1$; the corresponding orthonormal eigenvectors are $\eta_1 = \begin{pmatrix} \sqrt{\frac{1}{3}} & \sqrt{\frac{2}{3}}  \end{pmatrix}$, $\eta_2 = \begin{pmatrix} - \sqrt{\frac{2}{3}} & \sqrt{\frac{1}{3}}  \end{pmatrix}$. The general solution to the Schr\"{o}dinger equation for even $L+S$ is given by (eq.~\ref{eq:sumsolapp}):
\begin{equation} 
\Psi({\bf r}) =  \sum_i A_i  \eta_i \, \phi(\lambda_i \alpha_W; {\bf r}) 
\label{eq:sumsol}
\end{equation}
where $\phi(\lambda_i \alpha_W; {\bf r})$ is the scalar function solving the Schr\"{o}dinger equation for a Coulomb potential, with coupling $\lambda_i \alpha_W$ The $A_i$ constants are chosen to ensure the appropriate boundary conditions for the incoming plane wave; see appendix \ref{app:coulomb} for further discussion. 
In particular, bound states cannot be supported by a repulsive Coulomb potential, so all bound states with even $L+S$ will be of the form $\phi(\lambda_1 \alpha_W; {\bf r}) \eta_1$. In this case $\lambda_1=2$, so the states have binding energies corresponding to a Coulomb potential with coupling $2 \alpha_W$ and reduced mass $\mu = M_\chi/2$, i.e. $E_n = \alpha_W^2 M_\chi/n^2$. The bound states with odd $L+S$ form a separate tower with wavefunctions of the form $\phi(\alpha_W; {\bf r}) \begin{pmatrix} 0 & 1  \end{pmatrix}$. Accordingly, their binding energies are $E_n = \alpha_W^2 M_\chi/4 n^2$. 

\begin{figure}[h]
\begin{center}
\includegraphics[scale=0.8]{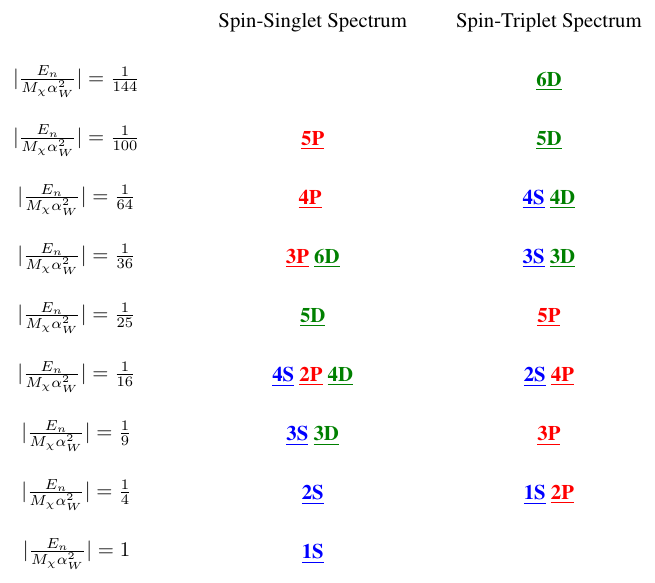}
\caption{The energy spectrum of bound states for each spin configuration in the large $M_\chi$ limit. The lowest four states for $L=0$ (\emph{blue}), $L=1$ (\emph{red}), and $L=2$ (\emph{green}) are included. For each spin configuration, the couplings in the $L$-even and $L$-odd potentials differ by a factor of two in the high-mass limit. This distorts the order of the bound states compared to a hydrogen atom. For the spin-singlet tower the $L=1$ bound states are pulled up to higher energies, while for spin-triplet they have been pushed down to lower energies. }
\label{fig:bs2}
\end{center}
\end{figure}

This means, for example, that the lowest-lying spin-singlet $L=1$ states are more weakly bound than the lowest-lying spin-singlet $L=2$ states; the former have odd $L+S$ and so have energy $E_2 = \alpha_W^2 M_\chi/16$, whereas the latter have even $L+S$ and so have binding energy $E_3 = \alpha_W^2 M_\chi/9$. Consequently, $n=2$ states may have multiple open decay channels, to $n=3$ states as well as $n=1$. The low-lying states for both spin-singlet and spin-triplet configurations are summarized in figure \ref{fig:bs2}.

%%%%%%%%%%%%%%%%%
\subsection{The bound state spectrum for all masses}
%%%%%%%%%%%%%%%%%

Beyond this high-mass limit, we must proceed numerically. We approximate the bound states as a linear combination of Coulombic wavefunctions, and solve for the coefficients of these basis states.  We exploit the fact that our bound-state potential (eq.~\ref{eq:potlone}) is rotationally symmetric, and thus $L$ is still a good quantum number.  This allows us to expand the solution for the full potential with fixed quantum numbers $(n,\,l)$ in terms of hydrogenic states with the same $L$, but summed over radial eigenvalues from $L-1$ up to some $n_{\rm max}$, beyond which the calculation is numerically stable.  Determining the coefficients of this expansion is a straightforward linear algebra exercise (cf.~eq.~\ref{eq:finproblem}).  Furthermore, in the limits $m_Z/M_\chi \rightarrow 0,\,\infty $, we recover a Coulombic potential with coupling $\alpha_W,\,\alpha$, respectively.  The details of our method are presented in appendix \ref{app:bs}.\footnote{We thank S.~Thomas for his help in developing this numerical procedure.}

\begin{figure}[h]
\begin{center}
\includegraphics[scale=0.5]{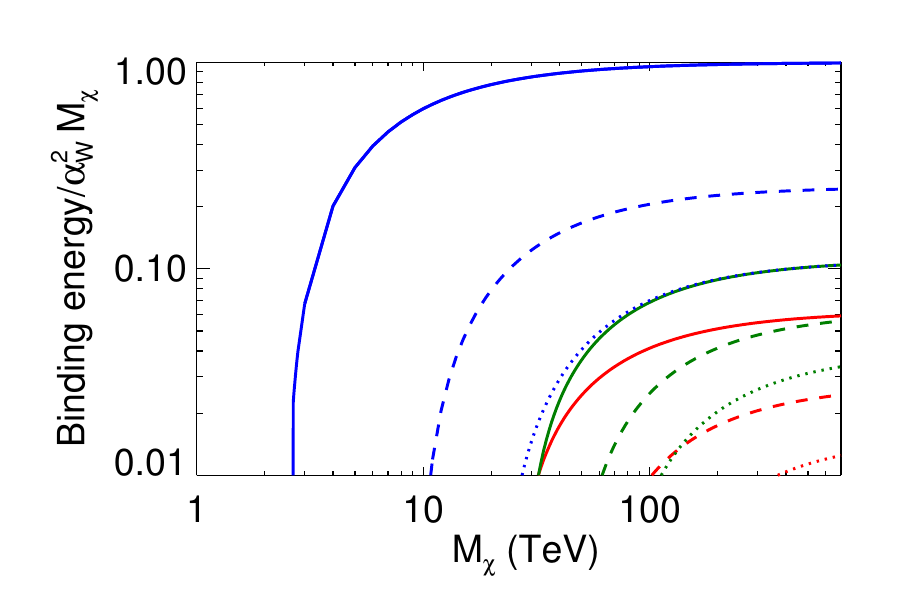}
\includegraphics[scale=0.5]{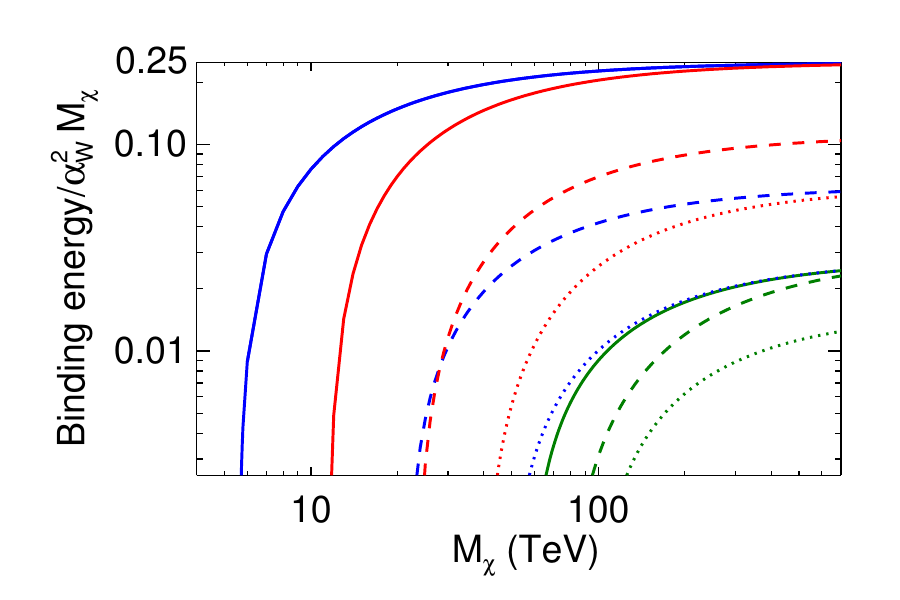}
\caption{SU(2)-triplet WIMPonium binding energies relative to $\alpha_W^2 M_\chi$.  {\bf Left:} Spectrum of spin-singlet states. {\bf Right:} Spectrum of spin-triplet states. \emph{Blue}, \emph{red}, and \emph{green} lines denote $s,\,p$ and $d$-wave bound states  respectively, with solid, dashed, dotted lines denoting the ranking in $n$ quantum number, where we have included only states with the three lowest $n$ for each partial wave.  In the high-mass limit, both potentials asymptote to Coulombic behavior, with effective coupling $2 \alpha_W$ for $L+S$-even states and $\alpha_W$ for $L+S$-odd states, and we recover the expected binding energies.}
\label{fig:bs1}
\end{center}
\end{figure}

We display the resulting spectrum of bound states in figure \ref{fig:bs1}. We will use these numerical wavefunctions to compute transition rates involving the bound states: between bound states, from bound states to SM particles, and from the initial free particles to the bound states.

We observe that the first negative-energy, spin-singlet bound state appears in the spectrum at $M_\chi \approx 2.6$ TeV, and the first negative-energy, spin-triplet bound state at $M_\chi \approx 5.6$ TeV. However, the  spin-singlet bound state cannot be accessed by single-dipole-photon capture from the initial state, since the $L$=1, $S$=0 continuum state is not populated by the identical fermionic DM particles (due to Fermi statistics). Spin-singlet configurations thus do not contribute to the single-photon capture rate until $M_\chi \gtrsim 25$ TeV, where the first accessible spin-singlet $p$-wave state appears.

At high DM masses, the spectrum of bound states in figure \ref{fig:bs1} converges to the limiting Coulombic case discussed above and displayed in figure \ref{fig:bs2}. At lower DM masses, however, the relative ordering of the states can shift.

%%%%%%%%%%%%%%%%%
\section{Formation, transitions and annihilation of WIMPonium}
\label{sec:decay}
%%%%%%%%%%%%%%%%%

Given an initial population of free neutralinos, bound states can form via radiative capture with the emission of a photon. Those bound states may subsequently decay to lower-energy states in the spectrum, or annihilate into SM particles. In this section we will develop the formalism for computing the relevant rates.

%%%%%%%%%%%%%%%%%
\subsection{Continuum-bound and bound-bound transitions}
\label{subsec:transition}
%%%%%%%%%%%%%%%%%

We calculate the rate for transitions between either continuum or bound states, with single photon emission, using time-ordered perturbation theory.  Our discussion parallels the treatment of radiative transition rates in \cite{weinberg2013lectures}. In the WIMP sector, our wavefunctions are eigenstates of the Hamiltonian constructed with $V(r)$ in eq.~\ref{eq:potltwo} for the initial state, and eq.~\ref{eq:potlone} for the final bound state:  
\bea
H^0_{L+S \, {\rm even}} \; \Psi_i &=& \frac{M_\chi v_\text{rel}^2}{4} \Psi_i \nn \\
H^0_{L+S \, {\rm odd}} \; \Psi_f \! \left[^{2S+1}L_J \right] &=& E_n \, \Psi_f \! \left[^{2S+1}L_J \right], 
\label{eq:schro}
\eea
where $E_n$ is the binding energy and $v_\text{rel}$ is the relative velocity of the two particles.  

Up to corrections that go like $M_\chi v_\text{rel}^4$, capture is kinematically possible if $E_n < M_\chi v_\text{rel}^2/4$. For the small velocities we consider, generally only bound states with $E_n < 0$ will be kinematically accessible. Accounting for the chargino's and $W$'s ability to radiate an on-shell photon, we obtain our full Hamiltonian in Coulomb gauge, 
\bea
H &=& H^0 + V_{\rm rad.} \nn \\
V_{\rm rad.} &=& \left( \sum_n \frac{e_n}{M_\chi} {\mathbf A}({\mathbf x}_n) \cdot {\mathbf p}_n + \sum_n \frac{e_n^2}{2 m_n} {\mathbf A}({\mathbf x}_n)^2 \right) \mathbb{P}_{CC} \nn \\
&&+ (-1)^{L_i + S}\,  \left(i \, \sqrt{2} \, e\, \alpha_W {\mathbf A}(0)  \cdot {\bf \hat{r}} \, e^{-m_W r} \right) \mathbb{P}_{NC},
\label{eq:fullhamil}
\eea
where $n$ labels the relevant chargino, with $e_n$ the signed EM coupling, and $e$ the coupling to a positive charge.\footnote{The relative sign in eq.~\ref{eq:fullhamil} between the contribution due to dipole emission off a single charged particle and that from radiation off of the potential has changed sign for processes with $(L_i +S)$-even compared to previous versions of the paper. This affects the rate for WIMP pairs to capture into bound states due to the interference between these terms.  This change has been propagated throughout the paper. We thank Kalliopi Petraki and Julia Harz for pointing out the earlier sign mistake.}  The sign convention for the coupling $e$ is chosen to be consistent with the covariant derivative convention in eq.~\ref{2.1}.  The factor in front of the $\mathbb{P}_{NC}$ term depends on the spin, $S$, and initial orbital angular momentum $L_i$ of the fermion pair. The relative spatial coordinate in our Hamiltonian, eq.~\ref{eq:hamil}, is given as
\be
{\mathbf r} = {\mathbf x}_1 - {\mathbf x}_2,
\ee
and in the CM frame, ${\mathbf x}_1 + {\mathbf x}_2 \,=\, 0$. The projectors $\mathbb{P}_{CC,NC}$ enforce that the interactions only couple the charged sector of the two-particle Hilbert space to itself, and neutral sector to the charged, respectively. For example, in the two-component Hilbert space of the $L+S$ even sector, the $\mathbb{P}_{CC}$ term only acts on the charged component, $\psi_{C}$ ({\it cf.}~eq.~\ref{eq:psidecomp}) of $\Psi_i$, which we will denote $\psi_{i,C}$.  This is just the standard, single-particle electric dipole coupling, familiar from atomic physics.  

\begin{figure}[h]
\begin{center}
\vspace{0.5cm}
\includegraphics[scale=1.25]{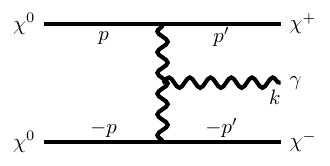}
\caption{WIMPs exchange ladder $W^\pm$ bosons, which can radiate ultrasoft, electric dipole photons.  This contributes to the $\Delta L = \pm 1$ capture rate to form WIMPonium.  Integrating out the $W\pm$ in this amplitude generates the effective operators eq.~\ref{eq:oppu}, \ref{eq:oppups}.  In the quantum mechanical, two-body Hamiltonian this gives rise to the $\mathbb{P}_{NC}$ term in eq.~\ref{eq:fullhamil}.}
\label{fig:potlus}
\end{center}
\end{figure}

The $\mathbb{P}_{NC}$ term accounts for the ability of the potential itself to emit electric dipole radiation.  The explicit $\alpha_W$ in this contribution makes it appear naively suppressed relative to the chargino dipole emission.  However, the ${\bf p}_n/M_\chi$ in the $\mathbb{P}_{CC}$ term brings in an expectation value of the WIMP velocity, $v \sim \alpha_W,$ where the matrix element is supported by the bound state wavefunction. Thus, both terms in eq.~\ref{eq:fullhamil} are {\it a priori} the same order and must be included.  
The dipole emission off the potential is an intrinsically nonabelian effect, known in the NRQCD literature, whose origin we now review \cite{Beneke:1999zr,Manohar:1999xd,Manohar:2000kr}.  It arises from the process shown in figure \ref{fig:potlus}, since our constituent WIMPs exchange charged force carriers.  
This contributes to the electric dipole transition, and thus the Fermi statistics and angular momentum considerations in section \ref{subsec:gencon} continue to hold.  The $W^\pm$ exchange connects the $\chi^0 \chi^0$ state to the $\chi^+ \chi^-$.  For the capture process, since the initial state contains both components, the amplitude for dipole emission off the potential involves only the neutral component of the initial wavefunction, $\psi_{i,N}$.  Unsurprisingly, in the nonrelativistic effective field theory (NREFT) description, integrating out the potential gauge boson in figure \ref{fig:potlus} gives a nonlocal operator that resembles the potential, but with an additional ladder propagator and a dipole coupling to the photon,  
\be
\mathcal{L}_{pu} \,=\, -\frac{4g^2}{\big ( ({\bf p^\prime - p})^2 + m_W^2 \big)^2} ({\bf p^\prime - p}) \cdot (e A_k) \left[ \left( \chi^{+\, \dag}_{p^\prime} \, \chi^{0}_{p}  \right) \left( \chi^{-\,\dag}_{-p^\prime} \, \chi^0_{-p} \right) \,+\, h.c. \right].
\label{eq:oppu}
\ee
The coupling $g$ is that of SU(2)$_L$, while $e$ is that of electromagnetism.  As these are nonrelativistic fields, each contains only creation or annihilation operators.\footnote{In NRQCD, the analogous operator describing gluon emission off the quark-antiquark potential is $\mathcal{L}^{QCD}_{pu} = \frac{2\,i\,g_s^2 f^{ABC}}{({\bf p^\prime - p})^4} ({\bf p^\prime - p}) \! \cdot \! (g_s A^C_k) \left[ \psi^{\dag}_{p^\prime} \,T^A \psi_{p} \right] \left[ \chi^{\dag}_{-p^\prime} \,\overline{T}^B \chi_{-p} \right],$ which in position space is $\mathcal{L}^{QCD}_{pu} = \alpha_s f^{ABC} \int d^3r \, \left[ \psi^{\dag}\,T^A \psi \right]\!(x+\vec{r}) \, \left[ \chi^{\dag}\,\overline{T}^B \chi \right]\!(x) \, \hat{r} \cdot \! \big(g_s A^C(t,0) \big)$ \cite{Beneke:1999zr,Manohar:1999xd,Manohar:2000kr}.}  The term explicitly written destroys two $\chi^0$s and creates a $\chi^\pm$ pair, while the conjugate term does the opposite.  One can find a more complete description of the field content and how it connects to two-particle, quantum mechanical states in the appendix of \cite{Baumgart:2014saa}.  This photon has $(k^0,\,\vec{k})\,\sim\, (M_\chi v^2,\, M_\chi v^2)$, and is thus ``ultrasoft'' in the NREFT terminology.  Since our scattering and bound-state wavefunctions are written in position space, it is easier to work with the Fourier transform of the operator in eq.~\ref{eq:oppu}, 
\be
\mathcal{L}_{pu} = -2i \,\alpha_W \int d^3r \left[ \chi^{+\,\dag}  \chi^0 \right]\!(x+\vec{r}) \, \left[\chi^{-\,\dag}\chi^0 \right]\!(x) \, \hat{r}\, e^{-m_W r} \big(e A(t,0) \big) \,+\, h.c.,
\label{eq:oppups}
\ee
where we note that the softness of the photon spatial momentum sets its position coordinate to the origin of space.  With the position-space operator, it is straightforward to use the quantum mechanical state definitions in the appendix of \cite{Baumgart:2014saa} to convert $\mathcal{L}_{pu}$ to a term in $V_{\rm rad.}$, the perturbative Hamiltonian that acts on our two-particle states, eq.~\ref{eq:fullhamil}.

We treat $V_{\rm rad.}$ as a perturbation, and capture from single-photon emission occurs at first order. 
We use nonrelativistic normalization for the initial and final states. 
We can act with the photon field in eq.~\ref{eq:oppups} to obtain an overlap integral in terms of the WIMP wavefunctions. Going to CM position, ${\mathbf X}$, and ${\mathbf r}$, the former trivially integrates to give a spatial-momentum $\delta$-function, which we evaluate in the CM frame. Together, these steps give an $S$-matrix element for capture, assuming an $L+S$-even initial state\footnote{Eq.~\ref{eq:smattwo} contains a mild abuse of notation as the photon field from the $\mathbb{P}_{NC}$ term in eq.~\ref{eq:fullhamil} is located at the spatial origin, and thus the ultrasoft photon spatial momentum does not give rise to the prefactor $\delta^{(3)}({\mathbf k} + {\mathbf P}_{\rm BS})$. Operationally though, this $\delta$-function just serves to remove the wimponium phase-space integral, $d^3 P_{\rm BS}$, in the cross section and the end result is the same with the formally correct factor for this term, $\delta^{(3)}({\mathbf P}_{\rm BS})$.} 
\bea
S_{i,\, f \gamma} &=& \frac{2\pi i}{\sqrt{2k(2\pi)^3}} \, \delta \! \left[ M_\chi v^2/4 - E_n - k \right] \delta^{(3)}({\mathbf k} + {\mathbf P}_{\rm BS})  \epsilon(\hat{k},\sigma)  \nn \\
&& \cdot \left( \frac{e}{M_\chi} \frac{1}{(2\pi)^{3/2}}  \int d^3 r \, \psi_{f,C}^*\left[{^{2S+1}L_J}\right]\!({\bf r}) \, (e^{-i {\mathbf k} \cdot {\mathbf r}/2} + e^{i {\mathbf k} \cdot {\mathbf r}/2}) (-i \nabla_{\mathbf r}) \, \psi_{i,C} ({\bf r}) \right. \nn \\
&& \left. + i \, \sqrt{2} \, e\, \alpha_W \frac{1}{(2\pi)^{3/2}} \int d^3 r \, \psi_{f,C}^*\left[{^{2S+1}L_J}\right]\!({\bf r}) \,  e^{-m_W r} \, {\bf \hat{r}} \, \psi_{i,N} ({\bf r}) \right).
\label{eq:smattwo}
\eea
where $P_{\rm BS}$ is the momentum of the bound state and ${\bf k}$ is the momentum of the emitted photon. 
We now have a factor of the photon polarization, $\epsilon(\hat{k},\sigma)$, that we will ultimately sum over upon squaring the amplitude and obtaining the capture rate.  The factor of $1/(2\pi)^{3/2}$ in front of the integral arises from our convention on wavefunction normalization.\footnote{In the free-theory limit, our continuum state would be a plane wave, $\psi_{i,C} ({\bf r}) = e^{i {\mathbf p} \cdot {\mathbf r}}$.  The $(2\pi)^{-3/2}$ we have pulled out of the integral in eq.~\ref{eq:smattwo}, is a factor giving the normalized continuum state $\Psi _i = (2\pi)^{-3/2}\psi_{i,C} ({\bf r})$.  The benefit of this convention is that we get a simple inner product for our continuum states, $\int d^3r \, \Psi_{i\,\mathbf{p^\prime}}^\dag({\bf r})  \, \Psi_{i\,\mathbf p}({\mathbf r}) = \delta^{(3)}({\bf p - p^\prime})$, which one can check trivially holds for the plane-wave case.}  We can make use of the dipole approximation, $e^{i {\mathbf k} \cdot {\mathbf r}/2} \approx$ 1, which holds in our regime of interest.  The bound state wavefunctions,  $\psi_{^{2S+1}L_J} (r)$ die off exponentially after a few Bohr radii, $\sim 1/(\alpha_W M_\chi)$, while the photon energy is set by the binding energy, $\sim \alpha_W^2 M_\chi$.  Thus, over the integral's domain of support, the exponent is small. 

To get the differential rate to capture to the two-particle final state of photon and WIMPonium, we strip the $\delta$-functions from the $S$ matrix in eq.~\ref{eq:smattwo} and integrate the bound-state phase space to get 
\bea
(d \sigma)v_\text{rel} &=& (2\pi)^2 \,  \mu_f \, k \, |M|_{i,\, f \gamma}^2 \, d\Omega_k, \quad  \mathrm{where} \nn \\
S_{i,\, f \gamma} &=& \delta \left( M_\chi v^2/4 - E_n - k \right)  \delta^{(3)}({\mathbf k} + {\mathbf P}_{\rm BS}) \, M_{i,\, f \gamma},
\label{eq:mmatdef}
\eea
and $\mu_f = k \, E_{\rm BS}/(k+E_{\rm BS}) \approx k$ is the final-state reduced energy, including the rest mass.
When computing the rate for decay from one bound state to another through emission of a single dipole photon, the calculation is identical, except that we replace $(d\sigma) v_\text{rel}$ with $d\Gamma$.  Additionally, if the initial bound state is $(L_i + S)$-odd, then one must flip the sign of the radiation-from-potential term, as indicated in eq.~\ref{eq:fullhamil}. For capture, the initial state wavefunction is dimensionless, and as mentioned in the above footnote, normalized so that $\int d^3 r \, \Psi_{\bf p^\prime}({\bf r})^\dagger \Psi_{\bf p}({\bf r}) = \delta^{(3)}({\bf p} - {\bf p}^\prime)$. For bound-bound state transitions, however, both initial and final state wavefunctions are normalized such that $\int d^3 r |\Psi({\bf r})|^2 = 1$, and thus the wavefunctions have units of (mass)$^{3/2}$. Thus the matrix element $M_{i, \, f \gamma}$ has units of (mass)$^{-2}$ in the case of capture into a bound state, and units of (mass)$^{-1/2}$ in the case of transitions between bound states. This yields the correct dimensions for $\Gamma$ and $\sigma v_\text{rel}$ (mass and mass$^{-2}$ respectively).

To summarize, in the dipole approximation we have: 
\begin{align} 
& \sigma v_\text{rel} \, \text{(continuum $\rightarrow$ bound) or} \, \Gamma \, \text{(bound $\rightarrow$ bound)} \nonumber \\
& =\frac{2\, \alpha}{\pi} \frac{k}{M_\chi^2}  \int d\Omega_k \left| \epsilon(\hat{k},\sigma) \cdot \int d^3 r \, \left( \psi_{E,C}^*({\bf r}) \nabla_{\mathbf r} \, \psi_{O,C} ({\bf r}) 
+ \frac{\alpha_W\, M_\chi \, e^{-m_W\, r}}{\sqrt{2}} \, \psi_{E,N}^*({\bf r}) \, {\bf \hat{r}} \, \psi_{O,C} ({\bf r}) 
\right) \right|^2, 
\label{eq:dipole}
\end{align}
where subscript $E$ ($O$) on the wavefunctions refers to states with even (odd) $L+S$, 
$\alpha = e^2/4\pi$, $k$ is the energy of the emitted photon ($k = -E_n + M_\chi v_\text{rel}^2/4$ in the case of capture, or the difference in binding energies in the case of a bound-bound state transition).  We note that eq.~\ref{eq:dipole} holds regardless of whether the $L+S$-even and $L+S$-odd states are initial or final.

For states of known initial and final angular momentum, we can perform the angular integral and reduce the necessary calculation to a one-dimensional integral over the radial wavefunctions, which we compute numerically as described in appendices \ref{app:init} and \ref{app:bs}. This procedure is particularly simple where either the initial or final state is $s$-wave, since (using integration by parts) we avoid the need to apply $\nabla_{\bf r}$ to a wavefunction with non-trivial angular dependence.\footnote{A common procedure in radiative transition calculations is to convert the expectation value of $\nabla_{\mathbf r}$ to ${\mathbf r}$ by the relation $\left[ H^0, {\mathbf r} \right] = -i \, {\mathbf p}/M$, which converts $\bra{f} -i \, \nabla_{\mathbf r} \ket{i} = M (E_f - E_i) \bra{f} {\mathbf r} \ket{i}$.  However, we cannot make use of this in a straightforward way in our capture or transition calculations as the Hamiltonian acting on our initial and final states is different.} In particular, for illustration, let us consider transitions between (continuum or bound) $s$-wave and $p$-wave states, where the integral in the first term of Eq.~\ref{eq:dipole} to be computed takes the form:
\begin{align} \int d^3 r \, \psi^*_{f,C}({\bf r}) \nabla_{\mathbf r} \, \psi_{i,C} ({\bf r})  & \rightarrow  \int d^3 r \, \phi^*_{L=1}(r) \, Y^*_{1m}(\theta,\phi) \, {\bf \hat{r}} \frac{\partial}{\partial r} \left[Y_{00}(\theta,\phi) \phi_{L=0}(r) \right] \nonumber \\
&= \frac{1}{\sqrt{4\pi}} \int d\Omega  \, Y_{1m}(\theta,\phi)^* \, {\bf \hat{r}} \int r^2 dr \, \phi^*_{L=1}(r) \phi_{L=0}'(r),
\label{eq:overlapsphharm}
\end{align}
where we have written the full wavefunctions $\psi_{C,N}(\vec{r}) = \phi_{C,N}(r) Y_{Lm}(\theta,\phi)$, using $\phi(r)$ to denote the radial wavefunctions, and $m=L_z$ labels the magnetic quantum number.  The second term, arising from dipole emission off the potential, follows from eq.~\ref{eq:overlapsphharm} by replacing $\partial_r \rightarrow \alpha_W M_\chi/\sqrt{2}$ and adjusting the (charged vs neutral) wavefunction components.

Since we are considering $p$-wave states, it is useful to write the unit vector ${\bf \hat{r}}$ in a basis of $L=1$ spherical harmonics,
\be
{\bf \hat{r}} = -\sqrt{\frac{4\pi}{3}} Y_{11} \, \hat{r}_{-1} -\sqrt{\frac{4\pi}{3}} Y_{1,-1} \, \hat{r}_1 + \sqrt{\frac{4\pi}{3}} Y_{10} \, \hat{r}_0,
\label{eq:rdecomp}
\ee
where $\hat{r}_0 =z$, $(\hat{r}_{-1} - \hat{r}_1)/\sqrt{2} = \hat{x}$, and $i\, (\hat{r}_{-1} + \hat{r}_1)/\sqrt{2} = \hat{y}$.  Thus, for a given $m$ in the $p$-wave wavefunction, only one of the terms in eq.~\ref{eq:rdecomp} will be nonvanishing.  Additionally, since we will be squaring the matrix element and summing over photon polarizations, we can make use of the identity
\be
\sum_\sigma \epsilon_i (\hat{k},\sigma) \epsilon_j^* (\hat{k},\sigma) = \delta_{ij} - \hat{k}_i \hat{k}_j.  
\label{eq:polsum}
\ee
Since the different $m$ states sum incoherently, the following angular overlap integrals will enter into the final cross section:
\begin{align}
(1-\hat{k}_0^2) \left[ \int d\Omega \sqrt{\frac{4\pi}{3}} Y_{10}^2 \right]^2 &= \frac{4\pi}{3} \sin^2 \theta_k & m &= 0 \nn \\
(1 + \hat{k}_{1} \hat{k}_{-1}) \left[ \int d\Omega \sqrt{\frac{4\pi}{3}} Y_{11} \, Y_{1,-1} \right]^2 &= \frac{4\pi}{3} \left( 1 - \frac{\sin^2 \theta_k}{2} \right) & m &= 1 \; {\rm or} \, -1,
\label{eq:angavg}
\end{align}
where we have used the fact that $Y^*_{1\pm1} = -Y_{1\mp 1}$ and $\hat{r}_{-1} \cdot \hat{r}_{1} = -1$.

Accordingly, when summing over $m$ states we obtain an overall factor of $8\pi/3$ from the angular integral including the insertion and sum over polarization vectors.  For initial states other than $s$-wave, a difference arises between capture and transition involving which $m$ states are included.  For the capture process, our initial state is asymptotically an incoming plane wave, $\Psi_i \propto e^{ikz}$.  This has no angular momentum about the direction of travel and therefore $m=0$.  Since our potential, eq.~\ref{eq:potltwo}, is spherically symmetric, the full wavefunction only has a $Y_{L0}$ component, and we do not average over initial polarizations.  The on-shell photon emission breaks the rotational symmetry and we can therefore capture into bound states with arbitrary $m$.  Thus, for any process with a WIMPonium initial state, we consider all $Y_{Lm}$ and average over $m$, dividing by 1/($2L+1$).  In practice though, both processes just give a factor of 1/($2L+1$) relative to the case of an initial $s$-wave state (in fact, the rate for transitions from a $p$-wave state to an $s$-wave state is independent of the initial value of $m$, so the average is trivial).  Consequently transitions between $s$- and $p$-wave states have rates given by: 
\begin{align} 
& \sigma v_\text{rel} \, \text{(continuum $\rightarrow$ bound) or} \, \Gamma \, \text{(bound $\rightarrow$ bound)}, \, \textbf{spin-triplet} \nonumber \\
& = \frac{16}{3} \frac{\alpha \, k}{M_\chi^2}  \left| \int r^2 dr \,  \left( \phi^*_{C, L=1}(r)\partial_r \, + \phi^*_{N, L=1}(r) \frac{\alpha_W M_\chi\, e^{-m_W\, r }}{\sqrt 2} \right) \phi_{C,L=0}(r) \right|^2 \times \left\{ \begin{array}{cc} 1 &  \text{initial $s$-wave} \\ 
1/3 & \text{initial $p$-wave} \end{array} \right. , \nonumber \\
& \sigma v_\text{rel} \, \text{(continuum $\rightarrow$ bound) or} \, \Gamma \, \text{(bound $\rightarrow$ bound)}, \, \textbf{spin-singlet} \nonumber \\
& = \frac{16}{3} \frac{\alpha \, k}{M_\chi^2}  \left| \int r^2 dr \,   \phi^*_{C, L=1}(r) \left(\partial_r \phi_{C,L=0}(r) - \frac{\alpha_W M_\chi\, e^{-m_W\, r }}{\sqrt 2} \phi_{N,L=0}(r) \right)  \right|^2 \times \left\{ \begin{array}{cc} 1 &  \text{initial $s$-wave} \\ 
1/3 & \text{initial $p$-wave} \end{array} \right. 
\label{eq:sigmastop} 
\end{align}
As a reminder, this rate includes a summation over all possible values of $m$ for the final state (this is the origin of the relative factor of 3 between the process with a $p$-wave final state and the one with an $s$-wave final state).

Repeating this calculation for transitions between $p$-wave and $d$-wave states yields: 
\begin{align} 
& \sigma v_\text{rel} \, \text{(continuum $\rightarrow$ bound) or} \, \Gamma \, \text{(bound $\rightarrow$ bound)}, \, \textbf{spin-singlet} \nonumber \\
& = \frac{16}{3} \frac{\alpha \, k}{M_\chi^2}  \left| \int r^2 dr \,  \left( \phi^*_{C, L=2}(r) \left( -\frac 1 r + \partial_r \right) \, + \phi^*_{N, L=2}(r) \frac{\alpha_W M_\chi\, e^{-m_W\, r }}{\sqrt 2} \right) \phi_{C,L=1}(r) \right|^2 \times \left\{ \begin{array}{cc} 1/3 &  \text{initial $p$-wave} \\ 
1/5 & \text{initial $d$-wave} \end{array} \right. , \nonumber \\
& \sigma v_\text{rel} \, \text{(continuum $\rightarrow$ bound) or} \, \Gamma \, \text{(bound $\rightarrow$ bound)}, \, \textbf{spin-triplet} \nonumber \\
& = \frac{16}{3} \frac{\alpha \, k}{M_\chi^2}  \left| \int r^2 dr \,   \phi^*_{C, L=2}(r) \left( \Big( -\frac 1 r + \partial_r \Big) \,  \phi_{C,L=1}(r) - \frac{\alpha_W M_\chi\, e^{-m_W\, r }}{\sqrt 2} \phi_{N,L=1}(r) \right)  \right|^2 \times \left\{ \begin{array}{cc} 1/3 &  \text{initial $p$-wave} \\ 
1/5 & \text{initial $d$-wave} \end{array} \right. 
\label{eq:sigmaptod} 
\end{align}
In this case, the transition rate does depend on $m$ for the initial and final states. To obtain the quoted $m$-independent rate/cross section we have summed over final $m$ and averaged over initial $m$ (note that after summing over final $m$ the transition rates are independent of initial $m$, and likewise after averaging over initial $m$ the transition rates are independent of final $m$). In appendix \ref{subapp:coulombtransition} we calculate the rate for a number of transitions, including $p \rightarrow d$ transitions, broken down by initial and final $m$.

These results all assume a specific spin state. This makes sense for bound-bound transitions, where states have definite total spin $S=0$ or $S=1$, but for the initial capture generically both spin-singlet and spin-triplet $\chi^0 \chi^0$ pairs will be present, in a ratio of 1:3 (singlet:triplet). As discussed above, the initial state must have even $L+S$ to admit a $\chi^0 \chi^0$ component, so once $L$ for the initial state is specified, there are contributions to the capture rate only from the spin-singlet pairs (even $L$) or the spin-triplet pairs (odd $L$). To obtain the overall spin-averaged capture rate, the rates above should therefore be multiplied by $1/4$ (even initial $L$) or $3/4$ (odd initial $L$).

Let us briefly discuss the boundary condition on the radial continuum wavefunctions. The asymptotic incoming state should be a plane wave with unit normalization, with support only in $\psi_N(r)$ at sufficiently large $r$.\footnote{As we discuss in appendix \ref{app:init} on calculating the positive-energy wavefunctions, in most of the parameter space we consider, only the neutral component of $\Psi$, $\psi_{N}$, scales like a Bessel function at large radii.  Because of the mass-shift, the charged component of the state $\psi_{C}$ is always off-shell and decays exponentially with distance.  It is straightforward to generalize to the case with non-decaying $\psi_C$, as the incoming, asymptotically plane-wave DM state is still purely in $\psi_N$.} However, because our initial condition corresponds to a pair of identical Majorana fermions, the incoming plane wave state must be antisymmetrized appropriately. For spin-singlet states, the spatial wavefunction must be symmetric, while for spin-triplet states, it must be antisymmetric. Using the asymptotic expansion of a plane wave propagating in the $z$-direction:
\begin{equation} e^{ikz} \rightarrow \frac{1}{2ikr} \sum_L (2L+1) P_L(\cos\theta) \left(e^{ikr} - (-1)^L e^{-ikr} \right), \end{equation}
we see that the appropriately (anti)symmetrized plane wave has the asymptotic form:
\begin{align} \text{spin-singlet:} \, \frac{1}{\sqrt{2}} \left( e^{ikz} + e^{-ikz} \right)  & \rightarrow \frac{1/\sqrt{2}}{2ikr} \sum_L (2L+1) P_L(\cos\theta) \left(e^{ikr} - e^{-ikr} \right) \left(1 + (-1)^L \right) \nonumber \\ & \rightarrow \sum_{L \, \text{even}} \sqrt{2} \sqrt{4 \pi (2L+1)} Y_{L0}(\theta,\phi) \frac{\sin(kr)}{kr},  \nonumber \\
\text{spin-triplet:} \, \frac{1}{\sqrt{2}} \left( e^{ikz} - e^{-ikz} \right)  & \rightarrow \frac{1/\sqrt{2}}{2ikr} \sum_L (2L+1) P_L(\cos\theta) \left(e^{ikr} + e^{-ikr} \right) \left(1 - (-1)^L \right) \nonumber \\ & \rightarrow \sum_{L \, \text{odd}} \sqrt{2} \sqrt{4 \pi (2L+1)} Y_{L0}(\theta,\phi) \frac{\cos(kr)}{ikr}, \end{align}
where we have used the fact that $P_L(\cos\theta) = \sqrt{\frac{4\pi}{2L+1}} Y_{L0}(\theta,\phi)$.

Thus at large $r$, the incoming piece of our continuum wavefunction for fixed $L$ should be normalized as 
\begin{equation} 
\psi_N({\bf r}) \rightarrow Y_{L0}(\theta,\phi) \left[ \sqrt{2} \sqrt{4\pi\, (2L+1)} \, \sin(p\,r)/(p \, r) \right], \quad r \rightarrow \infty, 
\label{eq:norm}
\end{equation} 
for even $L$, and with the same normalization except with a phase shift for odd $L$. Here $p=M_\chi v_\text{rel}/2$. Note this normalization is a factor of $\sqrt{2}$ higher than the standard normalization for the partial-wave components of the $e^{ikz}$ plane wave, because only half the partial waves are non-zero as a result of spin statistics.

%%%%%%%%%%%%%%%%%%%%%%%%%%%%%%%%%%
\subsection{Decay through annihilation to SM final states}
\label{subsec:SM decay}
%%%%%%%%%%%%%%%%%%%%%%%%%%%%%%%%%%

The bound states can also decay through annihilation to SM final states. We will proceed by writing the bound states in terms of free-particle states, but the normalization factor for the states depends on whether the particles involved are distinguishable or indistinguishable. In the center-of-mass frame we have:
\begin{align}
|\psi\rangle & = \sqrt{\frac{1}{2 \mu}} \int \frac{d^3 p}{(2\pi)^3} \psi(p) |{\bf p}, -{\bf p}\rangle \quad \text{(distinguishable particles)}, \nonumber \\
& = \sqrt{\frac{1}{4 \mu}} \int \frac{d^3 p}{(2\pi)^3} \psi(p) |{\bf p}, -{\bf p}\rangle \quad \text{(identical particles)}, 
\label{eq:decay.1} 
\end{align}
where $\mu =M_\chi/2$ is the reduced mass of the two-particle state. 

The tree-level annihilation cross sections for wino DM to SM final states have been computed previously in the literature, including the separate $s$-wave and $p$-wave contributions \cite{Hellmann:2013jxa}. The standard calculation assumes plane-wave initial states; in order to determine the decay rate of the bound states via annihilation, we will write the matrix element for the bound state decay in terms of the matrix elements for free-particle annihilation, following the standard procedure (e.g. \cite{Peskin:257493}). To whit, for a bound state $B$ and final state $f$, and working in the center-of-momentum frame, we write:

\begin{align} \mathcal{M}(B \rightarrow f) & = \sqrt{\frac{1}{2 \mu}} \int \frac{d^3 p}{(2\pi)^3} \psi(p) \mathcal{M}(\chi({\bf p}) \chi  (- {\bf p})\rightarrow f) \quad \text{(distinguishable particles)}, \nonumber \\
& = \sqrt{\frac{1}{4 \mu}} \int \frac{d^3 p}{(2\pi)^3} \psi(p) \mathcal{M}(\chi({\bf p}) \chi  (- {\bf p})\rightarrow f) \quad \text{(identical particles)}, 
\label{eq:decay.11}
\end{align}
where $ \mathcal{M}(\chi({\bf p_1}) \chi  ({\bf p_2})\rightarrow f)$ is the matrix element for annihilation of free particles with momenta ${\bf p}$, $- {\bf p}$ to final state $f$. The differing normalizations for identical and non-identical particles arise from the differing normalizations of the bound states (eq.~\ref{eq:decay.1}).

In the case of states with odd $L+S$, the bound state is composed purely of the $\chi^+ \chi^-$ two-particle state, and we need only use the result for distinguishable particles. For even $L+S$, the annihilation may proceed from either the $\chi^0 \chi^0$ or $\chi^+ \chi^-$ components of the bound state, and the matrix elements will add coherently. Thus, we should write:
\begin{align} \mathcal{M}(B \rightarrow f) & = \sqrt{\frac{1}{2 \mu}} \int \frac{d^3 p}{(2\pi)^3} \left[ \frac{1}{\sqrt{2}} \psi_N(p) \mathcal{M}(\chi^0({\bf p}) \chi^0  (- {\bf p})\rightarrow f) + \psi_C(p) \mathcal{M}(\chi^+({\bf p}) \chi^-  (- {\bf p})\rightarrow f) \right]. 
\label{eq:decay.12}
\end{align}
In the more general case where the bound state is composed of more than two distinct two-particle states, one should add all the matrix elements coherently, with normalizations determined by whether the particles are identical or not.  

Now let us consider the two simplifying cases where the bound state is $s$-wave or $p$-wave. In the case of $s$-wave annihilation, the matrix element for free-particle annihilation is independent of ${\bf p}$ in the small-$p$ nonrelativistic limit (and the wavefunction, which weights the integral, is suppressed for large $p$), and thus we can take it outside the integral. Since $\psi({\bf r}) = \int \frac{d^3 p}{(2\pi)^3} \psi({\bf p}) e^{i {\bf p} \cdot {\bf r}}$, it follows that $\int \frac{d^3 p}{(2\pi)^3} \psi({\bf p}) = \psi({\bf r} = 0)$. Thus we have, in the nonrelativistic limit: 
\begin{align}
\mathcal{M}^{L=0}(B \rightarrow f) & = \sqrt{\frac{1}{2 \mu}} \left[ \frac{1}{\sqrt{2}} \psi_N({\bf r} = 0) \mathcal{M}^{L=0}(\chi^0\chi^0 \rightarrow f) + \psi_C({\bf r} = 0) \mathcal{M}^{L=0}(\chi^+ \chi^- \rightarrow f) \right]. 
\label{eq:decay.13}
\end{align}

In the $p$-wave case, the matrix element for free-particle annihilation scales linearly with ${\bf p}$ in the limit of small $p$. Thus, the integrals over $d^3 p$ will take the form: 
\begin{equation}  \int \frac{d^3 p}{(2\pi)^3}  \psi({\bf p}) {\bf p} = \lim_{\bf r\rightarrow 0} \left(-i \nabla_{{\bf r}} \int \frac{d^3 p}{(2\pi)^3}  \psi({\bf p}) e^{i {\bf p} \cdot {\bf r}} \right) = -i \lim_{\bf r\rightarrow 0} \nabla_{{\bf r}} \psi({\bf r}).
\label{eq:plimit} 
\end{equation}
Furthermore, the $L=1$ wavefunctions have a universal form at small $r$: 
\bea
\psi_N({\bf r}) &=& \sqrt{\frac{4\pi}{3}} A_N \, r \, Y_{Lm}(\theta,\phi), \nn \\
\psi_C({\bf r}) &=& \sqrt{\frac{4\pi}{3}} A_C \, r \, Y_{Lm}(\theta,\phi). 
\label{eq:loneuniversal}
\eea
Accordingly, $\nabla_{{\bf r}}\psi_N({\bf r}) = A_N \, \hat{r}_m$, and similarly $\nabla_{{\bf r}}\psi_C({\bf r}) = A_C \, \hat{r}_m$.
Thus, we can write the matrix element for annihilation from the $p$-wave bound state in the form:
\begin{align}
\mathcal{M}^{L=1}(B \rightarrow f) & = \sqrt{\frac{1}{2 \mu}} \left[ \frac{1}{\sqrt{2}} \mathcal{M}^{L=1}(\chi^0({\bf p}) \chi^0(-{\bf p}) \rightarrow f)|_{{\bf p} \rightarrow i \nabla_{{\bf r}} \psi_N({\bf r})|_{{\bf r}=0} } \right. \nonumber \\
& \left. + \, \mathcal{M}^{L=1}(\chi^+({\bf p}) \chi^-(-{\bf p}) \rightarrow f)|_{{\bf p} \rightarrow i \nabla_{{\bf r}} \psi_N({\bf r})|_{{\bf r}=0} } \right],
\nonumber \\
 & = i \sqrt{\frac{1}{2 \mu}} \begin{pmatrix} \frac{1}{\sqrt{2}}  \hat{r}_m \cdot \mathcal{M}_0^{L=1}(\chi^0 \chi^0 \rightarrow f), \quad &  \hat{r}_m \cdot \mathcal{M}_0^{L=1}(\chi^+ \chi^- \rightarrow f) \end{pmatrix}  
 \begin{pmatrix} A_N \\  A_C \end{pmatrix}
\label{eq:lonedecay}
\end{align}
where the $\mathcal{M}_0$ matrix elements are vectorial but momentum-independent, and satisfy $\mathcal{M}^{L=1}(\chi \chi \rightarrow f)  = {\bf p} \cdot \mathcal{M}_0^{L=1}(\chi \chi \rightarrow f)$. We can use eq.~\ref{eq:plimit} to replace ${\bf p}$ in eq.~\ref{eq:lonedecay} as the dependence of the matrix element, $\mathcal{M}^{L=1}$  is linear in $\bf p$. 

The decay width for the bound state due to these annihilation processes is:
\begin{equation} \Gamma = \frac{1}{2 M_B} \int d\Pi_n \left|\mathcal{M}(B \rightarrow f) \right|^2, 
\label{eq:decay.16}
\end{equation}
where $M_B \approx 2 M_\chi$ is the mass of the bound state and $\Pi_n$ denotes the final state integral over phase space.

For $s$-wave annihilation we can therefore write:
\begin{align} \Gamma^{L=0} & = \frac{1}{4 M_\chi^2} \int d\Pi_n \left| \frac{1}{\sqrt{2}} \psi_N({\bf r} = 0) \mathcal{M}^{L=0}(\chi^0\chi^0 \rightarrow f) + \psi_C({\bf r} = 0) \mathcal{M}^{L=0}(\chi^+ \chi^- \rightarrow f)  \right|^2, \nonumber \\ 
& = \begin{pmatrix} \psi_N^*({\bf r} = 0) & \psi_C^*({\bf r} = 0) \end{pmatrix} \Sigma_{L=0}(f) \begin{pmatrix} \psi_N({\bf r} = 0) \\ \psi_C({\bf r} = 0) \end{pmatrix}, \nonumber \\
\Sigma_{L=0}(f) & \equiv \frac{1}{(2 M_\chi)^2} \begin{pmatrix} \frac{1}{2} \int d\Pi_n | \mathcal{M}(\chi^0\chi^0 \rightarrow f) |^2 & \Sigma_{L=0}^{12}(f) \\ 
(\Sigma_{L=0}^{12}(f))^* &  \int d\Pi_n | \mathcal{M}(\chi^+\chi^- \rightarrow f) |^2  \end{pmatrix}, \nonumber \\
\Sigma_{L=0}^{12}(f) & \equiv \frac{1}{\sqrt{2}} \int d\Pi_n  \mathcal{M}^*(\chi^0\chi^0 \rightarrow f)  \mathcal{M}(\chi^+\chi^- \rightarrow f)),
\label{eq:sann}
\end{align}
where in the last line all matrix elements are $s$-wave but we have omitted the $L=0$ superscripts for notational convenience.
 
Similarly, the decay rate corresponding to $p$-wave annihilation is:
\begin{align} \Gamma^{L=1} & = \frac{1}{M_\chi^2} \begin{pmatrix} A_N^* & A_C^* \end{pmatrix} \Sigma_{L=1}(f) \begin{pmatrix} A_N \\ A_C \end{pmatrix}, \nonumber \\
\Sigma_{L=1}(f) & \equiv \frac{1}{(2 M_\chi)^2} \begin{pmatrix} \frac{1}{2} \int d\Pi_n |  M_\chi \hat{r}_m \cdot \mathcal{M}_0(\chi^0 \chi^0 \rightarrow f) |^2 & \Sigma_{L=1}^{12}(f) \\  (\Sigma_{L=1}^{12}(f))^* &  \int d\Pi_n | M_\chi \hat{r}_m \cdot \mathcal{M}_0(\chi^+\chi^- \rightarrow f) |^2  \end{pmatrix}, \nonumber \\
\Sigma_{L=1}^{12}(f) & \equiv  \frac{1}{\sqrt{2}} \int d\Pi_n \left[ (M_\chi \hat{r}_m) \cdot \mathcal{M}_0(\chi^0 \chi^0 \rightarrow f) \right]^* \left[  (M_\chi \hat{r}_m) \cdot \mathcal{M}_0(\chi^+ \chi^- \rightarrow f) \right],
\label{eq:pann}
\end{align}
where all matrix elements are for $L=1$, but again we have omitted the superscripts for notational convenience. Note we have included factors of $M_\chi$ in $\Sigma_{L=1}(f)$ so that it retains the dimensions of a cross section.

The diagonal elements of $\Sigma(f)$ give the cross sections for free-particle annihilation for distinguishable particles, and the cross sections multiplied by a factor of $1/2$ for identical particles (as in the annihilation matrices of \cite{Beneke:2014gja}), except that in the $p$-wave case, in all cross sections ${\bf p}$ has been replaced with $M_\chi \hat{r}_m$. 
After integrating over the final-state phase space and performing all spin and polarization sums/averages, this amounts to multiplying all cross sections by $M_\chi^2/p^2$. If we set, for example, $A_C=p$ and $A_N=0$ (as appropriate for a plane wave purely in the $\chi^+ \chi^-$ state), then we recover the rate for free-particle annihilation from the chargino-chargino state.
These precise annihilation matrices $\Sigma$, up to trivial prefactors, have already been computed in the literature \cite{Hellmann:2013jxa, Beneke:2014gja} for general electroweakly interacting DM. To facilitate extension of our results to other models, in appendix \ref{app:anndictionary} we provide a general algorithm for determining the $\Sigma$ matrices from existing results. We have also independently derived several of the results presented below (all for the spin-singlet case, and the channels with the largest branching ratios for the $s$-wave spin-triplet case). 
 
In the particular case of the wino, we have for the $L=0$ spin-singlet bound states \cite{Hisano:2006nn, Beneke:2014gja}:
\begin{align} \Sigma(W^+ W^-) & = \frac{4 \pi \alpha_W^2}{M_\chi^2} \begin{pmatrix} 1 & \frac{1}{\sqrt{2}} \\ \frac{1}{\sqrt{2}} & \frac{1}{2} \end{pmatrix}, \quad \Sigma(ZZ) = \frac{4 \pi \alpha_W^2}{M_\chi^2} \begin{pmatrix} 0 & 0 \\ 0 & c_W^4 \end{pmatrix}  \nonumber \\
\Sigma(Z\gamma) & = \frac{4 \pi \alpha_W^2}{M_\chi^2} \begin{pmatrix} 0 & 0 \\ 0 & 2 c_W^2 s_W^2  \end{pmatrix}, \quad \Sigma(\gamma \gamma) = \frac{4 \pi \alpha_W^2}{M_\chi^2}\begin{pmatrix} 0 & 0 \\ 0 & s_W^4 \end{pmatrix},
\label{eq:decay.19} 
\end{align}
and for the $L=1$ spin-triplet bound states the similar result (see appendix \ref{app:anndictionary}):
\begin{align} \Sigma(W^+ W^-) & = \frac{28 \pi \alpha_W^2}{9 M_\chi^2} \begin{pmatrix} 1 & \frac{1}{\sqrt{2}} \\ \frac{1}{\sqrt{2}} & \frac{1}{2} \end{pmatrix}, \quad \Sigma(ZZ) = \frac{28 \pi \alpha_W^2}{9 M_\chi^2} \begin{pmatrix} 0 & 0 \\ 0 & c_W^4 \end{pmatrix}  \nonumber \\
\Sigma(Z\gamma) & = \frac{28 \pi \alpha_W^2}{9 M_\chi^2} \begin{pmatrix} 0 & 0 \\ 0 & 2 c_W^2 s_W^2  \end{pmatrix}, \quad \Sigma(\gamma \gamma) = \frac{28 \pi \alpha_W^2}{9 M_\chi^2}\begin{pmatrix} 0 & 0 \\ 0 & s_W^4 \end{pmatrix}.
\label{eq:tripletann} 
\end{align}
As discussed above, $p$-wave bound states in the spin-singlet configuration and $s$-wave bound states in the spin-triplet configuration (odd $L+S$) are only composed of $\chi^+ \chi^-$, with no $\chi^0 \chi^0$ component. Thus the $\Sigma$ matrix now only has one non-zero component, namely the diagonal entry corresponding to $\chi^+ \chi^-$ annihilation. Furthermore, for the $p$-wave state the Landau-Yang theorem forbids the decay into massless neutral vector bosons. However, for the $s$-wave bound state, the $s$-channel annihilation is open and permits decays into all SM final states.
As calculated in appendix \ref{app:anndictionary}, the non-zero annihilation matrices for these bound states are given by:
\begin{equation} 
\Sigma(W^+ W^-)  = \frac{2}{3} \frac{\pi \alpha_W^2}{M_\chi^2} \begin{pmatrix} 0 & 0 \\ 0 & 1 \end{pmatrix} 
\label{eq:singletWW} 
\end{equation}
for the spin-singlet $p$-wave states, and
\begin{align} 
\Sigma(W^+ W^-) & = \frac{1}{12} \frac{\pi \alpha_W^2}{M_\chi^2} \begin{pmatrix} 0 & 0 \\ 0 & 1 \end{pmatrix}, \quad
\Sigma(Z h^0)  = \frac{1}{12} \frac{\pi \alpha_W^2}{M_\chi^2} \begin{pmatrix} 0 & 0 \\ 0 & 1 \end{pmatrix}, \nonumber \\
\Sigma(q \bar{q}) & = \frac{1}{2} \frac{\pi \alpha_W^2}{M_\chi^2} \begin{pmatrix} 0 & 0 \\ 0 & 1 \end{pmatrix}, \quad 
\Sigma(l^+ l^-, \nu \bar{\nu})  = \frac{1}{6} \frac{\pi \alpha_W^2}{M_\chi^2} \begin{pmatrix} 0 & 0 \\ 0 & 1 \end{pmatrix}. 
\label{eq:tripletswave}
\end{align}
for the spin-triplet $s$-wave states.  

In principle, one could hope to detect an energetic, monochromatic photon line from the bound state's annihilation to $\gamma \gamma$ or $\gamma Z$.  However, only the $L+S\,$-even bound states have a sizable branching ratio to photons, but, as discussed above, we directly capture only to states with odd $L+S$.  Thus, line-photon annihilation events will require that capture occurs into an excited state that can decay by dipole emission to a state with even $L+S$, e.g.~the free winos capture into the spin-singlet 2$p$ state, which subsequently transitions to $d$ or $s$-wave, or into the spin-triplet 2$s$ state, which subsequently transitions to lower-lying $p$-wave states. As we show in section \ref{subsec:woniaform}, for winos in the Milky Way halo, capture into the excited states is dominated by the direct rate for WIMPs to annihilate to $\gamma + X$.  Thus we expect a small branching ratio for monochromatic gamma-ray annihilation lines from bound states.

Note that by dimensional analysis, we naively expect $|\Psi(0)| \sim (\alpha_W M_\chi)^{3/2}$ for $s$-wave bound states, and $A_{N,C} \sim (\alpha_W M_\chi)^{5/2}$ for $p$-wave bound states. Thus we expect the decay width for annihilations from $s$-wave bound states to scale as $\alpha_W^5 M_\chi$, and for annihilations from $p$-wave bound states to scale as $\alpha_W^7 M_\chi$. More generally (as also noted in \cite{An:2016gad}), the width for decay via annihilation will scale as $\Gamma \propto \alpha_W^{5 + 2L} M_\chi$.

%%%%%%%%%%%%%%%%%%%%%%%%
\section{Analytic and numerical results}
\label{sec:anr}
%%%%%%%%%%%%%%%%%%%%%%%%

In this section we apply the results of section \ref{sec:decay}; we first consider the fate of bound states once they form, and then move on to discuss the capture cross section, which primarily determines the overall importance of bound state formation relative to direct annihilation.

%%%%%%%%%%%%%%%%%%%%%%%%
\subsection{WIMPonium decays}
\label{subsec:woniadecay}
%%%%%%%%%%%%%%%%%%%%%%%%

As discussed above, the WIMPonium bound states may decay to lower-energy states in the spectrum by emission of photons, or annihilate to SM particles. As we will demonstrate, the former generally dominate for $L > 0$ if electric-dipole transitions are allowed, with widths scaling as $\Gamma \propto \alpha \, \alpha_W^4 M_\chi$ in the unbroken SU(2) limit. Let us begin by considering the circumstances under which such transitions can occur.

Unsuppressed (single-photon electric-dipole) transitions, either between continuum states or bound states, require $\Delta L = \pm 1$, e.g. 
$p$-wave states can decay to $s$-wave or $d$-wave bound states. As discussed above, bound states populated by single-photon capture will have odd $L+S$, and so will be purely comprised of chargino pairs. The states to which they can decay by single-photon emission will have even $L+S$; they will consequently tend to have larger binding energies for a given principal quantum number (since the potential for even $L+S$ has a larger eigenvalue for its attractive component, in the illustrative ``SU(2) positronium'' limit). This means that, for example, the $1s$ and $2s$ states with odd $L+S$ (i.e. spin-triplet states) may in some circumstances have available single-photon dipole decays to a $2p$ state with even $L+S$, and so need not be (meta)stable as they are in the hydrogen atom. Similarly, the $2p$ state with odd $L+S$ (spin-singlet) may have open decay channels to the $3d$ and $3s$ states as well as the $1s$ state. This point is illustrated in figure \ref{fig:bs2}.

As we will demonstrate in the next subsection, at low DM masses the dominant capture process populates the spin-triplet $1s$ state, which is generically the lowest-lying spin-triplet bound state; consequently this process gives rise to no subsequent transitions, which would require an $\alpha_W$-suppressed spin-flip.
At higher DM masses the most important capture processes either populate the lowest-lying spin-singlet states with odd $L$, i.e.~the $2p$ states, via capture from the $s$- and $d$-wave parts of the original plane wave, or the $2s$ spin-triplet state via capture from the initial $p$-wave component. These statements assume the typical velocity of the Milky Way halo $v \sim \mathcal{O}\left( 10^{-3} \right)$; at low velocities where $M_\chi v \lesssim m_W$, then the system is still in the Yukawa regime and the contributions from higher-$L$ partial waves have velocity suppression due to the short-range nature of the potential (see appendix \ref{app:hulthen} for a discussion of the scaling).  
The $2p$ states generically have open and unsuppressed decays to the spin-singlet $1s$, $2s$, $3s$ and $3d$ states. The $3d$, $3s$ and $2s$ spin-singlet states are themselves metastable, as there are no lower-lying spin-singlet states with odd $L$, so they can decay only through annihilation to SM states or:
\begin{itemize}
\item two-photon transitions to the $1s$ spin-singlet ground state, 
\item electric quadrupole transitions to the $1s$ spin-singlet ground state with even $L+S$ (available for $d$-wave states only),
\item magnetic dipole, spin-flip transitions to the spin-triplet states, induced by relativistic effects.
\end{itemize}

In order to develop intuition, let us first consider the allowed transitions from the spin-singlet $2p$ states in the high-energy limit where the SU(2) is unbroken.  From figure \ref{fig:bs2}, we see that the 1\textendash3$s$ and 3$d$ spin-singlet states are at lower energies.  Note that while we formally set the masses of all force carriers to zero for this analysis, we still examine only capture through emission of $W^3$, which will correspond to photon emission after SU(2) is broken. We obtain the total decay rates for each of the spin-singlet $nlm=21m$ as $\Gamma \approx 0.16 \, \alpha \, \alpha_W^4 M_\chi$ (appendix \ref{subapp:coulombtransition}). Approximately half the total branching ratio is to the ground state, followed closely by decays to the $2s$ excited state; decays to the $3s$ and $3d$ excited states contribute only $\sim 6\%$ of the total rate.

For annihilation, let us consider the same unbroken limit and compute the annihilation rates for the $2p$ and $1s$ spin-singlet states, and the $1s$ spin-triplet state, using the explicit form of the Coulombic bound state wavefunction (see discussion in appendix \ref{subapp:coulombbound}):
\begin{equation} \Psi^i_{nlm}({\bf r}) = Y_{lm}(\theta,\phi) \eta_i R_{nl}(\lambda_i \alpha; r), 
\label{eq:decay.21}     
\end{equation}
\begin{equation}
R_{nl} \left(\lambda_i \alpha; r \right) = \left[ \left( \frac{2\alpha \lambda_i \mu}{n} \right)^{3} \frac{ \left( n - l -1 \right)!}{2n  \left( n + l \right)! } \right]^{1/2} \! e^{-\mu \lambda_i \alpha r / n} \left( \frac{2 \alpha \lambda_i \mu r}{n} \right)^{l} \text{L}^{2l + 1}_{n-l-1} \left( \frac{2 \alpha \, \lambda_i \mu r}{n} \right).
\label{eq:decay.22}
\end{equation}
For the spin-singlet $1s$ state, which has $\lambda_i=2$ and $\eta_i = \begin{pmatrix} \sqrt{\frac{1}{3}} & \sqrt{\frac{2}{3}} \end{pmatrix}$, we obtain $\Psi^i_{100}(0) = \eta_i (\alpha_W M_\chi)^{3/2}/\sqrt{\pi}$, i.e. $\psi_C(0) = \sqrt{2/3} \, (\alpha_W M_\chi)^{3/2}/\sqrt{\pi}$, $\psi_N(0) = \sqrt{1/3} \, (\alpha_W M_\chi)^{3/2}/\sqrt{\pi}$. Thus the decay rates to different final states, $\Gamma(f)$, from the spin-singlet $1s$ state, are given by:
 \begin{align} \Gamma(W^+ W^-) & = 
 \frac{16}{3} \alpha_W^5 M_\chi,  
 &\Gamma(ZZ) =  
 \frac{8}{3} c_W^4   \alpha_W^5 M_\chi, \nonumber \\
 \Gamma(Z\gamma) & = 
 \frac{16}{3} c_W^2 s_W^2  \alpha_W^5 M_\chi, 
& \Gamma(\gamma \gamma) =  
 \frac{8}{3} s_W^4 \alpha_W^5 M_\chi. 
\label{eq:decay.23} 
 \end{align}  
For the spin-triplet $1s$ state we have $\psi_N(0) = 0$, $\psi_C(0) = (\alpha_W M_\chi/2)^{3/2}/\sqrt{\pi}$, while for the spin-triplet $2s$ state we have $\psi_N(0) = 0$, $\psi_C(0) = (\alpha_W M_\chi/2)^{3/2}/2 \sqrt{2 \pi}$. The decay rates for the spin-triplet $1s$ state then become:
\begin{align} \Gamma(W^+ W^-) & = \frac{1}{96} \alpha_W^5 M_\chi, 
&\Gamma(Z h^0)  = \frac{1}{96} \alpha_W^5 M_\chi, \nonumber \\
\Gamma(q \bar{q}) & = \frac{1}{16} \alpha_W^5 M_\chi,  
&\Gamma(l^+ l^-, \nu \bar{\nu})  = \frac{1}{48} \alpha_W^5 M_\chi, \end{align}
while for the spin-triplet $2s$ state the rates are uniformly smaller by a factor of 8.
For annihilation from the $2p$ spin-singlet states, we have $|\nabla \psi_C(0) |^2 = (\alpha_W \mu)^5/(32 \pi)$ (for all $m$), where $\mu=M_\chi/2$, so therefore:
\begin{equation} \Gamma(W^+ W^-) =\frac{1}{3 \times 2^9} \, \alpha_W^7 M_\chi 
\label{eq:decay.24} 
\end{equation}
Similarly, for annihilation from the $2p$ spin-triplet states, we have $\nabla \psi_{N,C}(0) = \eta_{N,C} ((\alpha_W \mu)^{5/2}/\sqrt{\pi}) \hat{r}_m$, where $\eta_N=\sqrt{1/3}$ and $\eta_C=\sqrt{2/3}$. Thus we obtain the decay rates:
 \begin{align} \Gamma(W^+ W^-) & = 
 \frac{7}{54} \alpha_W^7 M_\chi,  
 &\Gamma(ZZ) =  
 \frac{7}{108} c_W^4   \alpha_W^7 M_\chi, \nonumber \\
 \Gamma(Z\gamma) & = 
 \frac{7}{54} c_W^2 s_W^2  \alpha_W^7 M_\chi, 
& \Gamma(\gamma \gamma) =  
 \frac{7}{108} s_W^4 \alpha_W^7 M_\chi. 
\label{eq:decay.25} 
 \end{align}  
 For annihilation from the $3p$ spin-triplet states, $\nabla \psi_{N,C}(0)$ is multiplied by a factor  $16/27$ relative to the $2p$ state, leading to overall rates smaller by a factor $2^8/3^6$.

We see that, as claimed earlier, the annihilation rate for the $2p$ spin-singlet state is parametrically suppressed relative to the electric-dipole single-photon transitions to lower-lying $s$ and $d$ states, which have rates $\sim 10^{-1} \alpha\,\alpha_W^4 M_\chi$; consequently, we can safely approximate that any capture into the $2p$ spin-singlet state results in a transition to an $ns$ ($n$=1\textendash 3) or $3d$ spin-triplet state, followed by annihilation or a suppressed decay as appropriate. The principal decay channel will be to the $1s$ spin-singlet state, and so in this unbroken limit, we expect capture to the $2p$ state to eventually result in annihilation decays to the SM with approximately the branching ratios in eq.~\ref{eq:decay.23}. For capture to the $1s$ spin-triplet state, a wide range of SM final states can be produced due to the presence of an $s$-channel annihilation; most of the branching ratio is to hadronic channels (quarks), and thus the resulting decay annihilation signal would be rich in continuum photons and charged particles, but with no appreciable gamma-ray line at the DM mass. 

Capture to the $2s$ spin-triplet state followed by a decay to the $2p$ or 3$p$ spin-triplet states could lead to a non-zero gamma-ray line contribution. In the SU(2) symmetric limit, the $2p$ and 3$p$ spin-triplet states have a zero matrix element for transitions to the $1s$ spin-triplet state at leading order, so we expect their decays to be dominated by production of SM particles, including $\gamma \gamma$ and $Z\gamma$ final states. However, away from this unbroken limit, single-photon decays to the $1s$ triplet state are open -- albeit suppressed by the low energy of the transition photon -- and must be computed numerically. 

\begin{figure}[h]
\begin{center}
\includegraphics[scale=0.22]{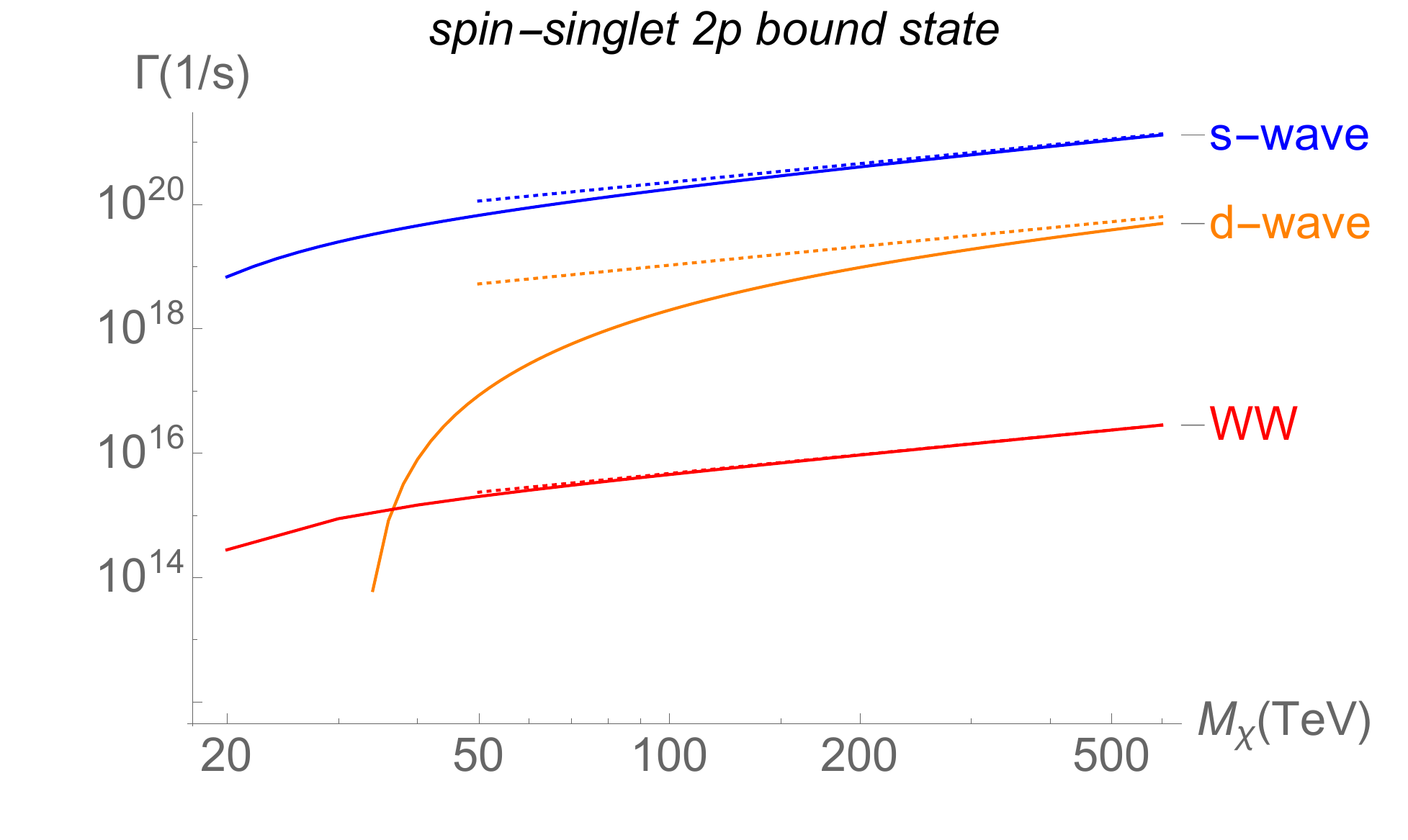}
\hspace{0.5cm}
\includegraphics[scale=0.18]{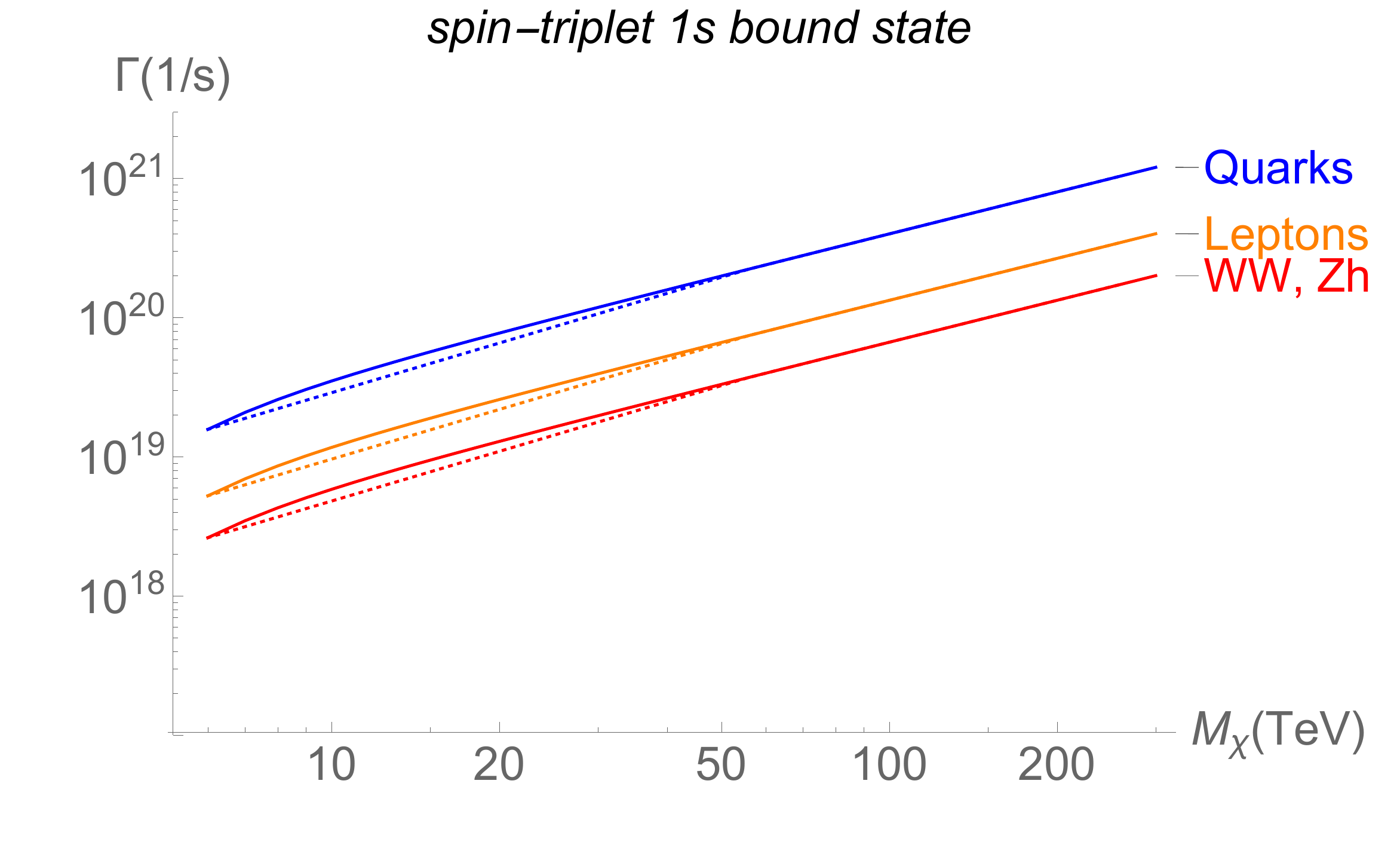}\\
\includegraphics[scale=0.37]{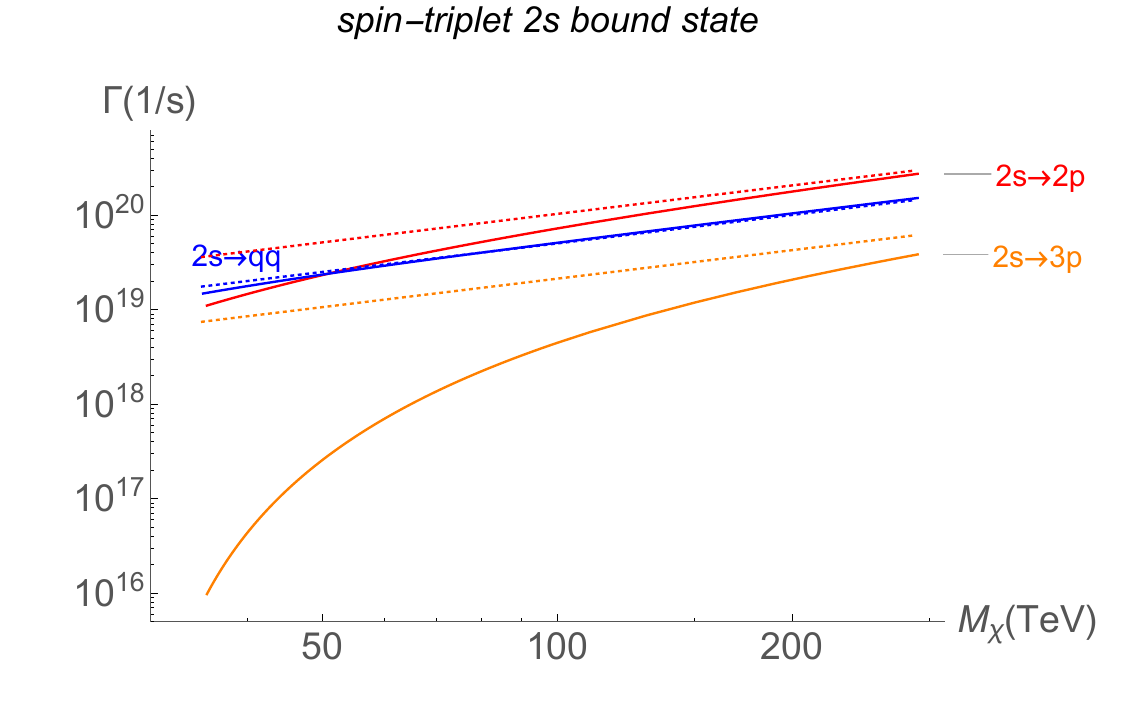}
\hspace{0.5cm}
\includegraphics[scale=0.37]{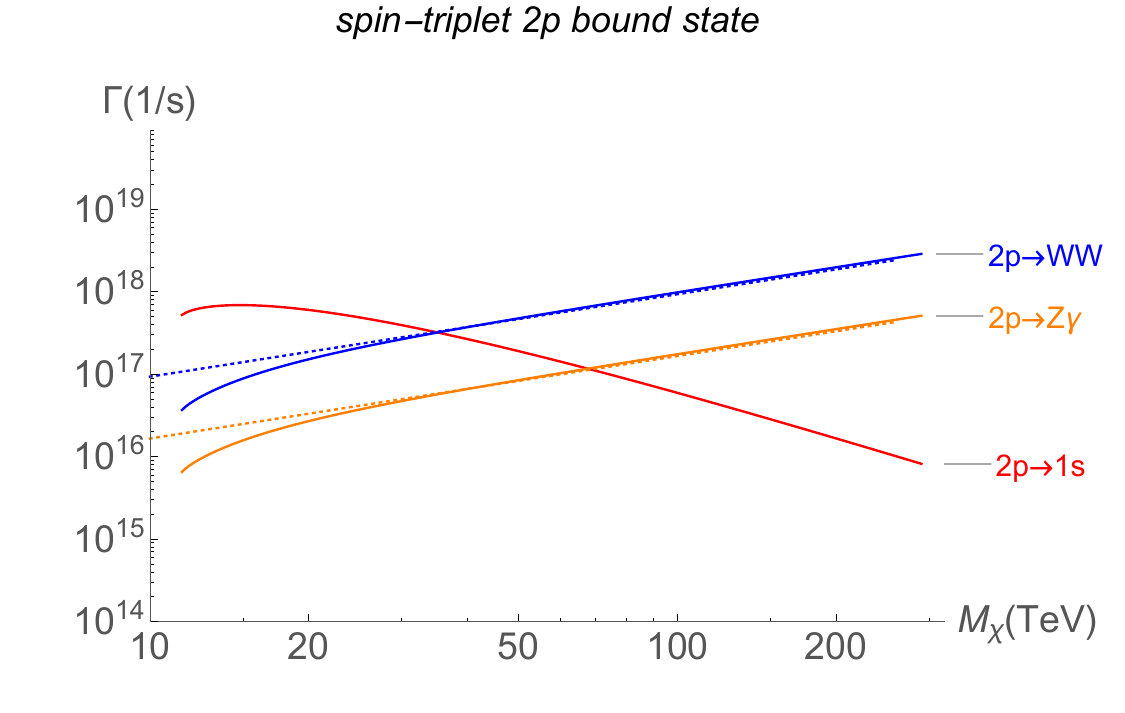}
\caption{Our numerical (solid) and exact results in the SU(2)-symmetric limit (dotted) for the decay channels of the lowest-energy bound states we capture into in each spin configurations. \textbf{Upper left:} Decay rates of the 2$p$ spin-singlet bound state. The \emph{blue} curve shows the inclusive decay rate to the three lower $s$-wave bound states available (1\textendash 3$s$); the \emph{yellow} (\emph{red}) curve is the rate to the 3$d$ bound state ($W^+W^-$). \textbf{Upper right:} Decay rates of the 1$s$, spin-triplet bound state, given by the annihilation matrices in eq.~\ref{eq:tripletswave}. \textbf{Lower left:} Selected decay rates of the 2$s$ spin-triplet bound state; we see there are comparable rates for transition to the 2$p$ state and decay via annihilation to $q\bar{q}$, which is the dominant channel for decay to Standard Model particles (the relative strengths of the Standard Model decay channels are similar to the {\it upper right} panel). \textbf{Lower right:} Selected decay rates of the 2$p$ spin-triplet bound state, populated by transitions from the 2$s$ spin-triplet state. Transitions to the 1$s$ spin-triplet state dominate at low DM masses while decays via annihilation to Standard Model gauge bosons dominate at high masses.}
\label{fig:decay.1}
\end{center}
\end{figure}

Moving beyond the SU(2) symmetric limit, we can use the numerical method introduced in appendix \ref{app:bs} to calculate the bound-state wavefunctions and then use eq.~\ref{eq:sigmastop} to find the spin-singlet, $2p$ to $ns$ transitions, and the decays of the spin-triplet $2s$ state. The results are shown in Figure \ref{fig:decay.1}, together with the analytic results presented above, which approach validity in the limit of high DM mass. As expected, for the $2p$ spin-singlet states, the decay via annihilation is suppressed by a few orders of magnitude compared to the transition to lower $s$- and $d$-wave bound states. For capture to the 2$s$ spin-triplet state, the situation is somewhat more complicated; decays to the 2$p$ spin-triplet state and to Standard Model quark-antiquark pairs dominate and have comparable rates across a wide range of masses. Depending on the DM mass, the 2$p$ spin-triplet state may decay dominantly via either transitions to the 1$s$ spin-triplet state (at low masses) or through annihilation to Standard Model gauge bosons (at high masses). This second decay pathway allows for production of gamma-ray line photons with a non-negligible branching ratio, but as we will see, the rate of formation of the 2$s$ spin-triplet state is always subdominant to the production of such line photons from direct annihilation of two DM particles.

Finally, since the bound-state Hamiltonian includes the positive mass-shift, some of the bound states (in the sense that their wavefunctions are exponentially suppressed at large $r$) will have positive energy, according to our definition of zero energy.  Such states could therefore decay into lower-energy unbound states (corresponding to free $\chi^0 \chi^0$ pairs at large $r$) through the emission of a photon, which changes $L+S$ from odd to even.  It is a quirk of our Hamiltonian with the mass-shift term that bound and continuum states overlap in the spectrum between $E=0$ and $E= \, +2\delta M$.  

However, we expect the impact of these positive-energy bound states to be small, and neglect them in our calculations. As we see in figure \ref{fig:bs1}, for $M_\chi >$ 6 TeV, there are negative-energy bound states in the $L+S$-odd Hamiltonian spectrum available for capture.  Capture rates are typically dominated by the deepest-available bound states, with the rates smoothly turning off as the binding energy approaches zero from below.  Furthermore, capturing into the full range of positive-energy bound states requires sufficient kinetic energy from the initial $\chi^0 \chi^0$.  For example, with our standard mean velocity, $v = 10^{-3}$, we would need $M_\chi >$ 1320 TeV to capture into all $L+S$-odd, $\chi^+ \chi^-$ bound states; if the kinetic energy is much smaller than this value, positive-energy bound states will only be available for capture in the fine-tuned case where their binding energy relative to the free $\chi^+ \chi^-$ state is very close to the mass splitting $2 \delta M$. 

If we do form such a WIMPonium, its ``fall-apart'' transition back to free $\chi^0\chi^0$ (with emission of another photon) is kinematically suppressed relative to standard dipole-emission decay to a negative-energy bound state, if such an accessible state exists in the spectrum.  As stated above, the bound-state to bound-state transition rate scales like $\alpha \, \alpha_W^4 M_\chi$.  We can estimate the rate of WIMPonium $\rightarrow \chi^0 \chi^0 \, \gamma$ from the capture rate in the Coulomb limit, which scales as $\sigma v \sim \alpha E_\gamma/v$, as the overlap integral and thus the squared matrix elements are the same.  However, to convert $\sigma v$ to $\Gamma$, we need an additional factor of the phase space for the relative WIMP momentum, $p = M_\chi v_\text{rel}/2$.  The positive powers of $v$ from this measure will (more than) cancel the $1/v$ that gave the Sommerfeld enhancement for capture.  We thus get a factor $(E_\gamma v_\text{rel}^2)/M_\chi$.  For the highest-energy bound states, $E_\gamma \sim \delta M$ and $v_\text{rel}^2 \sim \delta M/M_\chi$, for an overall scaling like $\alpha (\delta M/M_\chi)^2$.  Thus, making this process competitive with the dipole transition rate to another bound state would require $(\delta M/M_\chi) > \alpha_W^2$ and thus $M_\chi < 300$ GeV.  However, this is outside the regime where Sommerfeld and electroweak bound state effects occur, which requires $M_\chi \alpha_W/m_W  \gtrsim 1$. 

%%%%%%%%%%%%%%%%%%%%%%%%
\subsection{WIMPonium formation}
\label{subsec:woniaform}
%%%%%%%%%%%%%%%%%%%%%%%%

Again, we will begin by considering the symmetric limit where SU(2) is unbroken in order to build intuition, as in this limit the single-photon capture rates can be calculated analytically from the formulae derived in section \ref{subsec:transition}. The spin-averaged cross sections for radiative capture into the first few bound states from the full initial (asymptotically plane-wave) state, by single-photon emission in the dipole approximation and in the limit of small initial momenta, are given by (eq.~\ref{eq:capturexsec}):
\begin{align} \sigma v_\mathrm{rel} & = \frac{2^{13} \, \pi^2}{3^3} \frac{\alpha \, \alpha_W^2}{M_\chi^2 \, v_\mathrm{rel}} \frac{1 }{n^2 } e^{-8 n}   f_{nlm}. 
\label{eqxsec} 
\end{align}
where the $f_{nlm}$ coefficients are given by:
\begin{equation} 
f_{100} = 0, \quad f_{200} = 384, \quad f_{210} = 242, \quad f_{21\pm1} = 50.
\label{eq:accident}
\end{equation}
As discussed above, here we have multiplied the cross sections for even-$L$ final states by $3/4$ to account for the fact that the initial state must be odd-$L$ and hence spin-triplet, and likewise we have multiplied the cross sections for odd-$L$ final states by 1/4.

Consider the more general case where the incoming two-particle state experiences a Coulomb potential with coupling $\lambda_i \alpha_W$ and corresponding eigenvector $\eta_i$, and the final bound state is supported by a Coulomb potential with coupling $\lambda_f \alpha_W$ and corresponding eigenvector $\eta_f$. We find that (at least for these low-lying states) there is a generic accidental suppression in the cross section of the form $e^{-4 n \lambda_i/\lambda_f}$, arising from the overlap between the wavefunctions with different eigenvalues (see appendix \ref{subapp:coulombcapture} for the derivation). Since for the wino-like case, $\lambda_i=2$ for the attracted component (since the attracted component must have even $L+S$ to allow mixing between the $\chi^0\chi^0$ and $\chi^+\chi^-$ two-particle states), and $\lambda_f=1$, this suppression is $e^{-8n}$, and acts quite strongly to suppress capture into higher-$n$ bound states. For positronium, where there is only one relevant Coulomb potential, this factor is only $e^{-4n}$. As we will see, the single-photon capture cross section for the wino is generically well below the direct annihilation cross section, which is not the case for positronium (we discuss this point further in section \ref{subapp:coulombcompare}). 

In this regime, where the potential has infinite range, there is no velocity suppression of terms corresponding to higher partial waves in the incoming two-particle state. However, we expect such a velocity suppression to occur once the relative particle velocity is comparable to $m_W/M_\chi$. As a crude estimate of the effects on the cross section, we can separate out the contribution to eq.~\ref{eqxsec} originating purely from the $L=0$ partial wave, setting all other contributions to zero. For capture from the $s$-wave piece of the initial state to the $n=2$, $l=1$ bound states, we find (eq.~\ref{spwavem}):
\begin{equation} 
\sigma v_\mathrm{rel} = \frac{2^{12} \, \pi^2}{3^4} e^{-16} \frac{ \alpha \, \alpha_W^2}{M_\chi^2 \, v_\mathrm{rel}}.
\label{eq:spex}
\end{equation}
Note that the contributions to capture rates into the $nlm=210$, $211$ and 21-1 states are identical for this case; here we have summed the cross sections together. We have also averaged over the spin configuration, which amounts to dividing the cross section for the spin-singlet case by 4, since there is no $s$-wave component of the spin-triplet state due to Fermi statistics.
However, as we show in appendix \ref{app:hulthen}, the anticipated velocity suppression of the higher partial waves is of order $(M_\chi v/m_W)^{2L}$, rather than simply $v^{2L}$. Consequently, so long as $v$ is not too small compared to $m_W/M_\chi$, the contributions from higher partial waves may still be non-negligible and even dominate. However, even in the limit of unbroken $SU \left( 2 \right)$ there can be cases where an accidental cancellation sets the rate for a particular capture channel to zero. For example, for the wino this occurs for the spin-averaged capture rate from the $p$-wave piece of the initial state to the spin-triplet $1s$ state.  However, capture from $p$-wave to the triplet $2s$ state is parametrically similar, and modestly enhanced compared to eq.~\ref{eq:spex},
\begin{equation} 
\sigma v_\mathrm{rel} = \frac{2^{18} \, \pi^2}{3^2} e^{-16} \frac{ \alpha \, \alpha_W^2}{M_\chi^2 \, v_\mathrm{rel}}.
\label{eq:psex}
\end{equation}

To compute the full capture cross section, we numerically solve for the radial wavefunctions for both the continuum and bound states, using the methods presented in appendices \ref{app:init} and \ref{app:bs}. In figure \ref{fig:capture}, we plot the dominant capture rates, including the capture rate to the $2p$ spin-singlet bound state (split into contributions from initial $d$- and $s$-wave components, with the rates given in eqs.~\ref{eq:sigmastop}-\ref{eq:sigmaptod}, and the summed capture rates to the $1s$ and $2s$ spin-triplet bound states.\footnote{We note that for our numerical analysis, we have taken the parameters of the electroweak potential at their PDG $m_Z$ values \cite{Agashe:2014kda}.  Since the proper scale is given by value of order the momentum transfer, in the potential, this is $\max(m_W,\,M_\chi v_\text{rel})$ and for the photon emission is $\max(E_n,\,M_\chi v_\text{rel}^2)$, where $E_n$ is the binding energy, typically $\mo$(few $\times$ GeV).  Summing the logarithms associated with the hierarchy is beyond our scope.}
The capture to the $1s$ spin-triplet state is the only available channel at low DM masses, is competitive with capture to the $2p$ states at intermediate DM masses, and becomes subdominant at high DM masses (compared to the capture to the $2p$ and $2s$ states) because it vanishes in the Coulomb limit due to an accidental cancellation ({\it cf.}~eq.~\ref{eq:accident}). Generically the capture to excited states is suppressed by the $e^{-8n}$ factor (although as discussed above accidental cancellations can change this hierarchy).  In figures \ref{fig:sstates} and \ref{fig:pdstates}, we plot the rates to capture into a set of the deepest $s$ and $p$-wave bound states from $s$, $p$, and $d$-wave initial states.
\begin{figure}[ht]
\centerline{\scalebox{.7}{\includegraphics{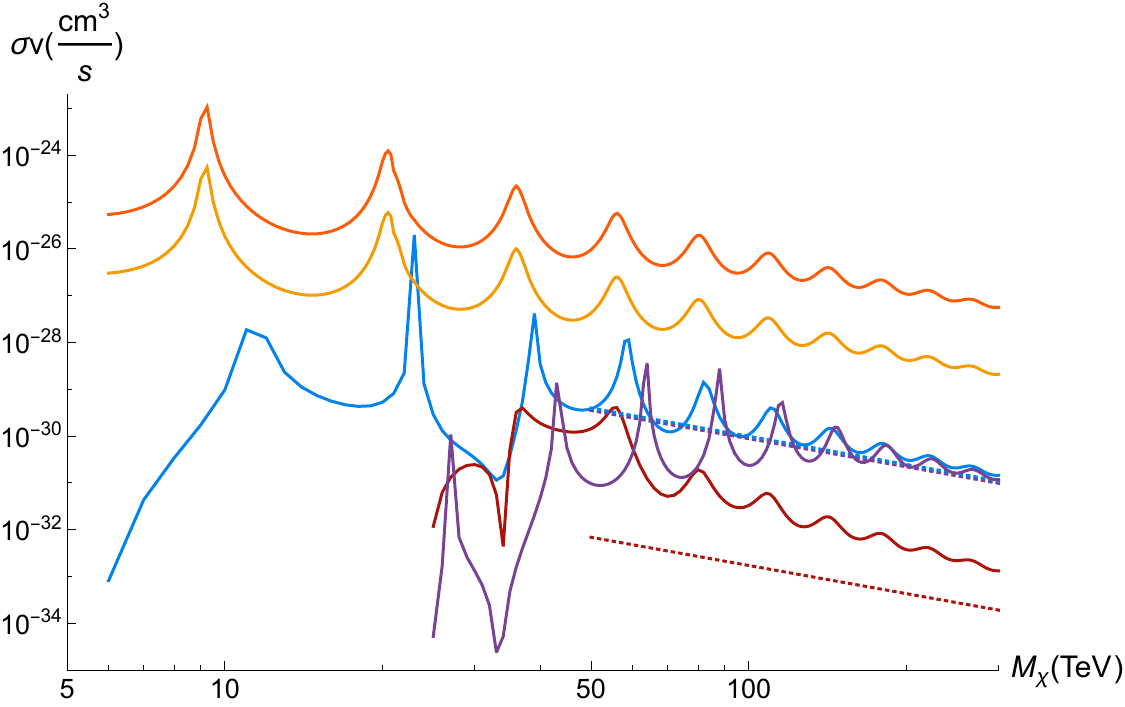}}}
\vskip-0.2cm
\caption[1]{Rates for initial wino dark matter state, $\chi^0 \chi^0$, to capture to WIMPonium or annihilate directly to SM bosons with $v_\text{rel} = 10^{-3}$.  \emph{Dark Orange}: Tree-level inclusive annihilation rate to $W^+ W^-,\, \gamma Z$, and $\gamma \gamma$.  \emph{Yellow}: Semi-inclusive observable to rate to a monochromatic photon, $\chi^0 \chi^0 \rightarrow \gamma + X$, assuming a HESS-like detector, using the results of \cite{Baumgart:2015bpa}.  \emph{Blue}: $\chi^0 \chi^0 (p$\textendash wave) $\rightarrow \,  ^3\!S_1+\gamma$, sum of bound states $n$=1 and 2. \emph{Purple}: $\chi^0 \chi^0 (d$\textendash wave) $ \rightarrow \,  ^1\!P_1+\gamma$ ($n=2$). \emph{Maroon}: $\chi^0 \chi^0 (s$\textendash wave) $\rightarrow \,  ^1\!P_1+\gamma$, $n=2$. \emph{Dashed} lines indicate 3$\times \sigma v_\mathrm{rel}$ computed analytically in the SU(2)-symmetric limit for $s \rightarrow p$ (\emph{maroon}, eq.~\ref{spwavem}), $p \rightarrow s$ (\emph{blue}, eq.~\ref{eq:psex}) and $d \rightarrow p$ (\emph{purple}), capturing to the lowest-$n$ bound states only (see the text for an explanation of the factor of 3).}
\label{fig:capture} 
\end{figure}

At low masses, the capture cross section experiences a pattern of resonances similar to that for Sommerfeld-enhanced direct annihilation, due to the enhancement of the continuum-state wavefunction close to the origin when a bound state in the spectrum passes through zero energy (note that these are bound states for the potential with \emph{even} $L+S$, whereas the bound states produced by the single-photon-mediated capture necessarily have odd $L+S$). At high masses, the resonance peaks diminish and the result approaches our analytic calculation for the unbroken SU(2) limit, up to an overall factor; when we test very high masses beyond the reach of Figure \ref{fig:capture} our numerical capture rate consistently exceeds the analytical prediction by a factor of 3 for $v \lesssim 10^{-3}$.  

We attribute this factor of 3 to a somewhat subtle effect discussed in appendix E2 of \cite{Schutz:2014nka}; the issue is that since the chargino states are not kinematically accessible, we are not truly in the limit of unbroken SU(2), as the mass splitting between the neutralino and chargino states is large compared to other energy scales in the problem (i.e. the kinetic energy of the particles). The transition between the large-$r$ regime, where the mass splitting dominates the potential, and the small-$r$ regime, where the potential is approximately Coulombic, can give rise to effects that do not appear in the unbroken-SU(2) limit.

Our cross section result in the unbroken SU(2) limit includes a factor of $1/3$ from the overlap between our initial condition (particles begin as neutralinos) and the eigenvector of the potential matrix that experiences an attractive interaction; in the language of appendix \ref{app:coulomb}, and particularly eq.~\ref{eq:coulombxsec}, this factor appears in the matrix element as ${\bf I} \cdot \eta_i$ ($=1/\sqrt{3}$ for the wino). More specifically, it appears in the contribution to the matrix element from each component of the continuum wavefunction that experiences an attractive interaction. In the low-velocity limit, it is these contributions that control the overall capture rate, since any component of the wavefunction that experiences a repulsive interaction is suppressed toward the origin and its contribution to the capture rate is exponentially suppressed (as we demonstrate in appendix \ref{app:coulomb}). In the unbroken SU(2) limit, the wavefunctions associated with the various eigenvectors of the potential obey decoupled Schr\"{o}dinger equations, and the eigenvectors themselves are independent of $r$; thus we can determine the wavefunctions associated with the various eigenvectors at some large $r$ (i.e. by setting an initial condition), and then evolve them straightforwardly for all $r$.

However, in the more general case where SU(2) is broken, the fraction of the wavefunction corresponding to each of the $r$-dependent eigenvectors will evolve with $r$ in a non-trivial way. In particular, when the $\chi^0 \chi^0$ and $\chi^+ \chi^-$ states have different masses -- that is, they have different energies as $r \rightarrow \infty$ -- this mass splitting defines the two eigenvectors of the potential matrix at large $r$, whereas at small $r$ Coulomb-like behavior is recovered and the eigenvectors of the matrix correspond to states experiencing attractive (lower energy) or repulsive (higher energy) Coulomb potentials. One particularly simple case occurs when the transition between the two regimes is sufficiently slow and adiabatic: then if the wavefunction is purely in the lower-energy eigenstate at large $r$ (i.e. the $\chi^0 \chi^0$ state), it will entirely populate the lower-energy (attracted) eigenstate at small $r$ also. Consequently, the $\chi^0 \chi^0$ state effectively feels a purely attractive interaction, and there is no suppression factor in the matrix element to account for the fraction of the state that experiences repulsion and does not contribute to the capture rate. 

If this adiabatic approximation is valid, then the factors of ${\bf I} \cdot \eta_i$ appearing in eq.~\ref{eq:coulombxsec} should be replaced by $1$ for the lowest-energy eigenstate at small $r$ -- i.e. the eigenstate $\eta_i$ corresponding to the largest value of $\lambda_i$ -- and by $0$ for all other eigenstates. However, caution is warranted when applying this naive estimate to systems with multiple eigenstates with $\lambda_i > 0$, as eigenstates with smaller values of $\lambda_i$ can yield exponentially larger contributions to the capture cross section (via the $e^{-2 n \lambda_i/\lambda_f}$ factor of eq.~\ref{eq:coulombxsec}), and to our knowledge this behavior has only been studied in systems with a single attracted eigenstate.  
In a simpler multi-state model, with only a single force carrier with mass $m_A$ and coupling $\alpha_A$, \cite{Schutz:2014nka} gave the criterion for this adiabatic rotation to occur as 
$m_A v_\text{rel} \lesssim 2 \delta$. In our model, the equivalent criterion would be $m_W v_\text{rel} \lesssim 2 \delta$, which is generically true for $v_\text{rel} \lesssim 2 \delta/m_W \sim 4 \times 10^{-3}$, independent of the DM mass.

\begin{figure}[ht]
\begin{center}
\includegraphics[scale=0.5]{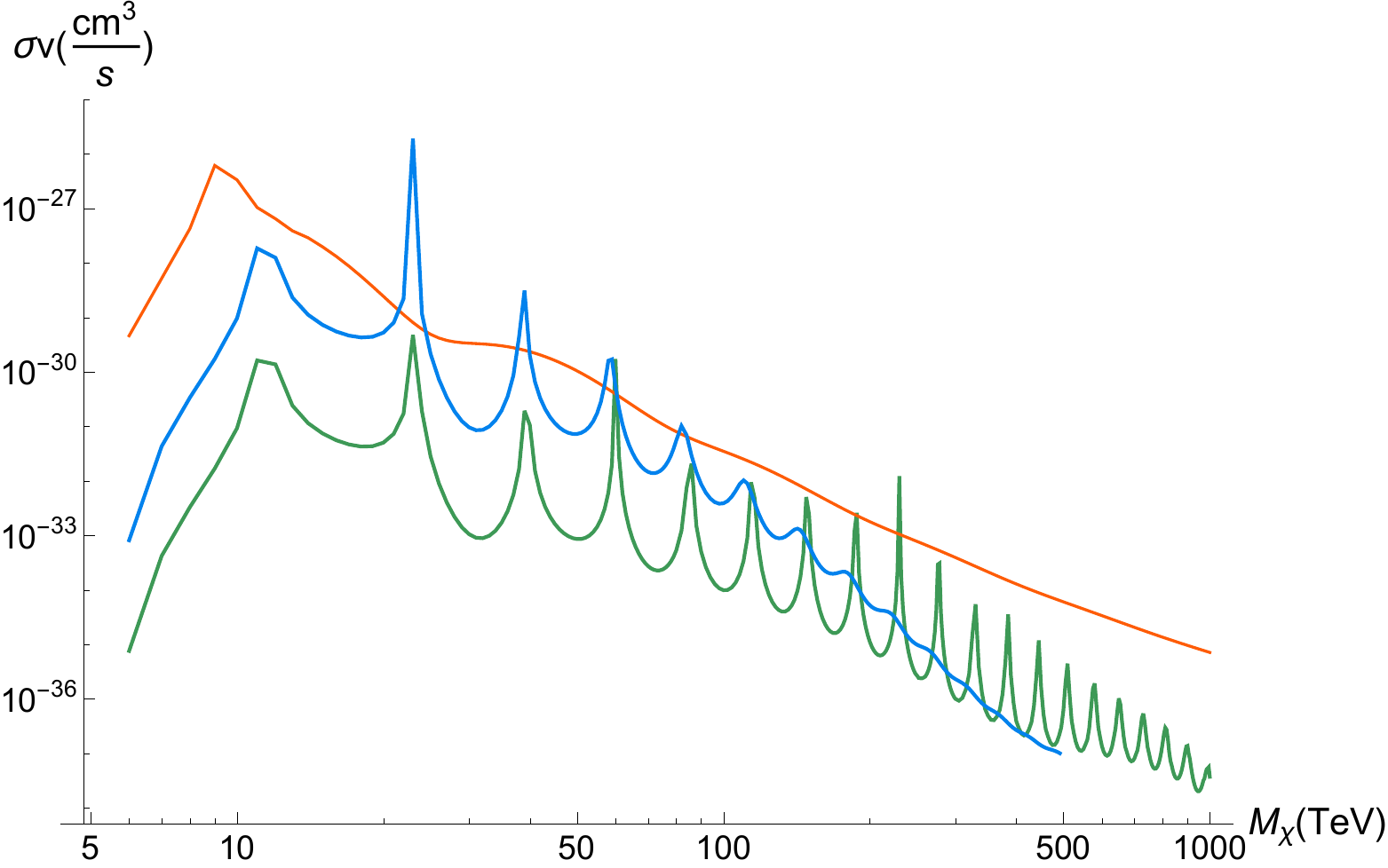}

\vspace{0.5cm}

\includegraphics[scale=0.5]{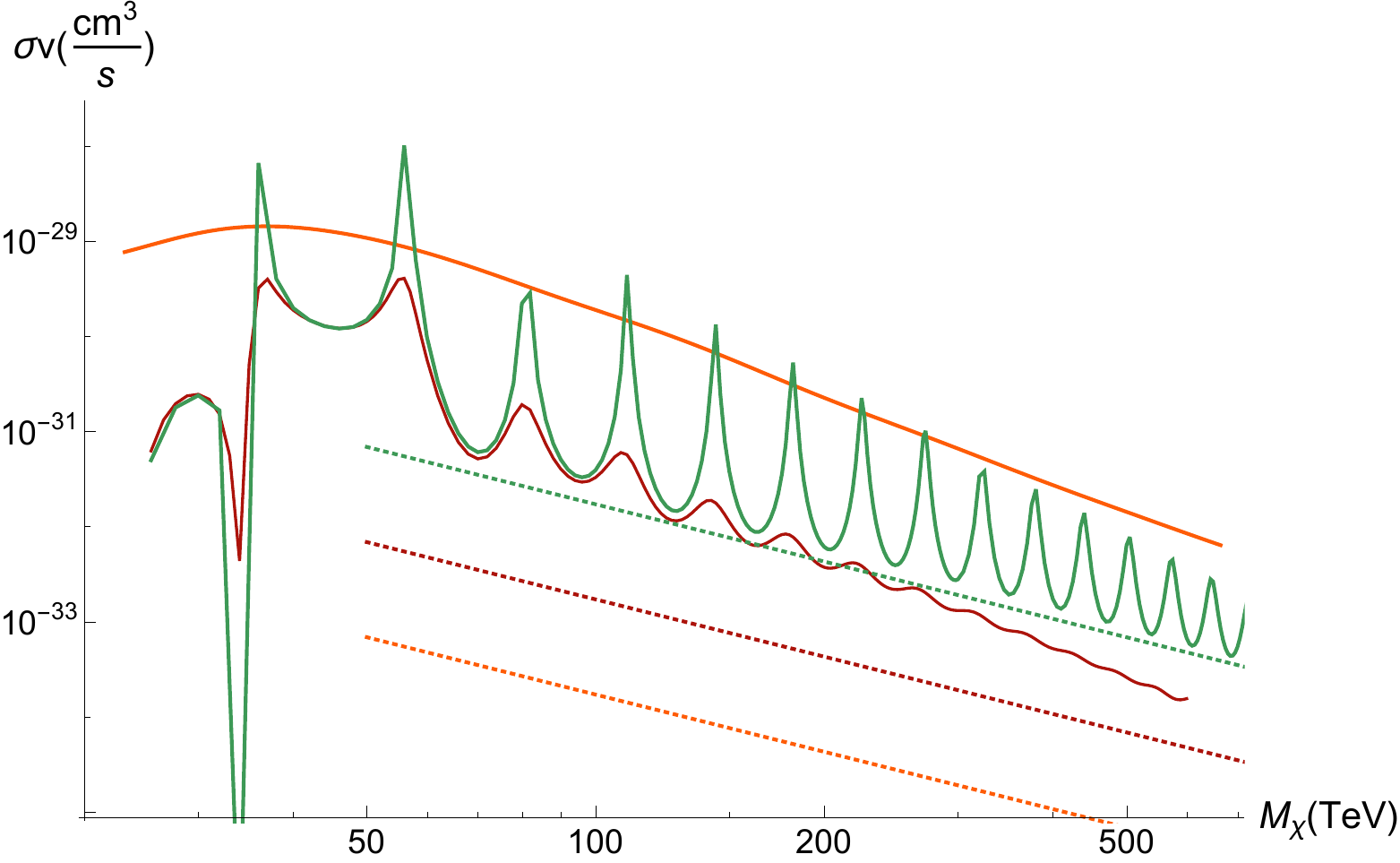}
\vskip-0.2cm
\caption[1]{{\bf Top:} Rate for $\chi^0 \chi^0 (p$\textendash wave) $\rightarrow \,  ^3\!S_1+\gamma$, considering only capture into the deepest bound state. {\bf Bottom:} Rate for $\chi^0 \chi^0 (s$\textendash wave) $\rightarrow \,  ^1\!P_1+\gamma$, considering only capture into the deepest bound states.  Velocities are $v_\text{rel} = 10^{-2}$ (\emph{orange}), $v_\text{rel} = 10^{-3}$ (\emph{blue}:$\, p$-wave; \emph{maroon}:$\, s$-wave), and $v_\text{rel} = 10^{-4}$ (\emph{green}). All colored \emph{dotted} lines display $3\times \sigma v$, the analytic predictions from the Coulomb limit (see the text for an explanation of the factor of 3).}
\label{fig:velos} 
\end{center}
\end{figure}

For the detailed analysis above, we have taken the relative WIMP velocity to be $v_\text{rel} = 10^{-3}$, typical of DM velocities in the Milky Way halo \cite{2009ApJ...704.1704B}.  However, the capture rates are not velocity-independent in general. It is interesting to scan in $v_\text{rel}$ both because the true WIMP velocity has a Maxwellian distribution and thus will have support at other values, and as a check on our expectations for scaling of the rates with velocity; the latter will be particularly important when considering signals from e.g.~clusters, dwarf galaxies, or substructure in the Milky Way halo.  In figure \ref{fig:velos} we plot the effects of varying $v_\text{rel}$ by an order of magnitude for the $s \, \rightarrow p$\textendash wave and $p \, \rightarrow s$\textendash wave capture rates, to the deepest bound states available in both cases.

In these figures, we only considered capture via photon emission, even though for some of the parameter space on both plots, on-shell $Z$-emission is also allowed.  We did however, take into account that for $\frac{8 \delta M}{M_\chi v^2}<1$, the charged component of the wavefunction, $\psi_C$, is no longer exponentially suppressed at large radii.  For example, this covers the $M_\chi > 13$ TeV range of $v_\text{rel} = 10^{-2}$ in both plots.  We see that as expected for a short-range potential, at lower $M_\chi$ we have a pronounced velocity suppression for the $p$-wave initial state.  However, at higher masses, where we are in the SU(2)-symmetric limit, the velocity suppression is lifted, and we expect the slower WIMPs to cross over to having a larger capture rate, scaling like $1/v_\text{rel}$, which saturates once $v_\text{rel} < m_W/M_\chi$. 

We observe that the $v_\text{rel}=10^{-2}$ case does not respect this scaling, with rates that can be much higher than the $v_\text{rel}=10^{-3}$ case. Our simple analytic results neglect the contribution from the continuum states that experience a \emph{repulsive} Coulomb potential, as discussed in appendix \ref{subapp:coulombcapture}, on the grounds that this contribution is exponentially suppressed at low velocities. The suppression scales as $e^{-2\pi\alpha_W/v_\text{rel}}$ in the cross section (see eq.~\ref{eq:expsuppress}), so when $v_\text{rel}$ becomes comparable to $\alpha_W$, this term can no longer be clearly neglected. More generally, we have worked in the limit of small $v_\text{rel} \ll \alpha_W$ throughout this calculation; as $v_\text{rel}$ becomes large our analytic results should not be expected to describe the full behavior of the system.

\subsection{Capture vs direct annihilation}
\label{subapp:coulombcompare}

From figure \ref{fig:capture}, we see that in the wino case the capture rate is quite suppressed relative to direct annihilation, across almost the whole range of possible DM masses. This contrasts with the case of $e^+ e^-$ annihilation, where capture into positronium dominates direct annihilation at low relative velocities. In this subsection we explore the origin of this difference, and how it might generalize to other complex dark sectors. As previously, we proceed by examining the limit where SU(2) is unbroken.

For positronium, where $\lambda_i = \lambda_f = 1$, all the gauge factors are trivial, diagrams of the form shown in figure \ref{fig:potlus} are forbidden (i.e. $\hat{C}_2=0$, in the notation of appendix \ref{app:coulomb}), and there is only a single relevant two-body state ($e^+ e^-$), we obtain the cross section for capture into the positronium ground state from eq.~\ref{eq:coulombxsec} as:
\begin{equation} \sigma v_\text{rel} = \frac{2^{10} \pi^2}{3} e^{-4} \, \alpha^3 \frac{1}{m_e^2}\frac{1}{v_\mathrm{rel}}.\end{equation} 

For the wino, as discussed above, the high-mass limit of the cross section for capture into the ground state is zero. The cross section for capture into the 2$p$ states is:
\begin{equation} \sigma v_\text{rel} = \frac{2^{12} \pi^2 \times 19}{3} e^{-16} \alpha \, \alpha_W^2 \frac{1}{M_\chi^2}\frac{1}{v_\mathrm{rel}}.\end{equation} 
We see that the numerical prefactor is smaller by a factor of $\sim 5\times10^{-4}$ for the wino compared to positronium; the factor of $e^{-16}$ vs.~$e^{-4}$ from the overlap integral suppresses the rate for the wino, and is not fully compensated by other numerical prefactors (including capture to the $2s$ state, which also experiences a $e^{-16}$ prefactor as it has $n=2$, would only change this ratio by an $\mathcal{O}(1)$ amount).

Now let us consider the rate for direct annihilation. The Sommerfeld enhancement at low velocities and for massless force carriers is $2 \pi \alpha/v_\mathrm{rel}$. The spin-averaged annihilation cross section for $e^+ e^- \rightarrow \gamma \gamma$ \emph{without} accounting for Sommerfeld enhancement is $\sigma v_\text{rel} = \pi \alpha^2/m_e^2$ (e.g. \cite{Pospelov:2008jd}). Thus the enhanced cross section is:
\begin{equation} \sigma v_\text{rel}= \frac{2 \pi^2 \alpha^3}{m_e^2 \, v_\text{rel}}. \end{equation}
 
 For direct annihilation (into all channels), on the other hand, the leading-order spin-averaged $s$-wave annihilation rate for the wino is:
 \begin{equation} \sigma v_\text{rel} = \frac{2 \pi \alpha_W^2}{M_\chi^2} \begin{pmatrix} \psi^*_N(0) & \psi^*_C(0) \end{pmatrix} \begin{pmatrix} 1 & \frac{1}{\sqrt{2}} \\ \frac{1}{\sqrt{2}} & \frac{3}{2} \end{pmatrix}  \begin{pmatrix} \psi_N(0) \\ \psi_C(0) \end{pmatrix}. \end{equation}
 
Assuming that only the attracted eigenstate gives a non-negligible contribution to the wavefunctions at the origin, and that this eigenstate has eigenvector $\eta$ and eigenvalue $\lambda$, we obtain:
\begin{align} 
\sigma v_\text{rel} & = \frac{2 \pi \alpha_W^2}{M_\chi^2} |{\bf I} \cdot \eta|^2 \, |\phi(\lambda \alpha_W;0)|^2 \, \eta^\dagger \begin{pmatrix} 1 & \frac{1}{\sqrt{2}} \\ \frac{1}{\sqrt{2}} & \frac{3}{2} \end{pmatrix} \eta \nonumber \\
& =  \frac{2 \pi \alpha_W^2}{M_\chi^2} \, \frac{1}{3} \, |\phi(2 \alpha_W;0)|^2 \, \begin{pmatrix} \sqrt{\frac{1}{3}} & \sqrt{\frac{2}{3}} \end{pmatrix} \begin{pmatrix} 1 & \frac{1}{\sqrt{2}} \\ \frac{1}{\sqrt{2}} & \frac{3}{2} \end{pmatrix}  \begin{pmatrix} \sqrt{\frac{1}{3}} \\ \sqrt{\frac{2}{3}} \end{pmatrix}. 
\end{align}
 We have $|\phi(\lambda \alpha;0)|^2 \approx 2 \pi \lambda \alpha/v_\text{rel}$ for small velocities, from our earlier results (this also cross-checks our Sommerfeld enhancement formula for positronium). Thus overall we obtain:
\begin{align} 
\sigma v_\text{rel}  & =  \frac{2^4 \pi^2 \alpha_W^3}{3 M_\chi^2 v_\text{rel}}. 
\end{align}
 
 We see that the cross section is slightly larger for the wino than one would expect from a naive extrapolation from positronium; the presence of multiple channels and the stronger coupling (since $\lambda=2$) outweighs the penalty factor from the non-trivial overlap between the initial conditions and the attracted state. While for positronium, the capture/annihilation ratio is $2^9 e^{-4}/3 \approx 3$, for the wino we expect it to be $2^{8} \times19 e^{-16} (\alpha/\alpha_W) \approx 5 \times 10^{-4} (\alpha/\alpha_W) \approx 10^{-4}$.
 
 With regard to general dark sectors, we see that a large attractive eigenvalue for the initial state boosts the rate for direct annihilation by a factor $\lambda$, but suppresses the capture rate by an exponential factor (that depends on the ratio of this eigenvalue to the attractive eigenvalue of the potential supporting the final state). Thus in general, smaller attractive eigenvalues for the initial state (and also larger attractive eigenvalues for the bound state) will tend to boost the capture/annihilation ratio.

%%%%%%%%%%%%%%%%%%%%
\subsection{Discussion}
\label{sec:disc}
%%%%%%%%%%%%%%%%%%%%

The capture rate we have derived for the wino is very small, consistently well below the direct annihilation cross section. Furthermore, at low masses the dominant capture mode (for velocities typical of the Milky Way halo) is to the $1s$ spin-triplet state, which is a pure- chargino bound state that subsequently decays dominantly through $s$-channel annihilation to SM quarks. Thus the presence of bound states will not directly enhance the annihilation rate by a significant fraction, and in particular will not enhance the gamma-ray line cross section. Previous calculations of the gamma-ray line signal, neglecting the impact of radiative capture into bound states, thus remain valid. 

One might ask to what degree this conclusion is generic to complex dark sectors, where the DM interacts through the exchange of multiple force carriers and may have nearly-degenerate partner particles. Compared to the case of positronium, where the capture rate dominates the direct annihilation rate by a factor of a few at low velocities, there are three principal sources of suppression of the capture cross section for the wino:
\begin{itemize}
\item Only some fraction of the propagating two-particle state couples to the radiated particle (the photon, in our case), leading to $\mathcal{O}(1)$ suppression factors. Thus, for example, the capture cross section scales as $\alpha \, \alpha_W^2$ in the high-mass limit, whereas the direct annihilation cross section scales as $\alpha_W^3$. Factors of this form will be generic in complex dark sector models, although their exact size will vary.
\item For positronium the capture into the ground state dominates but for the wino this capture rate is generically suppressed, as it vanishes in the Coulombic limit due to an accidental cancellation. This suppression is not universal to other dark sector models. For models where this term does not vanish in the Coulomb limit it still may be suppressed at low velocities since it involves a $p$-wave initial state; this suppression does not affect the $s$-wave direct annihilation cross section. This factor depends on the mass of the force carriers relative to the mass of the DM; in non-electroweakino DM models, there is much greater freedom to adjust the force carrier mass and hence the degree of velocity suppression. For example, lowering the force carrier masses will reduce the effect of the velocity suppression on the capture rate from higher-partial-wave components of the continuum wavefunction, since this velocity suppression scales as $(M_\chi v/m_W)^{2L}$.   Also, for $m_W \ll M_\chi$, we enter the Coulombic regime, where there is no velocity suppression for higher partial waves, and we recover a $1/v$ scaling in the capture rate.

\item There is also the apparently accidental $e^{-4n\lambda_i/\lambda_f}$ factor appearing in the cross section for capture, arising from the overlap integral between the continuum and bound states, which does not affect the direct annihilation cross section. For positronium and the case of capture to the ground state, $\lambda_i=\lambda_f$ and this factor is just $e^{-4}$. For the wino case, where $\lambda_i = 2 = 2 \lambda_f$, this factor is $e^{-8}$ at most, and it increasingly suppresses capture into bound states with higher principal quantum number. This factor is \emph{not} universal; for example, for a simple two-state model coupled to a single force carrier \cite{Slatyer:2009vg} we have $\lambda_i/\lambda_f=1$, for fermionic DM transforming as a SU(2) doublet (quintuplet) we find $\lambda_i/\lambda_f= 1$ ($\frac 6 5$ and $\frac 3 5$, as the quintuplet has two eigenvectors that experience an attractive potential, both of which can contribute to capture) \cite{Cirelli:2007xd}.

\end{itemize} 
Dark sectors where the bound states experience a stronger attractive potential than the continuum states, or where the ratio of force carrier mass to DM mass is not much larger than typical velocities in the Milky Way halo, are therefore more likely to have large cross sections for capture relative to direct annihilation.

One might also ask whether the photons radiated on capture \emph{themselves} constitute a detectable signal. In principle, detecting lines from capture and/or transitions between bound states could allow study of the quantum numbers of the DM. However, because the capture rate for wino-like DM is so low and the mass scales where bound states occur are quite high, for this particular toy model this would be an extremely challenging search. Assuming an NFW DM profile with local DM density $\rho(8.5 \text{kpc}) = 0.4$ GeV/cm$^3$ and scale radius 20 kpc, and (as a benchmark) 10 TeV DM with a capture cross section of $10^{-29}$ cm$^3$/s, we find that on average one would receive $\mathcal{O}(10^{-4})$ photons/m$^2$/yr at Earth from the whole Milky Way halo. From the region within 1 degree of the Galactic center, the rate is instead $\mathcal{O}(10^{-5})$ photons/m$^2$/yr. This rate is prohibitively small for any reasonable space-based telescope. Ground-based gamma-ray telescopes, on the other hand, can have effective areas of $\sim 10^{5-6}$ m$^2$ and so might be able to observe a very small number of capture photons -- but current and near-future ground-based telescopes have low-energy thresholds in the $10-20$ GeV range or higher, which would need to be lowered by an order of magnitude to observe capture and transition photons from $\mathcal{O}(10)$ TeV DM (for which the deepest bound states accessible by capture have $E_n \sim 1$ GeV), and would likely also need excellent energy resolution in order to isolate such a small line signal from substantial astrophysical backgrounds. Higher DM masses would produce capture line photons with higher energies -- e.g. 25 GeV for 100 TeV DM -- but would also correspond to a much lower DM number density, suppressing the already-low rate of possible detections. However, if an annihilation signal had already been detected, such a search would be well-motivated,  and might provide one of the only ways to probe the particle properties of the DM in the absence of a discovery at a collider. Detection of a high-energy annihilation signal would also open up other options in searching for the capture transition lines, for example by examining cross-correlations with the DM annihilation spatial distribution.

%%%%%%%%%%%%%%%%%%%%%%%
\section{Conclusions}
\label{sec:conclusion}
%%%%%%%%%%%%%%%%%%%%%%%

We have computed the rate for formation of wino-onium bound states, and their subsequent decays to lower-energy states or SM particles. We find that bound state formation by single photon emission is possible for large wino masses, $M_\chi \gtrsim$ 5.6 TeV, but in general, and in contrast to the case of positronium, the capture rate is subdominant to direct annihilation. Consequently, previous calculations of the detectability of e.g.~high-energy gamma-ray lines from wino DM should not require significant modification in most of parameter space.

This scenario has several novel features relative to the case of positronium, or dark-sector configurations where there is only one DM state and the potential is mediated by a single dark photon. Many of these features will generalize to any complex dark sector where the gauge group is nonabelian and the potential couples together several nearly-degenerate dark-matter-like states.

Spin statistics demands that only two-particle states with even $L+S$ can possess a $\chi^0 \chi^0$ component; states with odd $L+S$ must in this case be entirely comprised of $\chi^+ \chi^-$. Consequently, states with odd vs even $L+S$ experience different effective potentials and form distinct towers of bound states, which will generically be displaced from each other in energy. The unsuppressed decay channels to lower-energy bound states may thus be very different from the familiar case of hydrogen-like atoms. The annihilation channels of the two towers of states are also quite different; for the wino, states with even $L+S$ decay primarily to gauge bosons, whereas those with odd $L+S$ decay primarily through an $s$-channel diagram to quarks and leptons.

The presence of massive force carriers generically suppresses the capture cross section at low velocities, by suppressing all contributions from initial states with $L > 0$. However, the distortion of the continuum wave functions due to the presence of near-threshold bound states can lead to resonant enhancement of the capture cross section, in the same way that resonant Sommerfeld enhancement leads to a larger direct annihilation cross section. Furthermore, for the wino and for velocities typical of the Milky Way halo, the capture rate can have a significant velocity dependence, 
in contrast to direct annihilation.

Detection of the low-energy photon lines ($\mathcal{O}$(GeV) energies for 10 TeV+ DM) from radiative capture and transitions between bound states could potentially provide a unique probe into the gauge structure of the dark sector. However, for the heavy wino this search appears very challenging, due to the low number density of multi-TeV DM; experiments designed to search for high-energy gamma rays have large enough effective areas to observe these photons, but their energy threshold is presently too high to have sensitivity, and furthermore the gamma-ray backgrounds at these low energies are substantial.

In contrast to the features discussed above, the factors which suppress the wino-onium capture cross section are \emph{not} generic; they depend sensitively on the representation of the DM under the gauge group, and the relative masses of the DM and force carriers. Thus the formation of bound states cannot be safely ignored in models with non-trivial dark sectors. We have presented general analytic results for the capture rate into DM bound states in the limit where the force carriers are very light and the gauge symmetry is approximately unbroken, to facilitate estimates of whether the capture rate can be important for a given dark-sector model. In such models, the presence of bound states could enhance the capture rate, change the branching ratio to different SM final states, and perhaps generate non-negligible transition lines -- although if the dark gauge group is not the electroweak gauge group, the transition lines would presumably be comprised of ``dark photons'', and their observable signatures would depend on the coupling of those dark photons to the SM.

\section*{Acknowledgements}
We thank Eric Braaten, Maxim Pospelov, Ira Rothstein, Iain Stewart, and Scott Thomas for useful discussion. We particularly thank Julia Harz and Kalliopi Petraki for pointing out a sign error in an earlier version of this work. This work is supported by the U.S. Department of Energy under grant Contract Numbers DE-SC00012567 and DE-SC0013999.  MB and PA are supported by Contract Numbers DE-SC0003883.  MB thanks the Aspen Center for Physics for its hospitality where a portion of this work was completed.

\appendix 

%%%%%%%%%%%%%%%%%%%%%%
\section{Numerical method for computation of scattering states}
\label{app:init}
%%%%%%%%%%%%%%%%%%%%%%

Our initial-state wavefunctions are positive-energy solutions of the Schr\"{o}dinger equation,
\bea
H^0_{L+S \, {\rm even}} \; \Psi &=& E \, \Psi, \nn \\
E &=& \frac{M_\chi v^2}{4}
\label{eq:appschro}
\eea
with 
\bea
V^0_{L+S \, {\rm even}}(r) = \left( \begin{array}{cc}
0  & -\sqrt{2}\alpha_W \frac{e^{-m_W r}}{r}   \\
-\sqrt{2}\alpha_W  \frac{e^{-m_W r}}{r}  & \;\;  2\delta M -\frac{\alpha}{r} - \alpha_W c_W^2 \frac{ e^{-m_Z r}}{r}
\end{array} \right).
\label{eq:apppotltwo}
\eea
Asymptotically, we are describing a state of two, free neutral WIMPs, $\chi^0\chi^0$, and thus know the energy eigenvalue in eq.~\ref{eq:appschro}.  The fact that our state contains two identical Majorana fermions fixes $L+S$ to be even in order to have a globally antisymmetric wavefunction.  Since $V$ is spherically symmetric, we can expand the general solution in Legendre polynomials,
\be
\Psi({\bf r})_a \,=\, \sum_L \frac{\left[ u_L(r) \right]_{a}}{r} \, A_{L\, a} \, P_L(\cos\theta).
\label{eq:gensol}
\ee
The $A_{L \,a}$ will ultimately be fixed by normalization considerations, leaving the nontrivial task of determining the reduced wavefunctions, $u_L(r)$.\footnote{Although the potential has spherical symmetry, our asymptotic solution contains an incoming plane wave.  General scattering theory dictates that at large $r$, $\Psi({\bf r})_a = e^{i k_a z} + f_a(\theta)e^{i k_a r}/r$.  Thus, the solution still possesses cylindrical symmetry, justifying the independence of the general form, eq.~\ref{eq:gensol}, on the azimuthal angle, $\phi$.}  The behavior of $u_L(r)$ near the origin strongly deviates strongly from that of a plane wave.  This leads to the well-known Sommerfeld enhancement in the direct annihilation of $\chi^0 \chi^0 \rightarrow \gamma+X$ \cite{Hisano:2004ds}.  This nonperturbative effect can only be treated numerically and there is now a well developed literature on computing the wavefunction at the origin \cite{Slatyer:2009vg,Beneke:2014gja}. Since annihilation of the incoming state proceeds via a highly-off-shell WIMP, to leading power in the velocity expansion, only $\Psi(0)$ is needed.  As seen in section \ref{sec:decay}, the rate to capture to a bound state requires an overlap integral with the bound-state wavefunction (cf.~eqs.~\ref{eq:sigmastop} and \ref{eq:sigmaptod}).  Since the bound states are spatially compact, their wavefunctions will decay exponentially past some number of Bohr radii, and as a practical matter, we only need the initial, scattering states out to this distance.  Nonetheless, we are still responsible for determining the function $\Psi({\bf r})$ (or $u_L(r)$) over a range of values.  We cannot simply quantify the non-perturbative physics with a single number as in the annihilation problem.  

We will find it useful to work with a dimensionless radial variable, $x(\equiv p\,r = M_\chi v_\text{rel} \, r/2)$.  Thus, we are solving the following reduced-wavefunction problem:
\be
\left[ u_L^{\prime\prime}(x) \right]_a + \left[ \left( 1 - \frac{L(L+1)}{x^2} \right) \delta^{ab} - \frac{V^{ab}(x)}{E} \right] \, \left[ u_L^{\prime\prime}(x) \right]_b.
\label{eq:schrox}
\ee
For completeness, the rescaled potential term is
\bea
\frac{V(x)}{E} = \left( \begin{array}{cc}
0  & -2\sqrt{2}\alpha_W \frac{e^{-2 m_W x/(M_\chi v_\text{rel})}}{x\,v_\text{rel}}   \\
-2\sqrt{2}\alpha_W  \frac{e^{-2 m_W x/(M_\chi v_\text{rel})}}{x\,v_\text{rel}}  & \;\;  \frac{8\delta M}{M_\chi v_\text{rel}^2} -\frac{2\alpha}{x\,v_\text{rel}} - 2\alpha_W c_W^2 \frac{e^{-2 m_Z x/(M_\chi v_\text{rel})}}{x\,v_\text{rel}}
\end{array} \right).
\label{eq:potlresc}
\eea
Although $V(x)$ creates significant distortion at small $x$, at large $x$ we recover a free theory, but with the charged component exponentially decaying due to the mass-shift term, $\frac{8\delta M}{M_\chi v_\text{rel}^2}$.  It is straightforward to write the asymptotic solutions to eq.~\ref{eq:schrox} in terms of the normalized wavenumbers, $\hat{k}_a = k_a/p$, with
\bea
\hat{k}_1 &=& 1 \nn \\
\hat{k}_2 &=& i\, \sqrt{\frac{8\delta M}{M_\chi v_\text{rel}^2}-1}.
\label{eq:khats}
\eea
At the level of pure math, at large $x$ we get solutions that are linear combinations of $e^{\pm i \, \hat{k}_i\, x}$.  For the neutral component, we just need to determine the appropriate phase in the sinusoid.  The charged component, however, contains an intrinsic instability for numerical evaluation.  On physical grounds, by the (plane-wave) normalizability of the wavefunction, at long distances we will only get an exponentially-decaying term.\footnote{For a sufficiently massive wino, $M_\chi >$ 1320 TeV for $v=10^{-3}$, we get a real-valued $\hat{k}_2$.   Thus, the charged component can also propagate to spatial infinity.  This marks a qualitative change to the problem we are considering, as we must now consider the full SU(2) triplet throughout the calculation.  This is an interesting regime, and one much closer to the Coulomb-limit cases we have discussed.  Its full treatment is nonetheless beyond our scope.}  However, in practice it is intractable to set the boundary conditions for this second-order equation precisely enough to obtain a numerical solution that is purely-decaying.  Tiny errors will generate an exponentially-growing term that eventually dominates and spoils the wavefunction.  For this reason, in the current implementation of Mathematica's \texttt{\texttt{NDSolve}}, we can only obtain a solution to the direct Schr\"{o}dinger equation (\ref{eq:schrox}) for $M_\chi \lesssim$ 100 TeV.  We will thus employ a numerical technique developed in the nuclear community \cite{Ershov:2011} known as the Variable Phase Method.  Much of the setup follows \cite{Beneke:2014gja}, who developed this procedure in a DM context in order to compute the value of the wavefunction at the origin.  We will extend their approach to find the wavefunction in a range out to several hundred Bohr radii.  To our knowledge, the details of our method are novel and provide an efficient, powerful means to calculate positive-energy, multi-component DM wavefunctions.

The Variable Phase Method treats the reduced wavefunction as a spatially-varying linear combination of solutions to the free Schr\"{o}dinger equation with the appropriate normalized wavenumber,
\be
\left[ u_L^{\prime\prime}(x) \right]_a + \left[ \hat{k}_a^2 - \frac{L(L+1)}{x^2}  \right] \left[ u_L(x) \right]_a \,=\, 0. 
\label{eq:freeschrohat}
\ee
Following \cite{Beneke:2014gja}, we take the following free reduced wavefunctions,
\bea
f_a(x) &=& \sqrt{\frac{\pi x}{2}} J_{L+\frac1 2}(\hat{k}_a x) \nn \\
g_a(x) &=& -\sqrt{\frac{\pi x}{2}} \left[ Y_{L + \frac 1 2}(\hat{k}_a x) \,-\, i \, J_{L+\frac1 2}(\hat{k}_a x) \right],
\label{eq:freesols}
\eea
which have the appropriately-normalized Wronskian, $f^\prime \, g - f \, g^\prime =1$.  We can now write the solution to the full problem, eq.~\ref{eq:schrox}, as 
\be
\left[ u_L(x) \right]_a \,=\, f_a (x) \, \alpha_a(x) \,-\, g_a (x) \, \beta_a(x).
\label{eq:ufg}
\ee
Needing $\alpha_a(x)$ and $\beta_a(x)$ doubles the degrees of freedom in the solution, so we eliminate this redundancy by imposing the normalization
\be
f_a (x) \alpha_a^\prime (x) \,-\, g_a(x) \beta_a^\prime(x) = 0.
\ee
Instead of using the parametrization of the wavefunction in eq.~\ref{eq:ufg}, we define
\be
N_{ab} \,=\, f_a \, g_a \delta_{ab} - g_a O_{ab} \, g_b,
\label{eq:ndef}
\ee
where $\beta_a = O_{ab} \, \alpha_b$, and 
\be
\tilde{\alpha}_a = \frac{\alpha_a}{g_a}.
\ee
These combine so that 
\be
u_a(x) = N_{ab} \, \tilde{\alpha}_b.
\label{eq:unatilde}
\ee
The advantage of writing the reduced wavefunction in this way is that we can now determine it from a pair of nested, first-order equations,
\bea
N^\prime_{ab} &=& \delta_{ab} + \left( \frac{g^\prime_a}{g_a} + \frac{g^\prime_n}{g_b} \right) N_{ab} - N_{ac} \frac{\hat{V}_{cd}}{E} N_{db} \nn \\
\tilde{\alpha}^\prime &=& \left( -\frac{g^\prime_a}{g_a} \delta_{ab} + \frac{\hat{V}_{ac}}{E} N_{cb}\right) \tilde{\alpha}_b.
\label{eq:nalphaeqs}
\eea
We get $\hat{V}(x)$ from the full potential, eq.~\ref{eq:potlresc} by only keeping those terms which vanish as $x \rightarrow \infty$, ${\rm i.e.}~V(x) = V_{\rm inf.} + \hat{V}(x)$, and $\lim_{x \rightarrow \infty} V(x) = V_{\rm inf.}$.
Furthermore, we know that in the limit of very small $x$, our problem just becomes that of a Coulomb potential.  Thus, near the origin $u_a(x) \propto x^{L+1}$.  However, we do not know the appropriate prefactor.  In the $L=0$ case, this is just the Sommerfeld factor, or wavefunction at the origin, for which many numerical determinations exist.  We could employ such a method, and then take an appropriate linear combination of solutions determined with independent sets of boundary conditions to determine the full wavefunction.  However, we have found a way to impose a boundary condition on eq.~\ref{eq:nalphaeqs} that requires no additional inputs and finds the correct wavefunction.  Following the method of \cite{Beneke:2014gja} for finding the Sommerfeld factor, for some small $x$ (which we generally take as $x_0 = 10^{-6}$), we demand
\be
N_{ab}(x_0) = \frac{x_0}{2L+1} \delta_{ab}.
\label{eq:xbc}
\ee
We found that in practice, imposing $\tilde{\alpha}(x)$ boundary conditions also at $x_0$ led to numerical instability beyond $\mo(10{\rm s})$ of Bohr radii, which was insufficient for the overlap integral needed to compute the capture rate.  Just as in other numerical routines, the exponentially growing solution of the charged component eventually overwhelms the solution.

Instead, we impose the following boundary condition at relatively large $x_f$,\footnote{We found minimal sensitivity to the exact location of $x_f$.  Typical values taken were $x \in (50,\, 100)$. }
\bea
\tilde \alpha_1 (x_f) &=& 1, \nn \\
\tilde \alpha_2 (x_f) &=& 0. 
\label{eq:atbc}
\eea
Fixing the charged component to vanish at large $x$ eliminates the exponentially growing solution from our regime of interest.  In fact, we imposed a similar boundary condition on the charged component when solving for the full $u(r)$ directly with \texttt{NDSolve}.  The advantage of the Variable Phase Method over the direct approach is that the equations are first-order.  Thus, \texttt{Mathematica} is not solving a Boundary Value Problem, shooting solutions from one boundary to the other and attempting to line them up.  A key signal of breakdown in the direct method was the inability to satisfy the boundary conditions at both ends.  This issue simply does not arise with $\tilde \alpha(x)$.

The further $\tilde \alpha_1 (x_f) = 1$ condition, along with those imposed on $N_{ab}$ ({\it cf.}~eq.~\ref{eq:xbc}), is actually sufficient to determine the physical solution.  To see this it is useful to rewrite eq.~\ref{eq:nalphaeqs} in the limits of both small and large $x$, given the boundary conditions for $N_{ab}$ in eq.~\ref{eq:xbc}.  These decouple $N_{12(21)}$ at small $x$ and fix $N_{11(22)}(x_0) \sim x_0$ in the region of some arbitrarily small $x_0$.  Thus, we get 
\bea
N_{aa}^\prime &=& 1 + 2 \, \frac{g_a^\prime}{g_a} N_{aa} \nn \\
\tilde \alpha_a^\prime &=& -\frac{g_a^\prime}{g_a} \tilde \alpha_i,
\label{eq:nalphasmallx}
\eea
where there is no sum over $a$.  For our choice of $g_a$ in eq.~\ref{eq:freesols},
\begin{align}
\lim_{x \rightarrow 0} \, \frac{g_a^\prime}{g_a} &= \left\{ \begin{array}{cc} i\, \hat{k}_a &  L=0 \\ 
-\frac{L}{x} & L \neq 0 \end{array} \right. .
\end{align}
The functional dependence is different for $L=0$, but the scaling for $N_{aa}$ and $\tilde \alpha_a$ will be the same as for higher $L$.  We get for small $x$,
\bea
N_{aa} &\propto& x \nn \\
\tilde \alpha_a &\propto& x^L.
\eea
This sets $u(x) \propto x^{L+1}$, which is exactly the scaling of the regular solution to the Schr\"{o}dinger equation with a Coulomb potential.  Thus, we have obtained the correct physical solution up to an overall scaling factor.  However, we know that at sufficiently large $x$, the potential is negligible, and thus the charged component will be exponentially suppressed, and the neutral component will just be a unit amplitude sinusoid, up to known factors that depend only on $L$ and the number of identical particles in the initial state (see eq.~\ref{eq:norm}):
\be
u_1(x) = \sin \left( \! x -\frac{\pi \, L}{2} + \delta_L(x) \right),
\label{eq:ulimit}
\ee
with $\delta_L \ll x$.  Setting $\tilde \alpha_2(x_f) = 0$ is a good approximation to the exponential suppression for sufficiently large $x_f$.\footnote{Our method trivially generalizes to the case where $u_2$ is non-decaying at large $x$.  Using the normalization conditions spelled out in \cite{Schutz:2014nka}, we can still impose that the radial component of $u_2$ is purely outgoing.  Operationally, this is realized with the same boundary conditions as in eqs.~\ref{eq:xbc} and \ref{eq:atbc}.}  We will now show how $\tilde \alpha_1(x_f) = 1$ reproduces the normalization of eq.~\ref{eq:ulimit}.  To start, we get the $N,\, \tilde \alpha$ equations in the large-$x$ limit.  This behavior is determined by the log derivative of $g_a$,
\be
\lim_{x \rightarrow \infty} \, \frac{g_a^\prime}{g_a} \,=\, i\, \hat{k}_a.
\ee
This result is $L$-independent, because for sufficiently large $x$, the centrifugal term drops out of eq.~\ref{eq:freeschrohat}.  We therefore get the following equations for $N_{ab}$,
\bea
N_{11}^\prime &=& 1 + 2 \, i \, N_{11} \nn \\
N_{12}^\prime &=& - \left(\sqrt{\frac{8\delta M}{M_\chi v_\text{rel}^2}-1} - \,i \right) N_{12} \nn \\
N_{22}^\prime &=& 1 - \left( 2 \, \sqrt{\frac{8\delta M}{M_\chi v_\text{rel}^2}-1} \, \right)  N_{22},
\label{eq:nlargex}
\eea
and for our potential and $N_{ab}$ boundary conditions, $N_{21}=N_{12}$.  These have the following solutions,
\bea
N_{11} &=& \frac{i}{2} + A_{11} \, e^{2i \, x} \nn \\
N_{12} &=& A_{12} \exp \left[ \left(-\sqrt{\frac{8\delta M}{M_\chi v_\text{rel}^2}-1} + \,i \right)x \right] \nn \\
N_{22} &=& \frac{1}{2\sqrt{\frac{8\delta M}{M_\chi v_\text{rel}^2}-1}} + A_{22} \exp \left[ -\left( \sqrt{\frac{8\delta M}{M_\chi v_\text{rel}^2}-1} \right)x  \right].
\label{eq:nlargex2}
\eea
We see that $A_{12}$ and $A_{22}$ decay exponentially, the latter to a predetermined, nonzero value.  Since we are further setting $\tilde \alpha_2(x_f)$ = 0, it is just the solution to $N_{11}$ we need to examine in detail to determine the proper normalization of $u_1(x)$.  The $N_{ab}$ satisfy a first-order equation.  Therefore, having imposed the boundary condition at small $x$, the value of $A_{11}$ is predetermined, but its value is unknown based on the equation's behavior at large $x$.  To constrain it, we now examine the behavior of $\tilde \alpha_1$ at large $x$.  It obeys the trivial equation,
\bea
\tilde \alpha_1^\prime &=& -i \, \tilde \alpha_1. \;\;\;\; {\rm Thus,} \nn \\
\tilde \alpha_1 &=& C \, e^{-i\,x},
\label{eq:alphalargex}
\eea
where $C$ is the free parameter we can set with the boundary condition at large $x$.  Combining $N$ and $\tilde \alpha$ to get $u_1(x)$, we find
\bea
u_1(x) &=& C \left[ A_{11} \, e^{i\,x} - \frac{e^{-i\, x}}{2i}  \right] \nn \\
&=& B \, \sin \left( \! x -\frac{\pi \, L}{2} + \delta_L(x) \right),
\eea

\begin{figure}[ht]
\centerline{\scalebox{.7}{\includegraphics{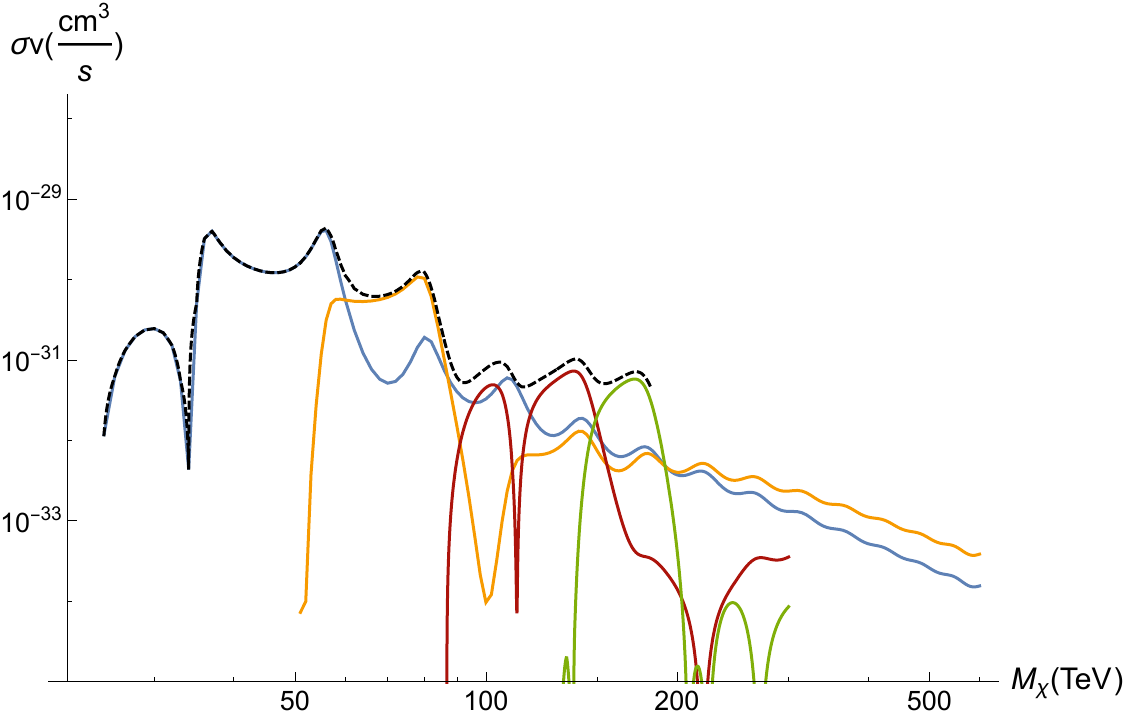}}}
\vskip-0.2cm
\caption[1]{The capture rate for $\chi^0 \chi^0 (s-{\rm wave}) \rightarrow \,  ^1\!P_1+\gamma$, summed over (\emph{black-dashed}) all accessible bound states, up to the first 4, (\emph{blue}) $n=2$ bound states only, (\emph{yellow}) $n$=3 bound states only, (\emph{red}) $n$=4 bound states only, (\emph{green}) $n$=5 bound states only.}
\label{fig:sstates} 
\end{figure}

\begin{figure}[ht]
\begin{center}
\includegraphics[scale=0.65]{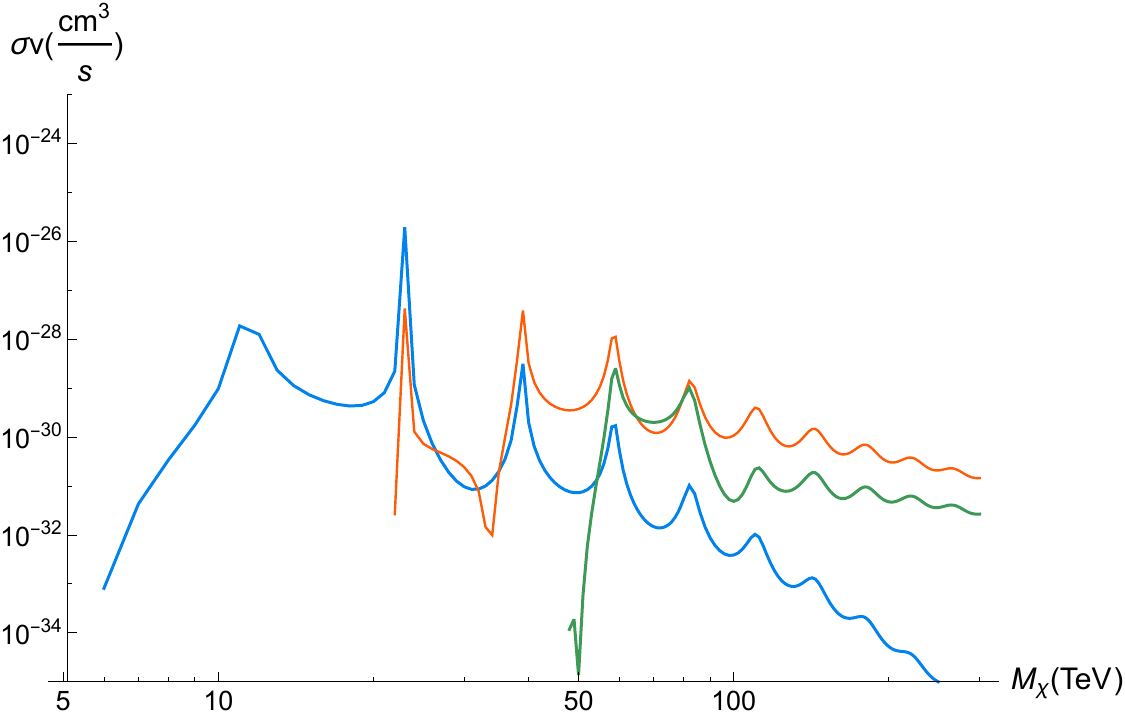}

\vspace{0.5cm}

\includegraphics[scale=0.65]{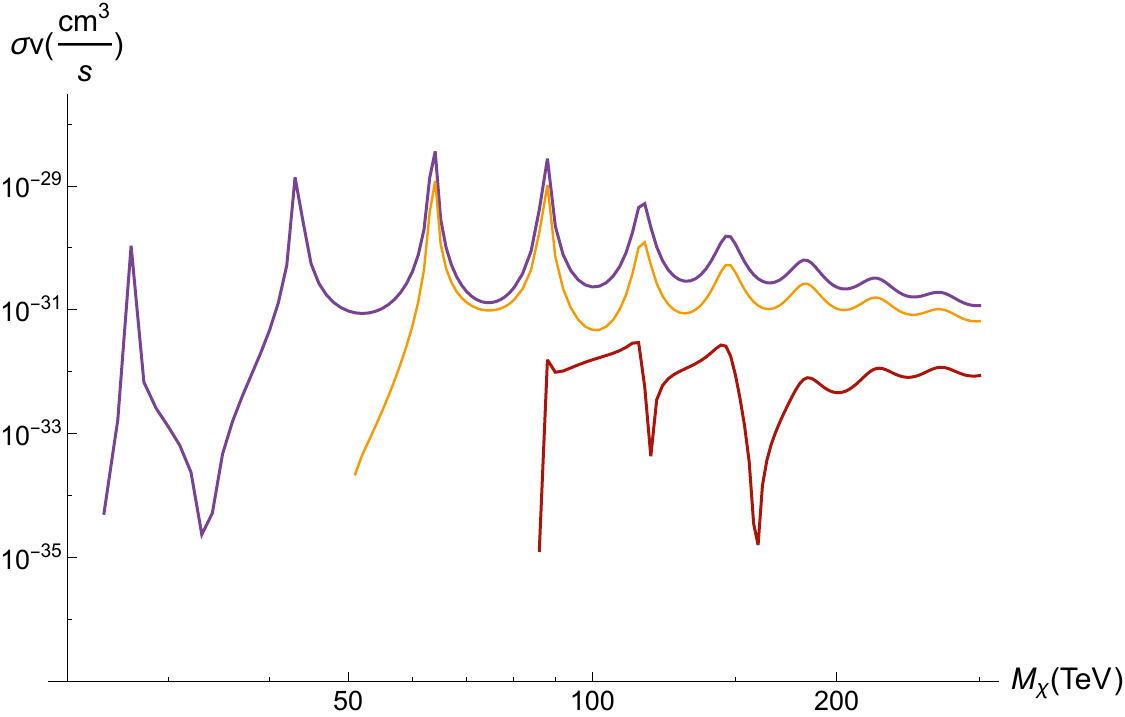}
\vskip-0.2cm
\caption[1]{{\bf Top}: the capture rate for $\chi^0 \chi^0 (p-{\rm wave}) \rightarrow \,  ^3\!S_1+\gamma$, for (\emph{blue}) $n$=1 bound states only, (\emph{orange}) $n$=2 bound states only, (\emph{green}) $n$=3 bound states only. {\bf Bottom:} The capture rate for $\chi^0 \chi^0 (d-{\rm wave}) \rightarrow \,  ^1\!P_1+\gamma$, (\emph{purple}) $n=2$ bound states only, (\emph{yellow}) $n=3$ bound states only, (\emph{red}) $n=4$ bound states only.}
\label{fig:pdstates} 
\end{center}
\end{figure}

Finally, in figure \ref{fig:capture}, we showed the spin-averaged capture rates computed with eqs.~\ref{eq:sigmastop} and \ref{eq:sigmaptod} for the bound states with the lowest principal quantum numbers $n$ (summed over $m$) for $s$, $p$, and $d$-wave initial states. 
For completeness, in figures \ref{fig:sstates}-\ref{fig:pdstates}, we show how these summed rates break down into capture into each of the individual lowest-lying bound states (while still summing over $m$).  For the $s$-wave component of the initial state, this includes the contributions of all accessible bound states for $M_\chi$ up to 300 TeV. For the $p$-wave ($d$-wave) components of the initial state, we consider $n$=1\textendash 3 ($n$=2\textendash 4).
%

%%%%%%%%%%%%%%%%%%%%%%%%%%%%%%%%%
\section{Numerical method for computation of bound states}
\label{app:bs}
%%%%%%%%%%%%%%%%%%%%%%%%%%%%%%%%%

The bound state wavefunctions solve a Schr\"{o}dinger equation with the appropriate potential depending on the $L+S$ of the DM pair.  Generically, the neutralino ($\chi^0\chi^0$) and chargino ($\chi^+\chi^-$) pairs can turn into one another by exchange of $W^\pm$ bosons.  However, since the neutralinos are identical fermions, they require a globally antisymmetric wavefunction and therefore $L+S$-even.  Thus, $V_{L+S \, \text{even}}$ is a matrix potential with off-diagonal entries between these pairs, while $V_{L+S \, \text{odd}}$ only evolves $\chi^+\chi^-$. As in the main text, we put the neutralino and the chargino components of the bound state's wavefunctions into a vector $\Psi^\top = \left(\psi_N \, \psi_C \right)$. The reduced radial wavefunction, $u_i(r)=r\, \psi_i(r)$, solves the Schr\"{o}dinger equation
\begin{equation}
-\frac{1}{M_\chi} 
\partial^2_r\left(\begin{matrix}
u_N (r)\\u_C (r)
\end{matrix} \right) + 
\left(\begin{matrix}
\frac{L(L+1)}{M_\chi r^2}+V_{11} & V_{12}\\
V_{21} & \frac{L(L+1)}{M_\chi r^2}+ \, 2\delta M + V_{22}
\end{matrix} \right)
\left(\begin{matrix}
u_N(r)\\u_C(r)
\end{matrix} \right)= \frac{M_\chi v_\text{rel}^2}{4} \left(\begin{matrix}
u_N (r)\\u_C (r)
\end{matrix} \right),
\label{bs.1}
\end{equation}
where the energy of the state is defined as $E=M_\chi v_\text{rel}^2/4$.

To simplify our analysis, we use the variable $z\equiv M_\chi \alpha_W \, r$, which normalizes the physical distance $r$ to half the Bohr radius of WIMPonium in the high-mass limit.  Solving the radial Schr\"{o}dinger equation for the reduced wavefunction $u(z)=z\psi (z)$ with $V_{L+S-\text{even}}$, we have 
\begin{equation}
\left(\begin{matrix}
-\partial_z^2  + \frac{L(L+1)}{z^2} & - \sqrt{2} \frac{e^{-\epsilon_\phi z}}{z}\\
- \sqrt{2} \frac{e^{-\epsilon_\phi z}}{z}& -\partial_z^2  + \epsilon_\delta^2 + \frac{L(L+1)}{z^2} - \frac{c_W^2 e^{-\frac{\epsilon_\phi}{c_W}z}}{z}-\frac{s_W^2}{z}
\end{matrix}
\right) \left( \begin{matrix}
u_N(z)\\u_C(z)
\end{matrix}\right) =\epsilon_v^2 \left( \begin{matrix}
u_N(z)\\u_C(z)
\end{matrix}\right),
\label{bs.2}
\end{equation}
where $c_W$ ($s_W$) stands for cosine (sine) of Weinberg angle, $\epsilon_v = v_\text{rel}/2\alpha_W$, $\epsilon_\phi = m_W/(\alpha_W M_\chi)$, and $\epsilon_\delta = \sqrt{2\delta M/M_\chi}/\alpha_W$ with $\delta M = M_{\chi^\pm}-M_\chi^0$. 

If $L+S$ is odd, our bound state only has a chargino component, but we keep the matrix notation, giving
\begin{equation}
\left(\begin{matrix}
0 & 0\\
0 & -\partial_z^2 + \epsilon_\delta^2 + \frac{L(L+1)}{z^2} - \frac{c_W^2 e^{-\frac{\epsilon_\phi}{c_W}z}}{z}-\frac{s_W^2}{z}
\end{matrix}
\right) \left( \begin{matrix}
0\\ u_C(z)
\end{matrix}\right)=\epsilon_v^2 \left( \begin{matrix}
0\\u_C(z)
\end{matrix}\right).
\label{bs.3}
\end{equation}
A detailed derivation of these potentials can be found in \cite{Hisano:2004ds,Baumgart:2014saa}. In both eqs.~\ref{bs.2} and \ref{bs.3} the 22 component has a Coulomb term corresponding to the photon.  Thus, the $\chi^+ \chi^-$ will form bound states at any value of $M_{\chi^\pm}$, in analogy to positronium; in the limit $M_{\chi^\pm} \rightarrow 0$, the charged component will dominate the $L+S$-even wavefunction, as the off-diagonal mixing terms become exponentially suppressed and the neutralinos decouple from the attractive potential.

We are dealing with a mixture of Coulomb and Yukawa potentials. In both the high and low-mass limits, the non-zero components of the potential will become purely Coulombic with coupling $\alpha$ (low-mass) and $2\alpha_W$ ($\alpha_W$) for the high-mass $L+S$-even (odd) potentials. Using this insight, we expand our wavefunctions for a given $L$ in a basis of solutions to the Coulomb Schr\"{o}dinger problem with the same $L$, exploiting the fact that we still have rotational symmetry in the full problem. This will turn the Schr\"{o}dinger equation into a finite-dimensional, matrix eigenvalue problem.  We denote the Coulombic wave functions in the basis by $|n,l\rangle$ and the eigenfunctions of the full problem by $|\phi_l\rangle$.  The Schr\"{o}dinger equations thus becomes
\begin{equation}
\hat{H} |\phi_l \rangle \equiv \sum_{n=1}^N c_{n,l} \, \hat{H}  |n,l \rangle,
\label{bs.6}
\end{equation}
where the $c$ coefficients are used to write the eigenfunctions of the Hamiltonian in terms of $N$ Coulomb wavefunctions and $\hat{H}$ is the Hamiltonian. Inserting a unit operator in the space of wavefunctions with fixed $L$, we have the following eigenvalue problem
\begin{equation}
\hat{H} |\phi_l \rangle = \sum_n  \langle m,0|\hat{H}|n,0\rangle c_{n,l} |m,l \rangle = E_a \, c_{m,l}| m,l  \rangle,
\label{bs.7}
\end{equation}
where the $a$ in $E_a$ refers to the radial quantum number.  Hence, the Schr\"{o}dinger equation reduces to the matrix equation,
\begin{equation}
H_{mn}c_{n,l} = E_a c_{m,l}.
\label{eq:finproblem}
\end{equation}
where $H_{mn}=\langle m,l|\hat{H}|n,l\rangle$. We thus have to find the overlap integrals $H_{mn}$ for the Hamiltonians corresponding to Eqs.~\ref{bs.2} and \ref{bs.3}. 

For odd $L+S$ configurations (eq.~\ref{bs.3}) we use a basis of functions solving the Coulomb potential $-\alpha_W/z$,  corresponding to the high-mass limit of $V_{L+S \text{odd}}$. For even $L+S$ configurations (eq.~\ref{bs.2}), the potential matrix is off-diagonal; however, in the $M_\chi \gg m_W$ limit we can perform an $r$-independent diagonalization of  the potential to obtain a repulsive Coulomb potential with $+\alpha_W/z$ (which does not contribute to the formation of bound states) and an attractive one with $-2\alpha_W/z$, as discussed in section \ref{subsec:himass}. As a result, to find the eigenfunctions for eq.~\ref{bs.2} we choose a basis of functions solving the Coulomb potential, $-2\alpha_W/z$.  We noted empirically that in order for this method to work satisfactorily, the coupling of the Coulomb potential used for the expansion in eq.~\ref{bs.7} needs to be as large or larger than that of the full problem.  The choices above satisfy this criterion as it is only in the $M_\chi \rightarrow \infty$ limit that the potential has full $\alpha_W$ or $2\alpha_W$ strength.  At any finite mass, the Yukawa terms lead to a weaker attractive force.  It makes sense that the expansion coupling needs to be larger than the true potential.  The true wavefunction will have support to a particular Bohr radius.  If every term in the Coulomb expansion is wider, it is difficult to recover the correct behavior.  Starting with Coulomb wavefunctions that are more compact than that of the full problem, we can build up the approximate wavefunction by including more and more radially-excited states.  We find stability across our mass regime of interest when the number of Coulomb eigenfunctions in the expansion of eq.~\ref{bs.7} satisfies $N \gtrsim$ 15, and we typically take $N$=30 for our capture rate plots.

\begin{figure}[h]
\begin{center}
\includegraphics[scale=0.30]{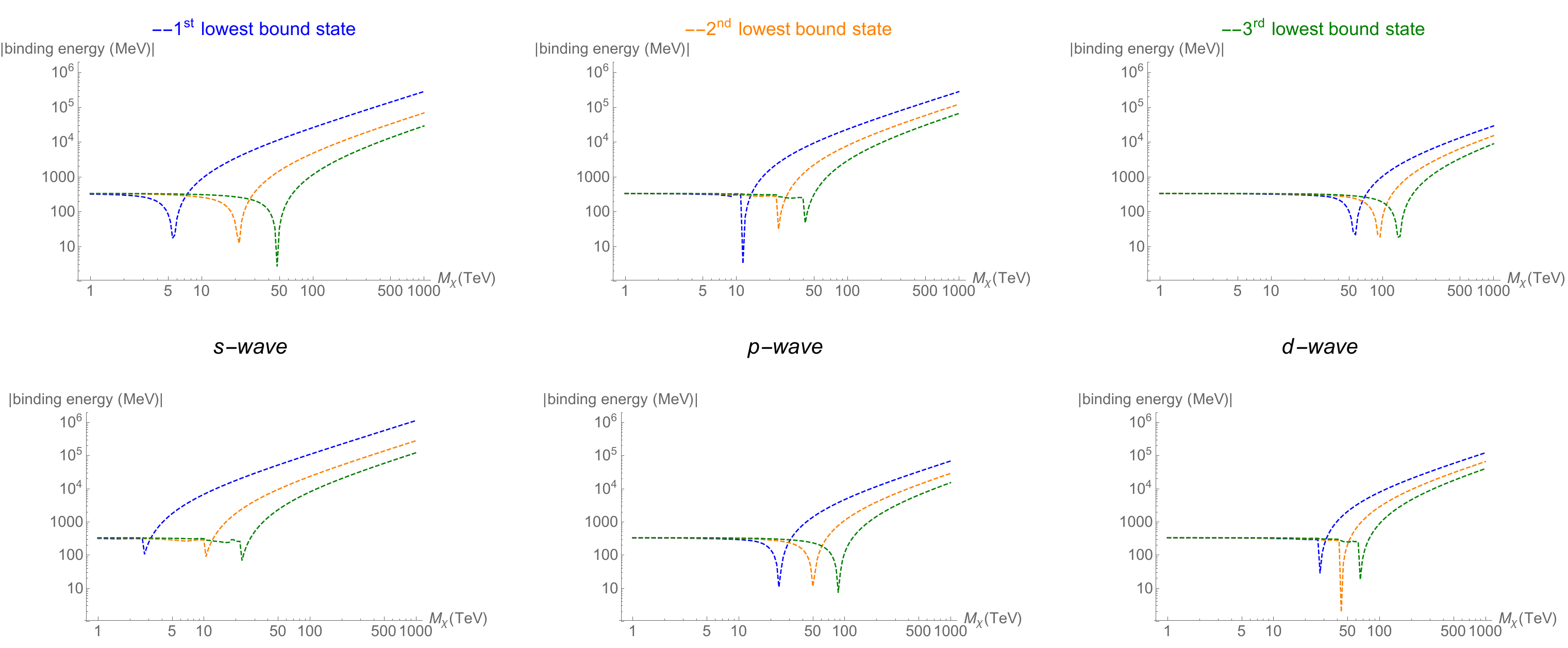}
\caption{The absolute value of the bound-state energy of the first three bound states, labeled by \emph{blue}, \emph{orange}, \emph{green} in order of increasing $n$. {\bf Top row:} spin-triplet configurations. {\bf Bottom row:} spin-singlet configurations. From {\bf left} to {\bf right}, panels correspond to $L=0,1,2$. All the diagrams follow a general trend: At low DM masses, where $\epsilon_\phi$ is large, the Yukawa potential is too short-range to hold the particles in a bound state.  The energy of the mostly $\chi^+\chi^-$ bound state -- which is held together by photon exchange down to $M_{\chi^\pm} \rightarrow 0$ -- is dominated by the mass splitting term, giving these states a positive binding energy in our convention. As we go to higher masses, corresponding to smaller $\epsilon_\phi$, the potential becomes long-range enough to form a negative-energy bound state. The dip on each line shows where the energy goes from being positive (compared to a neutralino pair at infinity) to being negative, as the DM mass increases.  At large enough masses, the potentials are effectively all Coulombic, and the binding energy becomes linearly dependent on DM mass, as expected.}
\label{fig:spectrum}
\end{center}
\end{figure}

We have performed several consistency checks of our numerical routine.  We recover the Coulomb limits at high and low masses with couplings $\alpha_W,\, \alpha$, respectively for $L+S$ odd, and $2 \alpha_W$ in the high-mass $L+S$ even limit.  We also recover some known results from the literature on the numerical study of solutions of Yukawa potentials \cite{PhysRevA.1.1577, PhysRevA.9.52}. The results therein suggest that for finite mediator mass, as we increase the ratio $\epsilon_\phi$, the bound states cease to exist.  Taking the limit $\delta M \rightarrow$ 0 and $\s_W^2 \rightarrow 0$, which corresponds to a pure-Yukawa problem with either a 2D (eq.~\ref{bs.2}) or 1D (eq.~\ref{bs.3}) potential and comparing the value of $\epsilon_\phi$ where this transition happens, we find an agreement between our results and \cite{PhysRevA.1.1577}. For instance, the maximum value of $\epsilon_\phi$ for which we found a $1s$ bound state in eq.~\ref{bs.3} is $\epsilon_\phi \approx 0.595$ which matches the values reported in Table IV of Ref.~\cite{PhysRevA.1.1577} with $Z=1$ in their notation. For other bound states, this upper-bound on $\epsilon_\phi$ is decreasing, in accord with Table III in Ref.~\cite{PhysRevA.1.1577}. For fixed mediator masses ($W$ and $Z$ bosons) and fixed mass splitting $\delta M$ = 0.165 GeV, the energy of the bound states as a function of DM mass is studied in figure \ref{fig:spectrum}.

%%%%%%%%%%%%%%%%%%%%%%%%%%%%%%%%%%%%%%%%%%%%
\section{The Coulombic limit: nonabelian ``positronium''}
\label{app:coulomb}
%%%%%%%%%%%%%%%%%%%%%%%%%%%%%%%%%%%%%%%%%%%%

In the case where the gauge symmetry is approximately unbroken, the force carriers can be treated as massless and the interacting DM states can be approximated as degenerate.  The matrix potential then takes a simple form and it is possible to solve analytically for the bound and continuum wavefunctions, along with the rates for bound state formation and decay. This limiting case provides intuition and furnishes a useful test of the detailed numerical results. In this appendix we leave the detailed structure of the potential matrix arbitrary, so that the results we present here can easily be generalized to other gauge groups or representations, facilitating quick estimates of when bound state formation is important in other dark-sector models.

 In this ``Coulombic'' limit, the potential matrix will have the general form $V(r) = - \frac{\alpha}{r} \bar{V}$ (e.g. \cite{ArkaniHamed:2008qn}), where $\bar{V}$ is a constant matrix determined from the generators of the symmetry group, and $\alpha \equiv g^2/4\pi$ is the gauge coupling.  For the case of the wino-like fermion triplet with zero hypercharge (transforming in the adjoint of SU(2)), in the unbroken limit in which $m_W, m_Z \rightarrow 0$, we have $\alpha = \alpha_W$ and $V(r)$ is given by eq.~\ref{eq:highmasslimit}.
 
 As well as generalizing the potential matrix $V(r)$, our expression for the capture and transition cross sections in the dipole approximation (eq.~\ref{eq:dipole}) should be generalized to: 
 \begin{align} 
 & \sigma v_\text{rel} \, \text{(continuum $\rightarrow$ bound) or} \, \Gamma \, \text{(bound $\rightarrow$ bound)} \nonumber \\
& =\frac{2\, \alpha_\text{rad}}{\pi} \frac{k}{M_\chi^2}  \int d\Omega_k \left|\epsilon(\hat{k},\sigma) \cdot \int d^3 r \left[ \Psi_{f}^*({\bf r}) \hat{C}_1 \nabla_{\mathbf r} \, \Psi_{i} ({\bf r}) +  \frac{\alpha M_\chi}{2} \Psi_{f}^*({\bf r}) \hat{C}_2 {\bf \hat{r}} \Psi_{i} ({\bf r}) \right]  \right|^2, 
\label{eq:dipolegeneral} 
\end{align}
where $\alpha_\text{rad}$ gives the coupling to the radiated particle (which may not be the full gauge coupling), $\hat{C}_1$ and $\hat{C}_2$ describe the gauge structure of the capture operator (respectively, how the radiated particle couples to the two-particle states directly, and how it couples to the potential line between any two two-particle states), and as in the main text, $\Psi({\bf r})$ is a vector describing the relevant two-particle states. For example, in the wino case, $\Psi({\bf r})$ is a two-component vector, $(\psi_N({\bf r}), \psi_C({\bf r}))$; $\alpha_\text{rad}$ is the coupling for electromagnetism, while the gauge coupling $\alpha$ becomes $\alpha_W$. The photon couples only to the $\chi^+ \chi^-$ component of the two-particle states, so $\hat{C}_1 = \begin{pmatrix} 0 & 0 \\ 0 & 1 \end{pmatrix}$; the photon also couples to the exchanged $W$ boson, which couples the $\chi^0 \chi^0$ component of one state to the $\chi^+ \chi^-$ component of the other state, and $\hat{C}_2 =  \sqrt{2} \begin{pmatrix} 0 & 1 \\ -1 & 0 \end{pmatrix}$.

The positronium case can be recovered by setting $\hat{C}_1=1$, $\hat{C}_2=0$, considering only the single two-particle state $e^+ e^-$, and setting $\alpha_\text{rad} = \alpha$.

Note that when the gauge symmetry is broken, radiation of massive gauge bosons is forbidden when the boson mass $m_V \gtrsim \alpha^2 M_\chi$, whereas the impact of the shortened range of the potential depends on the ratio of $m_V$ to $\alpha \, M_\chi$; thus there is parametrically a region where the main difference between the broken and unbroken cases is simply that radiation of particular (linear combinations of) gauge bosons is kinematically forbidden in the broken case. Thus it can be productive to explicitly separate the bound-state formation induced by radiation of different gauge bosons; for the wino, we will calculate the capture/decay/annihilation rates with Coulombic wavefunctions, but still only consider radiation of the gauge boson combination that maps to the photon in the unbroken theory.

The final ingredient we need in this generalized calculation will be the initial condition for the unscattered part of the continuum state, corresponding to a plane wave at large radii, which we will describe by the unit vector ${\bf I}$. For example, in the wino case where we are interested in the scattering of particles initially in the $\chi^0 \chi^0$ state, ${\bf I} = (1, 0)$. In the case where the potential has a finite range, the unperturbed plane-wave will then have the asymptotic form ${\bf I} e^{i {\bf p} \cdot {\bf r}}$. In the true Coulombic case where the potential has infinite range, the wavefunction is distorted by the potential even as $r \rightarrow \infty$; nonetheless, we can still describe the large-$r$ boundary condition by the same constant vector, as we will discuss in section \ref{subsec:bcs}. 

When the gauge symmetry is truly unbroken, the incoming state could be in any linear combination of the (degenerate) two-body states; however, once the gauge symmetry is broken, when considering the case of annihilation of DM in the present day, the incoming particles will all (to a good approximation) be in the DM state, rather than one of the heavier states in the multiplet. Thus for ease of comparison to the broken case, it will be particularly useful to study the capture rate from the initial state that will correspond to the DM-DM state after symmetry breaking.

\subsection{The scattering wavefunction}

Given a potential matrix $V(r) = -\frac{\alpha}{r} \bar{V}$, suppose that diagonalizing $\bar{V}$ yields eigenvalues $\lambda_i$, $i=1..n$, and corresponding orthonormal eigenvectors $\eta_i$. Then let us write $\Psi({\bf r}) = \sum_i \eta_i \phi_i({\bf r})$, where $\phi_i({\bf r})$ is a scalar function. Then the Schr\"{o}dinger equation separates into equations for each $i$:
\begin{equation} \frac{p^{2}}{2\mu}   \phi_i({\bf r})  = -\frac{1}{2 \mu} \bold{\nabla}^{2}  \phi_i({\bf r}) -  \frac{ \lambda_i \alpha}{r}  \phi_i({\bf r}),
\end{equation}
where $\mu$ is the reduced mass.  That is, each $\phi_i$ is a solution to the Schr\"{o}dinger equation with a Coulomb potential with coupling $\lambda_i \alpha$; if $\lambda_i$ is positive, the potential is attractive, while if $\lambda_i$ is negative, the potential is repulsive.

For a pair of distinguishable fermions with equal and opposite charges, interacting via an attractive Coulomb potential with coupling $\alpha$, the solution to the Schr\"{o}dinger equation with incoming wave corresponding to the plane wave $e^{i {\bf p} \cdot {\bf r}}$ is given by (e.g. \cite{1996JPhB...29.2135A}):
\begin{equation} \phi(\alpha; {\bf r}) = e^{\pi \zeta/2} \Gamma(1 - i \zeta) ~_1F_1(i\zeta, 1, i (p r - {\bf p} \cdot {\bf r})) e^{i {\bf p} \cdot {\bf r}}, \label{eq:coulombsol} \end{equation}
where $\zeta = \alpha \mu/p = \alpha/v_\text{rel}$ and $F$ is the hypergeometric function.  Then we can write the general solution to the Schr\"{o}dinger equation as:
\begin{equation} \Psi({\bf r}) =  \sum_i A_i \eta_i \phi(\lambda_i \alpha; {\bf r}) \label{eq:sumsolapp}\end{equation}
(Note: in the main text we define a radial wavefunction $\phi(r)$ by $\psi({\bf r}) = \phi(r) \, Y_{lm}$; this should not be confused with the Coulombic wavefunction we define and use in this appendix.)

Asymptotically, this solution will correspond to a plane wave and a scattered spherical wave, and by construction the plane wave will be given by $e^{i {\bf p} \cdot {\bf r}} \sum_i A_i \eta_i$.\footnote{There is a subtlety here, as mentioned previously, in that the asymptotic solution is not exactly a plane wave when the potential has infinite range. See section \ref{subsec:bcs} for discussion.} Thus to impose the boundary condition that determines the initial state of the two interacting particles, we must impose ${\bf I} = \sum_i A_i \eta_i$. By orthonormality of the $\eta_i$, this is equivalent to requiring $A_i = {\bf I} \cdot \eta_i$. Thus our full solution has the form:
\begin{equation} \Psi({\bf r}) =  \sum_i ({\bf I} \cdot \eta_i) \eta_i \phi(\lambda_i \alpha; {\bf r})
\label{fullsoltnm}
 \end{equation}
 
 For a pair of \emph{identical} fermions, as discussed in section \ref{sec:decay}, this spatial wavefunction must be symmetrized (antisymmetrized) for spin-singlet (spin-triplet) configurations. This corresponds to omitting the terms in the partial-wave expansion with odd (spin-singlet) or even (spin-triplet) $L$, and multiplying the remaining wavefunction by $\sqrt{2}$. To make the connection to positronium clearer, we will generally use the normalization for distinguishable fermions throughout this appendix, except when translating our general results to the specific case of the wino. When the initial state consists of indistinguishable fermions, as in the wino case, all cross sections should be multiplied by 2, but only contributions to the capture rate where the final state has odd $L+S$ should be included (since the photon emission changes $L+S$ by 1, and the initial state must be purely $L+S$-even by spin-statistics arguments).

\subsection{The bound state wavefunction}
\label{subapp:coulombbound}

A similar approach is valid for the bound state wavefunction, with the following exception: repulsive Coulomb potentials do not support bound states, so only positive eigenvalues will produce bound-state solutions. Again, in general we will have $\Psi({\bf r}) = \sum_{\lambda_i > 0} \eta_i \phi_i({\bf r})$, and so a given bound state will be characterized by the usual quantum numbers $n, l, m$ but also by $\lambda_i$. To compute the total capture rate, we should sum over all such possible final states, but we may also be interested in capture into a specific state. 

We can therefore write the (properly normalized) bound state wavefunctions as:
\begin{equation} \Psi^i_{nlm}({\bf r}) = Y_{lm}(\theta,\phi) \eta_i R_{nl}(\lambda_i \alpha; r),
\label{boundwavm} \end{equation}
where $R_{nl}(\lambda_i \alpha; r)$ is the hydrogenic radial wavefunction with $\alpha$ replaced by $\lambda_i \alpha$ and $m_e$ replaced by $\mu$, given in eq.~\ref{eq:decay.22}.

\subsection{The capture rate}
\label{subapp:coulombcapture}

For convenience, let us define the reduced matrix element $\bar{M}$ for radiative capture following \cite{1996JPhB...29.2135A}; in the dipole approximation we have:
\begin{align} \bar{M} & \equiv \frac{1}{\mu} \epsilon \cdot \int d^{3} r \left[ \Psi_{f}^*({\bf r}) \hat{C}_1 \nabla_{\mathbf r} \, \Psi_{i} ({\bf r}) + \alpha \mu \Psi_{f}^*({\bf r}) \hat{C}_2 {\bf \hat{r}} \Psi_{i} ({\bf r})  \right] \nonumber \\
& = \frac{1}{\mu} \epsilon \cdot \sum_i ({\bf I} \cdot \eta_i) \eta_f^\dagger \left[ \hat{C}_1   \int d^{3} r Y_{lm}^*(\theta,\phi) R_{nl}^*(\lambda_f \alpha; r)  \nabla_{\mathbf r} \phi(\lambda_i \alpha; {\bf r}) \right. \nonumber \\
& \left. + \alpha \mu \hat{C}_2  \int d^{3} r Y_{lm}^*(\theta,\phi) R_{nl}^*(\lambda_f \alpha; r)  {\bf \hat{r}} \phi(\lambda_i \alpha; {\bf r}) \right] \eta_i,
\label{rme}
\end{align}
where the final bound state is characterized by quantum numbers $nlm$ and eigenvalue $\lambda_f$, and $i$ sums over the eigenvalues of the potential experienced by the initial continuum state. This reduced matrix element is related to the capture cross section by:
\begin{align} \sigma v_\text{rel} \, \text{(continuum $\rightarrow$ bound)} =\frac{\alpha_\text{rad} k}{2 \pi} \int d\Omega_k \left|\bar{M} \right|^2. \end{align}

To evaluate $\bar{M}$, we can substitute the explicit wavefunctions for the continuum and bound states (Eqs.~\ref{eq:decay.22} and \ref{eq:coulombsol}) into eq.~\ref{rme}. Noting that:
\begin{align} \nabla ~_1F_1\left[i\zeta, 1, i (q r - {\bf q} \cdot {\bf r} ) \right]  
& =  - \zeta q  ( {\bf \hat{r}} - {\bf \hat{q}})  ~_1F_1\left[1 + i\zeta, 2, i (q r - {\bf q} \cdot {\bf r} ) \right],  \end{align}
we obtain:
\begin{equation}
\bar{M} = \sum_i ({\bf I} \cdot \eta_i) \eta_f^\dagger \left[\alpha e^{\frac{\pi \alpha \lambda_i \mu}{2 p}} \Gamma\left(1 - i \frac{\alpha \lambda_i \mu}{p} \right) {\bf \epsilon} \right] \cdot \left[- \lambda_i \hat{C}_1 {\bf K}^1 +  \hat{C}_2 {\bf K}^2 \right]\eta_i 
\label{mbarcapm}
\end{equation}
where ${\bf K}^1$, ${\bf K}^2$ are given by the integrals:
\begin{align}
\bold{K}^1 & = \int d^{3} r Y_{lm}^*(\theta,\phi) R_{nl}^*(\lambda_f \alpha; r)  e^{i {\bf p} \cdot {\bf r}} ({\bf \hat{r}} - {\bf \hat{p}})  ~_1F_1(1 + i\alpha \lambda_i \mu / p, 2, i (p r - {\bf p} \cdot {\bf r})), \nonumber \\
\bold{K}^2 & = \int d^{3} r Y_{lm}^*(\theta,\phi) R_{nl}^*(\lambda_f \alpha; r)  e^{i {\bf p} \cdot {\bf r}} {\bf \hat{r}}  ~_1F_1(i\alpha \lambda_i \mu / p, 1, i (p r - {\bf p} \cdot {\bf r})).
\end{align}

Let us choose a coordinate system where ${\bf p}$ points in the $z$-direction. Then the angular integral will set to zero any transitions to states with $|m| > 1$, since the initial continuum state has only $m=0$ components and the single-photon dipole transition requires $|\Delta m| \le 1$. Performing the angular integrals over $\phi$ for $m=0, \pm 1$ yields:
\begin{align}
{\bf K}^1_{m = 0} & =  2\pi \hat{z}  \sqrt{\frac{(2l+1)}{4\pi} } \nonumber \\
& \times  \int r^2 dr \sin\theta d\theta P_{l}^{0*}(\cos\theta) R_{nl}^*(\lambda_f \alpha; r)  e^{i p r \cos\theta} (\cos\theta - 1)  ~_1F_1\left(1 + i\frac{\alpha \lambda_i \mu}{p}, 2, i pr (1 - \cos\theta)\right), \nonumber \\
{\bf K}^1_{m = \pm 1} & = \pi \sqrt{\frac{(2l+1)(l-m)!}{4\pi(l+m)!}}  (\hat{x} \mp i  \hat{y}) \nonumber \\
& \times \int r^2 dr d\theta \sin^2\theta P_{l}^{m*}(\cos\theta) R_{nl}^*(\lambda_f \alpha; r)  e^{i p r \cos\theta}  ~_1F_1\left(1 + i\frac{\alpha \lambda_i \mu}{ p}, 2, i p r (1 - \cos\theta)\right), \nonumber \\
\bold{K}^2_{m=0} & = 2 \pi \hat{z} \sqrt{\frac{(2l+1)}{4\pi} } \nonumber \\
& \times \int r^2 dr \sin\theta d\theta P_{l}^{0*}(\cos\theta) R_{nl}^*(\lambda_f \alpha; r)  e^{i p r \cos\theta} \cos\theta  ~_1F_1\left(\frac{i\alpha \lambda_i \mu}{p}, 1, i p r (1-\cos\theta)\right), \nonumber \\
\bold{K}^2_{m=\pm 1} & = \pi \sqrt{\frac{(2l+1)(l-m)!}{4\pi(l+m)!}} (\hat{x} \mp i \hat{y}) \nonumber \\
& \times \int r^2 dr \sin^2\theta d\theta P_{l}^{m*}(\cos\theta) R_{nl}^*(\lambda_f \alpha; r)  e^{i p r \cos\theta}   ~_1F_1\left(\frac{i\alpha \lambda_i \mu}{p}, 1, i p r (1 - \cos \theta)\right).
\end{align}

If we restrict ourselves to $n=1,2$ bound states, which are the most deeply bound and generally have the largest capture cross sections, the required integrals can be evaluated in the low-$v_\text{rel}$ limit using the results of appendix \ref{app:integrals}. For the possible choices of $nlm$ we obtain the following results for the {\bf K} integrals:
\begin{align} {\bf K}^1_{100}: \quad & -8 \sqrt{\pi}  (\alpha \lambda_f \mu)^{-3/2} e^{- 2 \lambda_i/\lambda_f} \hat{z} \nonumber \\ 
{\bf K}^1_{200}: \quad & -32 \sqrt{2 \pi}  (\alpha \lambda_f \mu)^{-3/2} e^{-4 \lambda_i/\lambda_f} \left[\frac{2 \lambda_i}{\lambda_f} - 1  \right]\hat{z} \nonumber \\
{\bf K}^1_{210}:  \quad & - 16 \sqrt{2 \pi} (\alpha \lambda_f \mu)^{-3/2} e^{-4 \lambda_i/\lambda_f} \left[\frac{4 \lambda_i }{\lambda_f} - 1 \right]  \hat{z} \nonumber \\
{\bf K}^1_{21(\pm 1)}:  \quad & 16 \sqrt{2 \pi}  (\alpha \lambda_f \mu)^{-3/2}  e^{-4 \lambda_i/\lambda_f} \left(\frac{\mp \hat{x} + i  \hat{y}}{\sqrt{2}} \right).  \end{align}
\begin{align} {\bf K}^2_{100}: \quad & 8 \sqrt{\pi}  (\alpha \lambda_f \mu)^{-3/2} e^{- 2 \lambda_i/\lambda_f} \frac{\lambda_i}{\lambda_f} \hat{z} \nonumber \\ 
{\bf K}^2_{200}: \quad & 32 \sqrt{2 \pi}  (\alpha \lambda_f \mu)^{-3/2} e^{-4 \lambda_i/\lambda_f} \left[\frac{\lambda_i}{\lambda_f} \left( \frac{4 \lambda_i}{\lambda_f} - 3 \right)  \right]\hat{z} \nonumber \\
{\bf K}^2_{210}:  \quad & 16 \sqrt{2 \pi} (\alpha \lambda_f \mu)^{-3/2} e^{-4 \lambda_i/\lambda_f} \left[1 - \frac{4 \lambda_i }{\lambda_f} + 8 \left(\frac{\lambda_i}{\lambda_f} \right)^2 \right]  \hat{z} \nonumber \\
{\bf K}^2_{21(\pm 1)}:  \quad & 16 \sqrt{2 \pi}  (\alpha \lambda_f \mu)^{-3/2}  e^{-4 \lambda_i/\lambda_f} \left[1 - \frac{4\lambda_i}{\lambda_f} \right] \left(\frac{\mp \hat{x} + i  \hat{y}}{\sqrt{2}} \right).  \end{align}

Accordingly we can write the matrix element for capture into the $nlm$ bound state as:
\begin{align} \bar{M} & = 8 \sqrt{\pi} \alpha \epsilon \cdot (\hat{r}_{m})^* \sum_i  {\bf I} \cdot \eta_i (\alpha \lambda_f \mu)^{-3/2} e^{-2 n \lambda_i/\lambda_f} e^{\frac{\pi \alpha \lambda_i \mu}{2 p}} \Gamma\left(1 - i \frac{\alpha \lambda_i \mu}{p} \right) \nonumber \\
& \times \eta_f^\dagger \left[ \lambda_i \hat{C}_1 u_{nlm} + \hat{C}_2 v_{nlm} \right] \eta_i, \label{mbartotsm}  \end{align}
where $\hat{r}_m$ is defined immediately after eq.~\ref{eq:rdecomp}, and $u_{nlm}$, $v_{nlm}$ are given by:
\begin{align} & u_{100}:  1, \quad u_{200}:  4 \sqrt{2}   \left[2 \lambda_i/\lambda_f - 1 \right], \quad u_{210}: 2 \sqrt{2}  [4  \lambda_i /\lambda_f - 1],
\quad u_{21(\pm 1)}: -2 \sqrt{2}, \nonumber \\
& v_{100}:  \lambda_i/\lambda_f,  \quad v_{200}:  4 \sqrt{2}   \left[\frac{\lambda_i}{\lambda_f} \left( \frac{4 \lambda_i}{\lambda_f} - 3 \right)  \right], \nonumber \\
&  v_{210}: 2 \sqrt{2}  \left[1 - \frac{4 \lambda_i }{\lambda_f} + 8 \left(\frac{\lambda_i}{\lambda_f} \right)^2 \right], 
\quad v_{21(\pm 1)}: -2 \sqrt{2} \left[\frac{4\lambda_i}{\lambda_f} - 1\right].  \end{align}

It is useful to note how the gamma-function term in eq.~\ref{mbartotsm} scales in the small-${\bf p}$ limit. We use the relation:
\begin{equation}
\left| \Gamma \left( 1 - i \zeta \right) e^{\frac{\pi \zeta}{2}} \right|^{2} = \frac{\pi \zeta}{\sinh \left( \pi \zeta \right)} e^{\pi \zeta} \rightarrow \left\{ \begin{array}{c} 2 \pi \zeta, \, \zeta \gg 1 \\ -2\pi \zeta e^{2 \pi \zeta}, \, \zeta \ll -1 \end{array} \right.
\label{eq:expsuppress}
\end{equation}
It follows that at small momentum, where $\alpha \mu/p$ becomes large, all contributions to $\bar{M}$ from $i$ corresponding to $\lambda_i < 0$ are suppressed by a factor of order $e^{\pi \alpha \mu \lambda_i /p}$. Physically, this is because $\lambda_i < 0$ indicates a repulsive potential, and at low velocities the resulting wavefunctions have very little overlap with the bound states. In the calculations that follow we will neglect these exponentially suppressed contributions.

To compute the cross section from this matrix element, we need only to sum over the final photon polarization states, performing the integral $\int d\Omega_k |\bar{M}|^2$ for each. Writing $\bar{M} = A \epsilon \cdot \hat{r}_{-m}$, and following the same procedure as Eqs. \ref{eq:polsum}-\ref{eq:angavg} (and noting that $\int d\Omega_k \sin^2\theta_k = \int d\Omega_k \left(1 - \frac{1}{2} \sin^2\theta_k \right) = 8\pi/3$), we find that $\int d\Omega_k |\bar{M}|^2 = \frac{8 \pi}{3} |A|^2$.

Thus finally we obtain the capture cross section:
\begin{align} & \sigma v_\text{rel} \, \text{(continuum $\rightarrow$ bound)} =\frac{2^8 \pi \alpha_\text{rad} k}{3} \nonumber \\
& \times  \left|  \sum_i {\bf I} \cdot \eta_i  \alpha (\alpha \lambda_f \mu)^{-3/2} e^{-2 n \lambda_i/\lambda_f} e^{\frac{\pi \alpha \lambda_i \mu}{2 p}} \Gamma\left(1 - i \frac{\alpha \lambda_i \mu}{p} \right)  \eta_f^\dagger \left[ \lambda_i \hat{C}_1 u_{nlm} + \hat{C}_2 v_{nlm} \right] \eta_i  \right|^2. \label{eq:coulombxsec} \end{align}

Recall this is the capture cross section for distinguishable fermions in the initial state; if the initial state consists of identical fermions, this cross section should be multiplied by 2 where the final state has $L+S$ odd, and set to zero when the final state has $L+S$ even. Since $1/4$ of all particle pairs are spin-singlet ($S=0$) and $3/4$ are spin-triplet ($S=1$), this corresponds to a spin-averaged cross section related to eq.~\ref{eq:coulombxsec} by a factor of $3/2$ for final states of even $L$, and a factor of $1/2$ for final states of odd $L$.

\subsection{Capture cross sections for the wino case}

As discussed in the main text, single-photon dipole emission changes angular momentum by $\Delta L = 1$ from an even $L + S$ state to an odd $L+S$ state, leading to different eigenvalues for the initial and final states. Neglecting the exponentially suppressed contributions from the negative eigenvalue of the even-$L+S$ potential matrix, for single-photon capture we need only consider the case with $\lambda_i=2$, $\lambda_f=1$. The associated eigenvectors are given in section \ref{subsec:himass}.

 If we impose the initial condition $I = \begin{pmatrix} 1 \\ 0 \end{pmatrix}$  as discussed earlier (i.e. far from the interaction site, all particles are neutralinos rather than charginos), then $ ({\bf I} \cdot \eta_i) (\eta_f^\dagger \hat{C}_1 \eta_i)  = \begin{pmatrix} 1 \\ 0 \end{pmatrix} \cdot \begin{pmatrix} \sqrt{\frac{1}{3}} \\ \sqrt{\frac{2}{3}} \end{pmatrix} \begin{pmatrix} 0 & 1 \end{pmatrix}  \begin{pmatrix} 0 & 0 \\ 0 & 1 \end{pmatrix}  \begin{pmatrix} \sqrt{\frac{1}{3}} \\ \sqrt{\frac{2}{3}} \end{pmatrix} = \frac{\sqrt{2}}{3}$, and likewise $ ({\bf I} \cdot \eta_i) (\eta_f^\dagger \hat{C}_2 \eta_i)  = -\frac{\sqrt{2}}{3}$. 
 
Using eq.~\ref{eq:coulombxsec}, replacing $\alpha_\text{rad} \rightarrow \alpha$, $\alpha \rightarrow \alpha_W$, and including the factor of 2 discussed above for identical particles in the initial state, we then obtain for the capture cross sections:
\begin{equation}
\sigma v_\mathrm{rel} = \frac{2^{15} \pi^2}{3^3} \frac{\alpha \alpha_W^2}{M_\chi^2 v_\mathrm{rel}} \frac{1 }{n^2 } e^{-8 n}   f_{nlm} ,
\label{eq:capturexsec}
\end{equation}
 where the $f_{nlm}$ are given by: 
 \begin{equation} f_{100} = 0
 , \quad f_{200} = 128
 , \quad f_{210} = 242
 , \quad f_{21\pm1} = 50
 .\end{equation}
 
 These cross sections are for initial states of definite spin (singlet or triplet). When averaging over all possible spin states, this contribution should be divided by 4 before adding it to the total for capture into states with odd $L$ (where the leading contribution comes from even-$L$ initial states, which must be spin-singlet), and multiplied by $3/4$ for capture into states with even $L$ (where the leading contribution comes from odd-$L$ initial states, which must be spin-triplet). We expect the latter processes to be suppressed at low velocities, once the SU(2) symmetry is broken and the masses of the force carriers are non-negligible, as discussed in appendix \ref{app:hulthen}.

 \subsection{Capture cross sections separated by partial wave}
 \label{subapp:coulombxsecbyl}
 
We have so far computed the cross section for capture into the bound states from an incoming plane wave. However, for comparison to the regime where the force carrier masses are non-negligible, it is useful to separate out the contributions from different partial waves in the initial state, since we expect the higher partial wave contributions to be velocity-suppressed when the potential has a finite range (see appendix \ref{app:hulthen}).

For positronium, the properly normalized solution to the Schr\"{o}dinger equation is given by eq.~\ref{eq:coulombsol} (for distinguishable fermions; see the discussion following that equation for the case of indistinguishable fermions).  Taking ${\bf p}$ in the $z$-direction, we can separate this into partial waves as follows:
\begin{equation} \phi(\alpha; {\bf r}) = \sum_L (2L+1) P_L(\cos\theta) \frac{\Gamma(1 + L - i \zeta)}{2 i p r\Gamma(2(L+1))} e^{\pi \zeta/2} M\left[-i\zeta, \frac{1}{2} (1 + 2 L), 2 i p r \right],
\label{partialm1}
\end{equation}
where $M$ is the Whittaker function, which is related to the hypergeometric function by:
\begin{equation} M(a, b, z) = e^{-z/2} z^{b+1/2} ~_1F_1(1/2 + b - a, 1 + 2 b, z).\end{equation}
In terms of the confluent hypergeometric function we can write:
\begin{equation}
\phi(\alpha; {\bf r}) =  \sum_L (2L+1) P_L(\cos\theta) \frac{\Gamma(1 + L - i \zeta)}{\Gamma(2(L+1))} e^{\pi \zeta/2} e^{- i p r} (2 i p r)^{L} ~_1F_1\left[1 + L + i \zeta, 2 (L+1), 2 i p r \right] .
\end{equation}

 \subsubsection{Capture from $s$-wave incoming state into $p$-wave bound state}
 
 Let us now consider the contribution to the capture cross section from the $s$-wave part of the incoming wino pair state, where the final state is one of the $n=2$, $l=1$ states. We will use this to cross-check our numerical results., 
 
From eq.~\ref{partialm1} the $s$-wave contribution to the continuum state, for a given eigenvalue $\lambda_i$,  is given by:
\begin{equation} 
\phi_s(\lambda_i \alpha; r) = \frac{\Gamma(1 - i \lambda_i \alpha \mu/p)}{2 i p r} e^{\pi \lambda_i \alpha \mu/2 p} M\left[-i\lambda_i \alpha \mu/p, \frac{1}{2}, 2 i p r \right].
\end{equation}

Repeating our previous analysis but projecting out the $s$-wave piece (i.e. replacing $\phi(\lambda_i \alpha; {\bf r})$ with $\phi_s(\lambda_i \alpha; r)$), we find for the reduced matrix element:
\begin{align} 
\bar{M} & = \epsilon \cdot (\hat{r}_{m})^* \frac{16 \sqrt{2 \pi}}{3} \alpha^{-1/2} \mu^{-3/2} \lambda_f^{-3/2}  \sum_i ({\bf I} \cdot \eta_i) \Gamma(1 - i \lambda_i \alpha \mu/p) e^{\pi \lambda_i \alpha \mu/2 p} 
 e^{-4 \lambda_i/\lambda_f} \nonumber \\
 & \times \eta_f^\dagger \left[ \hat{C}_1 \lambda_i \left(\frac{4\lambda_i}{\lambda_f} - 3 \right)  + \hat{C}_2 \left(3 - 12 \frac{\lambda_i}{\lambda_f }+ 
   8 \left(\frac{\lambda_i}{\lambda_f}\right)^2 \right)  \right] \eta_i. 
 \label{aspm} 
 \end{align}
 
For the wino we have $({\bf I} \cdot \eta_i) \eta_f^\dagger \hat{C}_1  \eta_i = -  ({\bf I} \cdot \eta_i) \eta_f^\dagger \hat{C}_2  \eta_i  =  \sqrt{2}/3$, $\lambda_i = 2$, $\lambda_f = 1$.  Substituting these into eq.~\ref{aspm}, we obtain:
 \begin{align} \bar{M}_{s\rightarrow p} & = - \epsilon \cdot (\hat{r}_{m})^*  \frac{2^5 \sqrt{ \pi}}{3^2} e^{-8}  \alpha_{W}^{-1/2} \mu^{-3/2}  \Gamma\left(1 - \frac{2 i \alpha_{W} \mu}{p} \right) e^{\pi  \alpha_{W} \mu/ p}
 ,    \end{align}
 for capture into any one of the $n=2$, $l=1$ states. (Here we have not inserted the factor of $\sqrt{2}$ for the normalization of the initial state containing identical fermions, consistent with our previous approach of inserting it in the cross section.)
 
 Evaluating eq.~\ref{mbartotsm} for capture from an initial plane wave into the $n=2$, $l=1$ states, we obtain the matrix elements: 
 \begin{align}\bar{M}_{210} & =  - \epsilon \cdot (\hat{r}_{0})^* \,  \frac{2^5 \times 11}{3} \sqrt{ \pi} e^{-8}  \alpha_{W}^{-1/2} \mu^{-3/2}  \Gamma\left(1 - \frac{2 i \alpha_{W} \mu}{p} \right) e^{\pi  \alpha_{W} \mu/ p}
 , \nonumber \\
\bar{M}_{21\pm 1} & = \epsilon \cdot (\hat{r}_{\pm 1})^* \,  \frac{2^5 \times 5}{3} \sqrt{ \pi} e^{-8}  \alpha_{W}^{-1/2} \mu^{-3/2}  \Gamma\left(1 - \frac{2 i \alpha_{W} \mu}{p} \right) e^{\pi  \alpha_{W} \mu/ p}. 
\end{align}

 Comparing to eq.~\ref{mbartotsm} for capture from an initial plane wave into these states, for the 210 state the $s$-wave contribution is 1/33
 of the total, whereas for the $m = \pm 1$ states it is $-1/15$
 of the total (here the $d$-wave piece has the opposite sign, with magnitude $16/15$
 of the total, and they add destructively).  If the higher partial-wave components are suppressed, averaging over initial spin and summing over final states $m = 0, \pm 1$, and including the factor of 2 for the identical particles in the initial state, we obtain the total cross section for capture from the $s$-wave piece of the incoming state into the $n=2$, $l=1$ bound states:
\begin{align} 
\sigma v_\mathrm{rel} & = \frac{2^{12} \pi^2}{3^4} e^{-16} \frac{\alpha \, \alpha_W^2}{M_\chi^2 v_\mathrm{rel}}. 
\label{spwavem}
\end{align}

\subsubsection{Capture from $p$-wave incoming state into $s$-wave bound state}

Capture into an $s$-wave state through single-photon dipole emission occurs only from $p$-wave states. Therefore the rates for $p$-to-$s$ capture are just given by the full capture rates from the incoming plane wave into $s$-wave states (see Equation \ref{eq:capturexsec}). In particular, the capture rate into the ground state is zero through single-photon processes in the Coulomb limit. Thus it is small at high masses as we approach the Coulomb limit. One might wonder if two-photon processes can give an appreciable capture rate into the ground state. This has been studied by Refs. \cite{ejTalk,Johnson:2016sjs} and found to be negligible.

\subsection{Transition rates}
\label{subapp:coulombtransition}
  
  We can likewise compute bound$\rightarrow$bound rates using eq.~\ref{eq:dipolegeneral} and the bound-state wavefunctions presented earlier; the decay rates are much simpler to calculate and can be done analytically using e.g. \texttt{Mathematica}, since both initial and final-state wavefunctions have fairly simple radial dependence (exponential functions of $r$ multiplying polynomials). Accordingly,  we will not detail the calculation here, but simply present results. We will be primarily interested in decays of the $n=2$, $l=1$ spin-singlet bound states, since these are populated by capture from the $s$-wave part of the initial continuum state.
  
The dominant decay of the $n=2$, $l=1$ states is to the spin-singlet ground state with $nlm=100$. For general initial and final eigenvalues, we find a total decay rate of:
\begin{align} 
\Gamma = \frac{2^7}{3} \alpha \alpha_{W}^4 M_\chi \frac{\lambda_i (2 \lambda_f - \lambda_i) (\lambda_i/\lambda_f)^4}{\left(2 + \frac{\lambda_i}{\lambda_f} \right)^7}  \left| \lambda_f \eta_f^\dagger \hat{C}_1 \eta_i + \eta_f^\dagger \hat{C}_2 \eta_i \right|^2. 
\end{align}
  
For the wino, and assuming both states are spin-singlet, we have $ \eta_f^\dagger \hat{C}_1 \eta_i =  \eta_f^\dagger \hat{C}_2 \eta_i    = \sqrt{\frac{2}{3}}$, $\lambda_i = 1$, and $\lambda_f = 2$, which yields:
\begin{equation}
\Gamma = \frac{2^{11} \, \times \, 3}{ 5^7} \, \alpha \, \alpha_W^4 M_\chi,
\end{equation}
for decay from each of the initial states $n_{i}=2$, $l_{i}=1$, $m_{i}=0,\pm1$.

Considering more general decays from initial $n_{i}=2$, $l_{i}=1$ states, the rate is always proportional to $\alpha \, \alpha_{W}^{4} M_\chi$, with proportionality factors given in table \ref{tab:results}.

 \begin{table}
  \setlength\extrarowheight{6pt}
  \begin{center}
  \begin{tabular}{|l|ccc|ccc|}
    \hline
       final $nlm$  & $m_i = 0$ & $m_i = 1$ & $m_i = -1$ & $m_i = 0$ & $m_i = 1$ & $m_i = -1$ \\ \hline
    100 & & $  \frac{2^{11} \times 3}{5^7}  $ & & $   $ & $7.9 \times 10^{-2}$ & \\ \hline
      200 & & $ \frac{2^{8}\times 7^2}{3^{11}} $ & & & $7.1 \times 10^{-2}$ & \\ \hline
    300 & & $ \frac{2^{11} \times 89^2}{3 \times 7^{11}} $ & & & $2.7 \times 10^{-3}$ & \\ \hline
    320 & $\frac{2^{24}}{3 \times 7^{11}}$ & $\frac{2^{22}}{3 \times 7^{11}}$ & $\frac{2^{22}}{3 \times 7^{11}}$ & $ 2.8 \times 10^{-3} $ & $7.1 \times 10^{-4}$ & $7.1 \times 10^{-4}$  \\ \hline
    321 & $\frac{2^{22}}{7^{11}} $ & $\frac{2^{22}}{7^{11}} $ & - & $2.1 \times 10^{-3}$ & $2.1 \times 10^{-3}$ & - \\ \hline
    32-1 & $\frac{2^{22}}{7^{11}} $ & - & $\frac{2^{22}}{7^{11}} $& $2.1 \times 10^{-3}$ & - & $2.1 \times 10^{-3}$ \\ \hline
    322 & - & $\frac{2^{23}}{7^{11}} $ & - & - & $4.2 \times 10^{-3}$ & - \\ \hline
    32-2  & - & - & $\frac{2^{23}}{7^{11}} $ & - & - & $4.2 \times 10^{-3}$ \\ \hline
  \end{tabular}
    \caption{Exact (left) and approximate (right) numerical prefactors for the decay rate $\Gamma$ of spin-singlet bound states with $n=2$ and $l=1$, which is given by $\alpha \, \alpha_W^4 M_\chi$ multiplied by this prefactor. Different columns indicate different initial states ($nlm = 21m_i$) while the rows label different final states. Rows with only one number indicate that the same result applies to different values of $m_i$. Processes forbidden via a single photon emission are marked with ``-". }
    \label{tab:results}
\end{center}
\end{table}

The overall decay rates for each of the three initial states are thus, 
\begin{equation} 
\Gamma_{210} = \Gamma_{211} = \Gamma_{21-1} = 0.16 \, \alpha \, \alpha_W^4 M_\chi, 
\end{equation}
main decays being to the ground state and the first excited $s$-wave state ($0.15/0.16 \approx 94\%$ of the total, with a $49\%$ branching ratio to the ground state and a $44\%$ branching ratio to the first excited state).

We can also evaluate the decay rate for the $2s$ spin-triplet state, which has open single-photon decays to the $2p$ and $3p$ spin-triplet states (these states have $L+S$ even and hence are more tightly bound than the $L+S$-odd $2s$ spin-triplet state). We find the decay to the $2p$ state (summing over $m$ values in the final state) is the dominant channel, with rate:
\begin{equation}\Gamma = \frac{2^{12}}{3^{10}} \alpha \alpha_W^4 M_\chi \approx 0.069 \alpha \alpha_W^4 M_\chi ,\end{equation}
while the decay rate to the $3p$ state is given by:
\begin{equation}\Gamma = \frac{2^{20} \times 3^3}{7^{11}} \alpha \alpha_W^4 M_\chi \approx 0.014 \alpha \alpha_W^4 M_\chi.\end{equation}

\subsection{Subtleties in imposing plane-wave boundary conditions on an infinite-range potential}
\label{subsec:bcs}

When we earlier imposed the boundary condition that the original plane wave should be described by the vector ${\bf I}$, we tacitly assumed that the asymptotic solution \emph{was} a plane wave, or at least that it could be characterized by a single $r$-independent vector multiplying a (scalar) modified plane wave. This is not correct for potentials with infinite range.

The true plane-wave-like solution to the Schr\"{o}dinger equation at large range, in the presence of the matrix potential we consider, has the form:
\begin{equation} \Psi(r) = e^{i {\bf p} \cdot {\bf r}} \sum_i A_i \eta_i e^{-i \frac{\alpha \lambda_i \mu}{p} \ln(p r - {\bf p} \cdot {\bf {r}})}. \end{equation}
 This follows directly from the large-$r$ solution of the scalar Schr\"{o}dinger equations corresponding to the various eigenvalues. We can equivalently write this expression, more compactly, as:
 \begin{equation} \Psi(r) = e^{i {\bf p} \cdot {\bf r}} e^{-i \frac{\mu}{p} \alpha \bar{V} \ln (p r - {\bf p} \cdot {\bf r})} \sum_i A_i \eta_i.
 \label{psi2ndm}
 \end{equation}
From eq.~\ref{psi2ndm}, it is apparent that the asymptotic coefficient of the distorted plane wave in this state should be considered to be ${\bf I} = \sum_i A_i \eta_i$, as in eq.~\ref{fullsoltnm}. This is the constant quantity on which we can impose our boundary condition that the incoming state is in the DM-DM two-body state. But the relative probability assigned to the various two-body states (in the default basis) at large $r$ will not be described by ${\bf I}$ alone, but by $e^{-i \frac{\mu}{p} \alpha \bar{V} \ln (p r - {\bf p} \cdot {\bf r})} {\bf I}$, where the exponential phase has non-trivial gauge structure through the $\bar{V}$ matrix.  

It remains to check that the solution given by eq.~\ref{eq:coulombsol} has the exact asymptotic plane-wave-like component we assumed above (in addition, it includes a scattered spherical wave):
\begin{equation} \phi(\alpha; {\bf r}) \rightarrow  e^{i {\bf p} \cdot {\bf r}} e^{-i \frac{\mu}{p} \alpha \ln (p r - {\bf p} \cdot {\bf r})} + \text{spherical scattered wave}. 
\label{assumed}
\end{equation}
To show this we use the well-known asymptotic form of the hypergeometric function:
\begin{equation} \lim_{|z| \rightarrow \infty} ~_1F_1(a,b,z) = \frac{e^{z} z^{a-b}}{\Gamma(a)} + \frac{e^{i \pi a} z^{-a}}{\Gamma(b-a)} , \end{equation}
to obtain the asymptotic form of eq.~\ref{eq:coulombsol}:
\begin{equation} \phi(\alpha; {\bf r}) \rightarrow e^{\pi \zeta/2} \Gamma(1 - i \zeta) \left[\frac{e^{i (p r - {\bf p} \cdot {\bf r})} (i (p r - {\bf p} \cdot {\bf r}))^{1-i\zeta}}{\Gamma(i\zeta)} + \frac{e^{- \pi \zeta} (i (p r - {\bf p} \cdot {\bf r}))^{-i\zeta}}{\Gamma(1-i\zeta)}\right] e^{i {\bf p} \cdot {\bf r}}. \end{equation}
We identify the first term as the scattered wave and the second term as the plane-wave-like part; focusing on this second term, we find:
\begin{align} \phi(\alpha; {\bf r}) & \rightarrow e^{- \pi \zeta/2}  (e^{i \pi/2} (p r - {\bf p} \cdot {\bf r}))^{-i\zeta} e^{i {\bf p} \cdot {\bf r}} + \text{spherical scattered wave}, \nonumber \\
& = e^{- \pi \zeta/2}  e^{(i \pi/2)(-i\zeta)} e^{-i \zeta \ln (p r - {\bf p} \cdot {\bf r})} e^{i {\bf p} \cdot {\bf r}} + \text{spherical scattered wave}, \nonumber \\
 & = e^{-i \frac{\alpha \mu}{p} \ln (p r - {\bf p} \cdot {\bf r})} e^{i {\bf p} \cdot {\bf r}} + \text{spherical scattered wave}, \end{align}
as required.

%%%%%%%%%%%%%%%%
\section{Effects of a massive force carrier}
\label{app:hulthen}
%%%%%%%%%%%%%%%%

To understand the qualitative effects of giving a mass to (some of) the force carriers, so that the potential has a finite range, it is useful to consider the simpler case of a single DM state, with DM particles (of mass $M_\chi$ as previously) interacting via the attractive Hulth\'en potential,
\begin{equation} V(r) = - \frac{\alpha_H m_H}{e^{m_H r} - 1}. \end{equation} 
This potential has been considered as an approximation to the Yukawa potential $V(r) = -\frac{\alpha}{r} e^{-m_A r}$, in the context of the Sommerfeld enhancement \cite{Cassel:2009wt,0256-307X-29-8-080302}. It has the advantage that the Schr\"{o}dinger equation is exactly solvable for the $s$-wave states, and approximately solvable for higher partial waves; both continuum-state solutions \cite{Cassel:2009wt} and bound-state solutions \cite{0256-307X-29-8-080302} have been presented in the literature, for arbitrary $l$. In the calculation of the Sommerfeld enhancement, good agreement between the analytical results for the Hulth\'en potential and numerical results for the Yukawa potential is obtained with the identifications $\alpha = \alpha_H$, $m_H = \frac{\pi^2}{6} m_A$ \cite{Cassel:2009wt}.

In this appendix we briefly summarize the key results for the Hulth\'en potential, and then estimate the scaling of the bound state formation rate with $m_H$.

\subsection{Wavefunctions for the Hulth\'en potential}

In our notation, the radial wavefunctions for bound states of the Hulth\'en potential are approximately given by \cite{0256-307X-29-8-080302}:
\begin{equation}R_{nl}(r) = \frac{N_{nl}}{r} e^{-\kappa r} (1 - e^{-m_H r})^{l+1} P_{n-l-1}^{\left(2 \kappa/ m_H, 2l + 1 - \kappa/m_H\right)}\left(1 - 2 e^{-m_H r} \right),\end{equation}
where the bound state energy is $E_n = -\kappa^2/M_\chi$, with $\kappa$ for a particular choice of principal quantum number $n$ ($n=1, 2, 3, ...$) given by:
\begin{equation} \kappa_{n}  = \frac{1}{2} \left( \frac{\alpha_H M_\chi - n^2 m_H}{n} \right). \end{equation}
Here $N_{nl}$ is a (dimensionful) normalization factor, chosen to give the correct normalization for the bound state (it is straightforward to calculate for any given state, but the expressions are cumbersome), and $P_n^{(a,b)}(x)$ is the Jacobi P-function. These results are exact when $l=0$ or $m_H=0$; otherwise, they require use of an approximate form for the centrifugal term in the radial Schr\"{o}dinger equation (see \cite{0256-307X-29-8-080302} for details).

Note that with the definitions above, we must have $\kappa_n > 0$ for bound states with principal quantum number $n$ to exist, i.e. we must have $n < \sqrt{\alpha_H M_\chi/m_H}$. Thus when $m_H > \alpha_H M_\chi$, the potential supports no bound states; when $\alpha_H M_\chi/4 < m_H < \alpha_H M_\chi$, only the $n=1$ bound state exists; and so on. In the Coulomb limit where $m_H \rightarrow 0$, we recover the usual expression for the hydrogen-like bound state energies (recalling that the reduced mass of the system is $M_\chi/2$).

The partial-wave wavefunctions for continuum states with $m=0$ are given by \cite{Cassel:2009wt}:
\begin{align} \psi_L({\bf r}) &= \frac{P_L(\cos\theta) }{(2 L)!} \frac{\left(1 - e^{-m_H r} \right)^{L+1}}{M_\chi r} e^{- i (M_\chi v_\text{rel}/2) r}~_2 F_1 \left[a^-,a^+,2(L+1),1 - e^{-m_H r}\right] \nonumber \\
& \times \sqrt{ \frac{\frac{2 \pi}{\alpha_H v_\text{rel}} \sinh\left(\frac{\pi v_\text{rel} M_\chi}{m_H}\right)}{\cosh\left( \frac{\pi v_\text{rel} M_\chi}{m_H}\right) - \cosh\left(\frac{ \pi  v_\text{rel} M_\chi}{m_H} \sqrt{1 - \frac{4 \alpha_H m_H}{M_\chi v_\text{rel}^2}} \right)} } \nonumber \\
& \times \sqrt{ \prod_{k=0}^{L} \left(k^4 + \left(\frac{\alpha M_\chi}{m_H}\right)^2 + k^2 \left( \frac{M_\chi v_\text{rel}}{m_H} \right)^2 \left(1 - \frac{2 \alpha_H m_H}{M_\chi v_\text{rel}^2} \right)\right)} \nonumber \\
a^{\pm} & = 1 + L + \frac{i M_\chi v_\text{rel}}{2 m_H} \left(1 \pm \sqrt{1 - \frac{4 \alpha_H m_H}{M_\chi v_\text{rel}^2}} \right). \end{align}
In deriving these results from those of  \cite{Cassel:2009wt}, we have used the identity $\Gamma(i z) \Gamma(- i z) = \pi/(z\sinh(\pi z))$. These wavefunctions are normalized to correspond to the $L$-wave components of a unit-normalized plane wave propagating in the $z$-direction. (Some caution must be taken in adding wavefunctions of different $L$ together to re-form a plane wave, as the normalization condition does not fix the relative phases, but these wavefunctions will suffice for considering the scaling of the contributions to the capture rate by states of different $L$.)

There is an additional subtlety in this case; the approximate centrifugal term used by \cite{Cassel:2009wt} to obtain the correct wavefunctions is exponentially suppressed at large $r$, whereas the real centrifugal term scales as $1/r^2$, and consequently the wavefunctions do not properly match onto the correct asymptotic solution at large $r$. (This is not a problem for the bound state wavefunctions, because the bound state wavefunctions are exponentially suppressed at large $r$ independent of the centrifugal term.) This issue is independent of the details of any short-range potential (although it depends on $m_H$ because the approximate form for the centrifugal term involves $m_H$), and so can be compensated by examining the effect of using the ``wrong'' centrifugal potential for plane waves; the prescription presented in \cite{Cassel:2009wt} multiplies the short-range annihilation rate derived using these wavefunctions by the factor:\begin{equation} C = \frac{w^{2L}}{\Pi^{L-1}_{k=0} [(L - k)^2 + w^2]}, \quad w=M_\chi v_\text{rel}/m_H. \label{eq:correc} \end{equation}
When using these approximate continuum wavefunctions to estimate scaling relationships, we will multiply the final capture rate by this correction factor. We do not expect this approach to be accurate in detail (for $L > 0$), because the capture process is not localized at very short distances in the same way as annihilation -- indeed, the scale of the potential is the same as the scale of the (incorrect) centrifugal potential. However, it will suffice to estimate scaling relations for the capture rate; we will then confirm these scaling relations using numerically computed continuum wavefunctions. 
 
 \subsection{Estimating the capture rate}

Let us focus here on the cases of $s\rightarrow p$ and $p \rightarrow s$ capture, into the most deeply-bound available states in both cases; this will serve to illustrate the essential points. From eq.~\ref{eq:sigmastop}, the capture rate in this case will be given by (omitting the emission-from-the-potential term in this single-component scenario): 
\begin{align} 
& \sigma v_\text{rel} \, \text{(continuum $\rightarrow$ bound)}& = \frac{16}{3} \frac{\alpha E_n}{M_\chi^2}  \left| \int r^2 dr \, \phi^*_{L=1}(r) \phi_{L=0}'(r) \right|^2 \times \left\{ \begin{array}{cc} 1 &  \text{initial $s$-wave} \\ 
1/3 & \text{initial $p$-wave} \end{array} \right. ,
\label{eq:sigmastopapp} 
\end{align}
where $\psi({\bf r}) = Y_{lm}(\theta,\phi)\phi(r)$. To make it easier to extract the scaling of $\sigma v_\text{rel}$ with $\alpha_H$, $M_\chi$ and $m_H$, let us define the dimensionless radial coordinate $x = \alpha_H M_\chi r$. Then $\sigma v_\text{rel}$ scales as $(\alpha E_n/M_\chi^2) (\alpha_H M_\chi)^{-4} \left| \int x^2 dx \, \phi^*_{L=1}(x) \phi_{L=0}'(x) \right|^2$.

We will be especially interested in the regime of low velocity; the usual $1/v_\text{rel}$ divergence as $v_\text{rel} \rightarrow 0$ will in this case be regulated by the non-zero $m_H$.

Let us define the parameter $\xi = \alpha_H M_\chi/m_H$; note $\xi \ge 1$ if any bound states exist. In terms of the $x$ coordinate and this parameter, the radial wavefunctions we need are given by:
\begin{itemize}
\item $nlm=100$ bound state: 

$\phi(x) = (\alpha_H M_\chi)^{3/2} \sqrt{\frac{1}{2} \left(1 - \frac{1}{\xi^2} \right)} \frac{(1 - e^{-x/\xi})}{x/\xi} e^{-(1 - 1/\xi) x/2} $
\item $nlm=210$ bound state (the scaling is identical for the $m=\pm 1$ states):

$\phi(x) = (\alpha_H M_\chi)^{3/2} \frac{\sqrt{1 - \frac{20}{\xi^2} + \frac{64}{\xi^4} }}{16\sqrt{3}}  x e^{-(1/4 -  1/\xi) x}\left(\frac{1 - e^{-x/\xi}}{x/\xi} \right)^2$

\item $L=0$ contribution to continuum wavefunction in the low-velocity limit:
\begin{align} \phi(x) &= \sqrt{ \frac{(2 \pi)^3 \xi}{1 - \cos\left( 2 \pi \sqrt{\xi} \right)} } \left( \frac{1 - e^{-x/\xi} }{x/\xi} \right) ~_2 F_1 \left[1 + \sqrt{\xi},1 - \sqrt{\xi},2,1 - e^{-x/\xi}\right]. \end{align}

\item $L=1$ contribution to continuum wavefunction in the low-velocity limit:
\begin{align} \phi(x) &= 
\sqrt{ \frac{\frac{2}{3} \pi^3 \xi}{1 - \cos\left( 2 \pi \sqrt{\xi} \right)} }\left(\xi - 1 \right) \frac{\left(  1 - e^{-x/\xi} \right)^2 }{x/\xi} ~_2 F_1 \left[2 + \sqrt{\xi},2 - \sqrt{\xi},2,1 - e^{-x/\xi}\right]. \end{align}
\end{itemize}

Let us consider the limit where $\xi \gg 1$. In this case, the bound state wavefunctions have support for (and only for) $x \lesssim 1$. Thus in the region which gives a non-negligible contribution to the overlap integral that determines the capture rate, $\int x^2 \phi^*_{L=1}(x) \phi_{L=0}'(x) dx$, it follows that $x/\xi \ll 1$, and so we can safely approximate $1 - e^{-x/\xi} \approx x/\xi$ in computing this integral. 

Using the identity:
\begin{equation} ~_2 F_1(a,b,c,z) = \sum_{k=0}^\infty \frac{(a)_k (b)_k}{(c)_k k!} z^k, \end{equation}
and the approximation $(a)_k \sim a^n$ for $|a| \gg 1$, it follows that in the range of $x$ relevant to the overlap integral,
\begin{equation}~_2F_1[1 + L + \sqrt{\xi}, 1 + L - \sqrt{\xi}, 2(L+1), 1 - e^{-x/\xi}] \approx  \sum_{k=0}^\infty \frac{(-\sqrt{\xi})^k (\sqrt{\xi})^k}{(2(L+1))_k k!} (x/\xi)^k \approx \sum_{k=0}^\infty \frac{(-x)^k}{(2(L+1))_k k!},  \end{equation}
and consequently, this term in the continuum wavefunctions is approximately $\xi$-independent.

It follows that within the range where the integrand is non-negligible, provided $\xi \gg 1$, the bound-state wavefunctions $\phi(x)$ scale as $(\alpha_H M_\chi)^{3/2} \times$ (function of $x$ only), with no leading-order $\xi$-dependence. The continuum wavefunctions scale approximately as $\sqrt{\xi/(1- \cos(2\pi \sqrt{\xi}))} \times$ (function of $x$ only). Thus $\int x^2 \phi^*_{L=1}(x) \phi_{L=0}'(x) dx \propto (\alpha_H M_\chi)^{3/2}  \sqrt{\xi/(1- \cos(2\pi \sqrt{\xi}))}$, where the proportionality factor arises from an integral over functions of $x$ only, and is independent of all the parameters of the problem.

We can now estimate the scaling of the capture rate, for the limit as $v_\text{rel} \rightarrow 0$ and assuming $m_H \ll \alpha_H M_\chi$:
\begin{align} \sigma v_\text{rel} \, \text{(continuum $\rightarrow$ bound)}& \propto  \frac{\alpha E_n}{M_\chi^2}(\alpha_H M_\chi)^{-4} (\alpha_H M_\chi)^{3} \frac{\xi}{1- \cos(2\pi \sqrt{\xi})} \nonumber \\
& \propto  \frac{\alpha E_n}{m_H} \frac{1}{M_\chi^2} \frac{1}{1 - \cos(2 \pi \sqrt{\alpha_H M_\chi/m_H})}.\end{align}

However, recall that -- as discussed above -- we must add a correction factor in the case where our initial continuum wavefunction has $L > 0$, to compensate for the incorrect asymptotic behavior of the centrifugal term and hence the continuum wavefunction. In the same limit as the one we are currently working in, where $v_\text{rel}$ is taken to zero, the correction factor $C = (1/L!)^2 (M_\chi v_\text{rel}/m_H)^{2L}$  (eq.~\ref{eq:correc}). This is the \emph{only} source of direct $L$ dependence in the scaling. Thus our final scaling estimate becomes:
\begin{align} \sigma v_\text{rel} \, \text{(continuum $\rightarrow$ bound)}& \propto \frac{\alpha E_n}{m_H} \frac{1}{M_\chi^2} \left( \frac{M_\chi v_\text{rel}}{m_H} \right)^{2L} \frac{1}{1 - \cos(2 \pi \sqrt{\alpha_H M_\chi/m_H})},\end{align}
where $L$ is the angular momentum quantum number for the initial state.

For comparison, in the Coulombic low-velocity limit we have:
\begin{align} \sigma v_\text{rel} \, \text{(continuum $\rightarrow$ bound)}& \propto \frac{E_n}{M_\chi} \frac{1}{M_\chi^2} \frac{\alpha}{v_\text{rel}}.\end{align}

If the prefactors in both cases are similar, and away from the resonance regions where the $1/(1 - \cos \theta)$ term can lead to a large enhancement in the case with $m_H \ne 0$, we would expect the two curves to intersect where $ \frac{1}{M_\chi v_\text{rel}} \sim \frac{1}{m_H} \left(\frac{M_\chi v_\text{rel}}{m_H}  \right)^{2L}$, i.e. when $\frac{M_\chi v_\text{rel}}{m_H} \sim 1$.

Thus we expect to observe Coulomb-like behavior, and in particular the universal (independent of partial wave) $1/v_\text{rel}$ velocity scaling in the capture rate, down to a ``saturation velocity'' $v_\text{rel} = m_H/M_\chi$. At this point, the capture rate $\sigma v_\text{rel}$ instead begins scaling as $v_\text{rel}^{2L}$; accordingly, capture from higher partial waves is expected to be suppressed relative to capture from the $s$-wave part of the initial continuum state, by a factor of order $(v_\text{rel} M_\chi/m_H)^{2L}$.

This is very similar to the parametric scaling of the Sommerfeld-enhanced annihilation rate \cite{Cassel:2009wt}; the enhancement factor scales as $1/v_\text{rel}^{2L+1}$ for velocities larger than the saturation velocity, but saturates at a constant value below this velocity. Since the bare un-enhanced annihilation rate ($\sigma v_\text{rel}$) scales as $v_\text{rel}^{2L}$, the enhanced annihilation rate scales as $1/v_\text{rel}$ above the saturation velocity $m_H/M_\chi$, but as $v_\text{rel}^{2L}$ below this velocity scale, just as we find for the capture process. The resonance peaks in the capture rate that we expect for $v_\text{rel} \lesssim m_H/M_\chi$ are also observed in the Sommerfeld enhancement. At these resonance locations, where $\sqrt{\alpha_H M_\chi/m_H}$ is close to an integer value, both Sommerfeld enhancement and capture rates pick up an extra scaling of $1/v_\text{rel}^2$, down to a saturation velocity that depends on proximity to the resonance. 

\cite{Braaten:2013tza} presents an expression for a universal capture rate into near-threshold $s$-wave states, analogous to the result for capture of a proton and neutron into a deuteron bound state:
\begin{equation} \sigma_{c} \sim \frac{p}{\gamma^2 + p^2},\end{equation}
where the binding energy is $\gamma^2/M_\chi$, and $p \propto M_\chi v_\text{rel}$ is the incoming momentum. This result is applicable to our analysis when the range of the potential is short relative to the wavelength of the incoming state (i.e. $m_H \lesssim M_\chi v_\text{rel}$), and the binding energy is small compared to $m_H^2/M_\chi$. It predicts that $\sigma_c v_\text{rel} \propto p^2$ when $\gamma \gg p$, and $\sigma_c v_\text{rel}$ roughly $p$-independent when $\gamma \ll p$. These scalings agree with our results above; they correspond to non-resonant and resonant capture into the $s$-wave ground state (from the $p$-wave part of the initial state), respectively, in the saturation regime where $m_H$ cannot be neglected.

\begin{figure}[h]
\begin{center}
\vspace{0.5cm}
\includegraphics[scale=0.31]{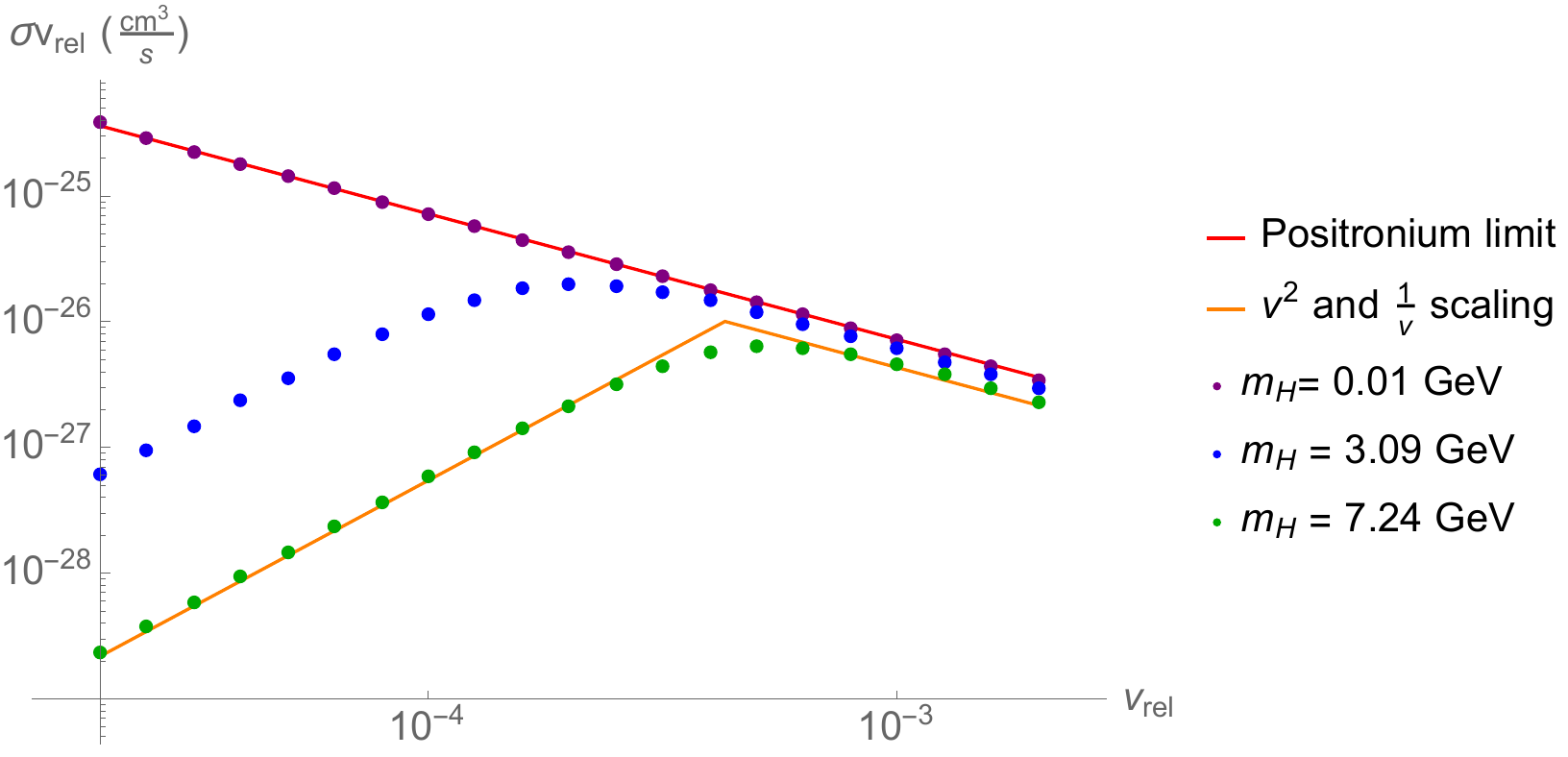}
\includegraphics[scale=0.28]{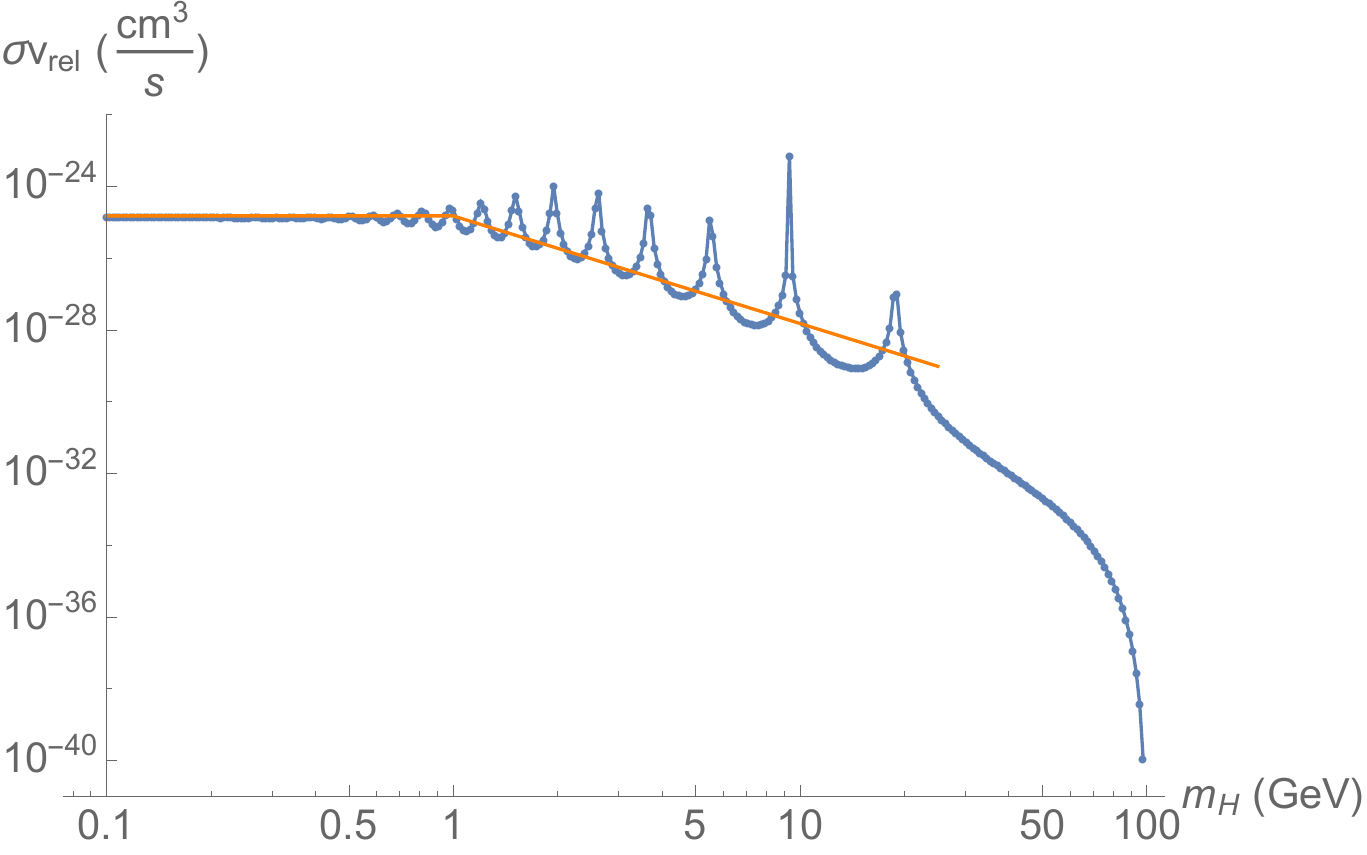}
\caption{$\sigma v_\text{rel}$ for capture into the $s$-wave ground state in the Hulth\'en potential from a $p$-wave initial state, for $M_\chi = 10$ TeV, $\alpha_H = 0.01$. {\bf Left:} $\sigma v_\text{rel}$ as a function of $v_\text{rel}$, for $m_H = 10$ MeV (\emph{purple dots}), $3.09$ GeV (\emph{blue dots}) and $7.24$ GeV (\emph{green dots}). The \emph{red} line indicates the analytic positronium-like limit where $m_H \rightarrow 0$; the \emph{orange} line is proportional to $v_\text{rel}^{2}$ ($v_\text{rel}^{-1}$) below (above) the break, and is to guide the eye. {\bf Right:} $\sigma v_\text{rel}$ as a function of $m_H$, for $v_\text{rel} = 15$ km/s $= 5 \times 10^{-5} c$ (\emph{blue dots and line}). The \emph{orange} line is proportional to $m_H^0$ ($m_H^{-3}$) below (above) the break, and is to guide the eye. Note we have chosen a small $v_\text{rel}$ in order to display a large region with saturation and resonance effects, before $m_H$ becomes too large to support a bound state.}
\label{fig:hulthen}
\end{center}
\end{figure}

To test our arguments above (and especially the scaling of the $p$-wave continuum states, since this relied on a correction factor that strictly only applies for short-range processes), we solved the Schr\"{o}dinger equation with the Hulth\'en potential numerically (since this is a single-state problem, the numerical issues discussed in appendix \ref{app:init} are not a concern), computing the $p$-wave part of the initial plane wave. We used this numerical wavefunction and the analytic wavefunction for the ground state (for which the equations above are exact) to compute the capture rate into the ground state. Figure \ref{fig:hulthen} demonstrates the scaling of the resulting cross section with both $v_\text{rel}$ and $m_H$, for sample parameters $\alpha_H = 0.01$, $M_\chi = 10$ TeV. We see the expected convergence to the positronium-like limit, with scaling $1/v_\text{rel}$, for $m_H \ll M_\chi v_\text{rel}$, and the resonance behavior and overall $v_\text{rel}^2/m_H^3$ scaling for $m_H \gtrsim M_\chi v_\text{rel}$. As $m_H$ approaches $\alpha_H M_\chi = 100$ GeV, the cross section is suppressed and eventually vanishes as the bound state energy goes to zero.

%%%%%%%%%%%%%%%%%%%%%%%%%%%%%%%%%%%%%%%%%%%%
\section{Conversion from Wilson coefficients to annihilation rates}
\label{app:anndictionary}
%%%%%%%%%%%%%%%%%%%%%%%%%%%%%%%%%%%%%%%%%%%%

In this section we demonstrate how to use the results in the literature on Wilson coefficients for processes of the form $\chi_{e_1} \chi_{e_2} \rightarrow XX \rightarrow \chi_{e_4} \chi_{e_3}$, as derived in \cite{Hellmann:2013jxa, Beneke:2014gja}, in order to compute annihilation rates for WIMPonium bound states.

Hellmann and Ruiz-Femenia (\cite{Hellmann:2013jxa}, hereafter ``HRF'') write the perturbative, spin-averaged annihilation cross section in the form:
 \begin{align} \sigma^{\chi_{e_1} \chi_{e_2} \rightarrow X_A X_B} v_\mathrm{rel} & = a + (b_P + b_S) \frac{4 p^2}{M_\chi^2} + \mathcal{O}(v_\mathrm{rel}^4), \nonumber \\
 a & = \hat{f}(^1S_0) + 3 \hat{f}(^3S_1) \nonumber \\
 b_P & = \frac{1}{16} \left(\hat{f}(^1P_1) + \hat{f}(^3P_J) \right), \end{align}
where the $\hat{f}$ terms are the Wilson coefficients derived in that work, we have set the reduced mass to $M_\chi/2$ and the mass of the two-particle state to $2 M_\chi$, and we have written $v_\text{rel} = 2 p / M_\chi$, where $p$ is the momentum of one of the initial particles in the COM frame. As discussed above, the non-spin-averaged leading-order perturbative annihilation rates from spin-triplet and spin-triplet initial states are then given by:
\begin{align} \text{singlet} \, L=0: \sigma^{\chi_{e_1} \chi_{e_2} \rightarrow X_A X_B} v_\mathrm{rel}  & = 4 \hat{f}(^1S_0), \nonumber \\
 \text{triplet} \, L=0: \sigma^{\chi_{e_1} \chi_{e_2} \rightarrow X_A X_B} v_\mathrm{rel}  & = 4 \hat{f}(^3S_1), \nonumber \\
 \text{singlet} \, L=1:  \sigma^{\chi_{e_1} \chi_{e_2} \rightarrow X_A X_B} v_\mathrm{rel} & = \frac{p^2}{M_\chi^2}  \hat{f}(^1P_1)  \nonumber \\
 \text{triplet} \, L=1: \sigma^{\chi_{e_1} \chi_{e_2} \rightarrow X_A X_B} v_\mathrm{rel} & = \frac{1}{3} \frac{p^2}{M_\chi^2}   \hat{f}(^3P_J).\end{align}

These Wilson coefficients are derived from the processes $\chi_{e_1} \chi_{e_2} \rightarrow XX \rightarrow \chi_{e_1} \chi_{e_2}$. In terms  of the matrix element for $\chi_{e_1} \chi_{e_2} \rightarrow XX$, in our case where we can assume the masses of the initial particle to be identical, they are proportional to $\int d\Pi_2 \left|\mathcal{M} (\chi_{e_1}({\bf p}) \chi_{e_2}(-{\bf p}) \rightarrow XX) \right|^2$ (with proportionality factors that can be trivially extracted from the expressions for the cross section just given). However, to compute the annihilation rates for bound states that mix the $\chi^0 \chi^0$ and $\chi^+ \chi^-$ states, we also need the terms of the form $\int d\Pi_2 \mathcal{M}^*(\chi_{e_1}({\bf p}) \chi_{e_2}(-{\bf p}) \rightarrow XX) \mathcal{M}(\chi_{e_4}({\bf p}) \chi_{e_3}(-{\bf p}) \rightarrow XX)$, where the initial states $\chi_{e_4} \chi_{e_3}$ and $\chi_{e_1} \chi_{e_2}$ need not be identical. These terms are given by the (absorptive part of the) off-diagonal Wilson coefficients extracted from the processes $\chi_{e_1} \chi_{e_2} \rightarrow XX \rightarrow \chi_{e_4} \chi_{e_3}$. We can thus promote the Wilson coefficients $\hat{f}$ for a given final state to a matrix $\hat{f}_{\{ e_1 e_2 \} \{ e_4 e_3 \}}$, where $\{ e_1 e_2 \}$ and $\{ e_4 e_3 \}$ label the relevant two-particle states. For the bound states we are interested in, the possible two-particle states are $\chi^0 \chi^0$ and $\chi^+ \chi^-$, and $\hat{f}$ is a matrix akin to the potential matrix defined in eq.~\ref{eq:potltwo}. 

Our situation is almost identical to the one we face in computing the Sommerfeld-enhanced annihilation cross section from the Wilson coefficients, which is discussed in detail in \cite{Beneke:2014gja}; the only difference is a subtlety in how one normalizes the initial state (which in this case is a bound state, rather than a continuum state). For the $s$-wave case with Sommerfeld enhancement, the annihilation matrix -- which we denote $\Sigma$ -- is to be contracted with the vector wavefunction at the origin to obtain the enhanced annihilation rate:
\begin{align} (\sigma v)_i = c_i \Psi^\dagger(0) \Sigma \Psi(0). \end{align}
The prefactor $c_i$ is 2 if the particles in the initial state are identical, and 1 otherwise. For the bound-state calculation, there is no such prefactor, as the different normalization of the bound state for identical particles cancels it out.

 As discussed in \cite{Beneke:2014gja}, the annihilation matrix $\Sigma$ is built from the Wilson coefficients, supplemented by factors of $1/\sqrt{2}$ if either the $\{ e_1 e_2 \}$ or $\{ e_4 e_3 \}$ states are comprised of identical particles, or by a factor of $1/2$ if both two-particle states are comprised of identical particles. In our analysis, we see that these factors of $1/\sqrt{2}$ arise naturally from the differing normalization of the bound states comprised of identical vs distinguishable particles (see section \ref{subsec:SM decay}). There is also an overall prefactor relating the Wilson coefficients to the annihilation cross section, as discussed above.
 
 Consequently, our algorithm for defining the general annihilation matrix (without spin averaging) is:
 \begin{itemize}
 \item Write down the Wilson coefficients $\hat{f}_{\{ e_1 e_2 \} \{ e_4 e_3 \}}$ for a specific final state, as calculated in \cite{Hellmann:2013jxa} (these are given explicitly for the pure wino in appendix C3 of that work).
 \item Construct the annihilation matrix by:
\begin{equation}
\Sigma_{\{ e_1 e_2 \} \{ e_4 e_3 \}} = (1/\sqrt{2})^{n_i} \, c(L,S) \hat{f}_{\{ e_1 e_2 \} \{ e_4 e_3 \}}, 
\end{equation}
where $n_i=0$ if both $\{e_1 e_2 \} $ and $\{ e_4 e_3 \}$ correspond to pairs of distinguishable particles, $n_i=1$ if one pair is identical and the other distinguishable, and $n_i=2$ if both pairs are comprised of identical particles (although the pairs may be different from each other). The constant prefactor $c(L,S)$, as discussed above, is 4 for $s$-wave states, 1 for spin-singlet $p$-wave states, $1/3$ for spin-triplet $p$-wave states.
 \end{itemize}
 
For $L=0$, the diagonal elements of this matrix give the annihilation cross sections $\sigma v_\mathrm{rel}$ for particles initially in the appropriate two-particle continuum state in the absence of any potential, up to the factors of $c_i$ discussed above. For $L=1$, the diagonal matrix elements must be multiplied by an additional factor of $(p^2/M_\chi^2)$ to obtain the annihilation cross sections. We have stripped this latter factor out of the matrix $\Sigma$ because it is precisely this factor that will be altered when the initial state is a bound state, rather than a free-particle continuum state.

The resulting annihilation matrix is precisely equivalent to the annihilation matrices defined in the main text in Eqs. \ref{eq:sann}-\ref{eq:pann}. To check the normalization, note that for free-particle annihilation of nonrelativistic particles of equal mass $M_\chi$, we have:
\begin{equation} \sigma v_\mathrm{rel} = \frac{1}{(2 M_\chi)^2} \int d\Pi_n \left|\mathcal{M} (\chi({\bf p}) \chi(-{\bf p}) \rightarrow f) \right|^2. 
\label{eq:decay.18}
\end{equation}

Using the pure wino case as an example, from \cite{Hellmann:2013jxa} we have the non-zero Wilson coefficients for the various cases:

\emph{Spin-singlet $L=0$} ($\,^1S_0$)
\begin{itemize}
\item Final state $W^+ W^-$: $\hat{f}_{\{\chi^0\chi^0\}\{\chi^0\chi^0\}} = \frac{2 \pi \alpha_W^2}{M_\chi^2}, \, \hat{f}_{\{\chi^+\chi^-\}\{\chi^+\chi^-\}} = \frac{\pi \alpha_W^2}{2 M_\chi^2}, \, \hat{f}_{\{\chi^+\chi^+\}\{\chi^0\chi^0\}} = \frac{\pi \alpha_W^2}{M_\chi^2}$,
\item Final state $ZZ$: $\hat{f}_{\{\chi^+\chi^-\}\{\chi^+\chi^-\}} = c_W^4 \frac{\pi \alpha_W^2}{M_\chi^2}$,
\item Final state $Z\gamma$: $\hat{f}_{\{\chi^+\chi^-\}\{\chi^+\chi^-\}} = 2 c_W^2 s_W^2 \frac{\pi \alpha_W^2}{M_\chi^2}$,
\item Final state $\gamma \gamma$: $\hat{f}_{\{\chi^+\chi^-\}\{\chi^+\chi^-\}} = s_W^4 \frac{\pi \alpha_W^2}{M_\chi^2}$,
\end{itemize}

Thus by the recipe above we obtain:
\begin{align} \Sigma(W^+ W^-)  & = \frac{4 \pi \alpha_W^2}{M_\chi^2} \begin{pmatrix} 1 & \frac{1}{\sqrt{2}} \\  \frac{1}{\sqrt{2}} & \frac{1}{2} \end{pmatrix}, \quad \Sigma(ZZ)  =  \frac{4 \pi \alpha_W^2}{M_\chi^2} c_W^4 \begin{pmatrix} 0 & 0 \\ 0 & 1 \end{pmatrix}, \nonumber \\
 \Sigma(Z\gamma)  & = \frac{4 \pi \alpha_W^2}{M_\chi^2} 2 c_W^2 s_W^2 \begin{pmatrix} 0 & 0 \\ 0 & 1 \end{pmatrix}, \quad \Sigma(\gamma \gamma)  = \frac{4 \pi \alpha_W^2}{M_\chi^2} s_W^4 \begin{pmatrix} 0 & 0 \\ 0 & 1 \end{pmatrix}.\end{align}

\emph{Spin-triplet $L=0$ ($~^3S_1$)}
\begin{itemize}
\item Final state $W^+ W^-$: $\hat{f}_{\{\chi^+\chi^-\}\{\chi^+\chi^-\}} = \frac{1}{48} \frac{\pi \alpha_W^2}{M_\chi^2}$,
\item Final state $Z h^0$: $\hat{f}_{\{\chi^+\chi^-\}\{\chi^+\chi^-\}} = \frac{1}{48} \frac{\pi \alpha_W^2}{M_\chi^2}$,
\item Final state $q \bar{q}$ (for each individual quark species; multiply by 6 to get total rate): $\hat{f}_{\{\chi^+\chi^-\}\{\chi^+\chi^-\}} = \frac{1}{8} \frac{\pi \alpha_W^2}{M_\chi^2}$,
\item Final state $l^+ l^-, \nu \bar{\nu}$ (for each individual lepton or neutrino flavor; multiply by 6 to get total rate): $\hat{f}_{\{\chi^+\chi^-\}\{\chi^+\chi^-\}} = \frac{1}{24} \frac{\pi \alpha_W^2}{M_\chi^2}$,
\end{itemize}

By the recipe above we have:
\begin{align} \Sigma(W^+ W^-) & = \frac{1}{12} \frac{\pi \alpha_W^2}{M_\chi^2} \begin{pmatrix} 0 & 0 \\ 0 & 1 \end{pmatrix}, \quad \Sigma(Z h^0) = \frac{1}{12} \frac{\pi \alpha_W^2}{M_\chi^2} \begin{pmatrix} 0 & 0 \\ 0 & 1 \end{pmatrix}, \nonumber \\
\Sigma(q \bar{q}) & = \frac{1}{2} \frac{\pi \alpha_W^2}{M_\chi^2} \begin{pmatrix} 0 & 0 \\ 0 & 1 \end{pmatrix}, \quad \Sigma(l^+ l^-, \nu \bar{\nu}) = \frac{1}{6} \frac{\pi \alpha_W^2}{M_\chi^2} \begin{pmatrix} 0 & 0 \\ 0 & 1 \end{pmatrix}. \end{align}

\emph{Spin-singlet $L=1$ ($~^1P_1$)}
\begin{itemize} 
\item Final state $W^+ W^-$: $\hat{f}_{\{\chi^+\chi^-\}\{\chi^+\chi^-\}} = \frac{2}{3} \frac{\pi \alpha_W^2}{M_\chi^2}$
\end{itemize}

The only non-zero annihilation matrix is thus $ \Sigma(W^+ W^-)  = \frac{2}{3} \frac{\pi \alpha_W^2}{M_\chi^2} \begin{pmatrix} 0 & 0 \\ 0 & 1 \end{pmatrix}$.

\emph{Spin-triplet $L=1$ ($~^3P_J$)}
\begin{itemize}
\item Final state $W^+ W^-$: $\hat{f}_{\{\chi^0\chi^0\}\{\chi^0\chi^0\}} = \frac{56}{3} \frac{\pi \alpha_W^2}{M_\chi^2}, \quad \hat{f}_{\{\chi^+\chi^-\}\{\chi^+\chi^-\}} = \frac{14}{3} \frac{\pi \alpha_W^2}{M_\chi^2}, \quad \hat{f}_{\{\chi^+\chi^+\}\{\chi^0\chi^0\}} = \frac{28}{3} \frac{\pi \alpha_W^2}{M_\chi^2}$,
\item Final state $ZZ$: $\hat{f}_{\{\chi^+\chi^-\}\{\chi^+\chi^-\}} = c_W^4 \frac{28}{3} \frac{\pi \alpha_W^2}{M_\chi^2}$,
\item Final state $Z\gamma$: $\hat{f}_{\{\chi^+\chi^-\}\{\chi^+\chi^-\}} = 2 c_W^2 s_W^2 \frac{28}{3} \frac{\pi \alpha_W^2}{M_\chi^2}$,
\item Final state $\gamma \gamma$: $\hat{f}_{\{\chi^+\chi^-\}\{\chi^+\chi^-\}} = s_W^4 \frac{28}{3} \frac{\pi \alpha_W^2}{M_\chi^2}$,
\end{itemize}

The annihilation matrices for this case are given by:
\begin{align} \Sigma(W^+ W^-) & = \frac{28}{9} \frac{\pi \alpha_W^2}{M_\chi^2} \begin{pmatrix} 1 & \frac{1}{\sqrt{2}} \\  \frac{1}{\sqrt{2}} & \frac{1}{2} \end{pmatrix}, \quad\Sigma(ZZ)  = \frac{28}{9} \frac{\pi \alpha_W^2}{M_\chi^2} c_W^4 \begin{pmatrix} 0 & 0 \\ 0 & 1 \end{pmatrix}, \nonumber \\
\Sigma(Z\gamma) & = \frac{28}{9} \frac{\pi \alpha_W^2}{M_\chi^2} 2 c_W^2 s_W^2 \begin{pmatrix} 0 & 0 \\ 0 & 1 \end{pmatrix}, \quad \Sigma(\gamma \gamma)  = \frac{28}{9} \frac{\pi \alpha_W^2}{M_\chi^2} s_W^4 \begin{pmatrix} 0 & 0 \\ 0 & 1 \end{pmatrix}.\end{align}

%%%%%%%%%%%%%%%%%%%%%%%%%%%%%%%%%%%%%%%%%%%%
\section{Useful integrals}
\label{app:integrals}
%%%%%%%%%%%%%%%%%%%%%%%%%%%%%%%%%%%%%%%%%%%%

In computing the continuum$\rightarrow$bound capture rates to the lowest-lying $s$- and $p$-wave bound states in the Coulombic limit, we need to evaluate several non-trivial integrals; we collect the required results here for reference. Similar calculations have been presented elsewhere in the literature; e.g.~\cite{Petraki:2015hla} computed the integrals needed for capture to the ground state for both scalar and vector mediators, and for capture to the first excited $p$-wave state in the case of scalar mediators.

The integrals in question are:
\begin{align} I_1 = & \int_0^\infty r^2  dr \int_{-1}^{1} dx \left[ (x - 1) e^{-\eta r} e^{i p r x} ~_1F_1\left(1 + i\zeta, 2, i pr (1 - x)\right) \right] \nonumber \\
I_2 = & \int_0^\infty r^2 dr  \int_{-1}^{1} dx \left[ r (x-1) e^{-\eta r} e^{i p r x} ~_1F_1\left(1 + i\zeta, 2, i pr (1 - x)\right) \right] \nonumber \\
I_3 = & \int_0^\infty r^2 dr \int_{-1}^{1} dx \left[ r x (x - 1) e^{-\eta r} e^{i p r x} ~_1F_1\left(1 + i\zeta, 2, i pr (1 - x)\right) \right]. 
\end{align}
We will be particularly interested in the limit where $p\rightarrow 0$ while $p \, \zeta$ is held fixed, since this corresponds to the low-velocity limit of the capture rates.

Our starting point is the identities \cite{1996JPhB...29.2135A}:
\begin{align}  \int_0^\infty r^2  dr  \int_{-1}^{1} dx \left[ x e^{i p r x - \eta r} ~_1F_1(i \zeta, 1, i p r (1 - x)) \right] & = 4 i p (1 - i \zeta) \frac{(\eta - i p)^{-2 i \zeta}}{(p^2 + \eta^2)^{2 - i \zeta}}, \nonumber \\
\int_0^\infty r^2  dr  \int_{-1}^{1} dx \left[ \frac{1}{r} e^{i p r x - \eta r} ~_1F_1(i \zeta, 1, i p r (1 - x)) \right] & = 2 \frac{(\eta - i p)^{-2 i \zeta}}{(p^2 + \eta^2)^{1 - i \zeta}} .
 \end{align}

 Differentiating these expressions with respect to $p$, we find:
\begin{align}  &  \int_0^\infty r^2  dr  \int_{-1}^{1} dx e^{i p r x - \eta r} \left[\zeta r(x - 1) x  ~_1F_1(1 + i \zeta, 2, i p r (1 - x)) + i r x^2  ~_1F_1(i \zeta, 1, i p r (1 - x)) \right] \nonumber \\
 & = 4 i \frac{d}{dp} \left[ p (1 - i \zeta) \frac{(\eta - i p)^{-2 i \zeta}}{(p^2 + \eta^2)^{2 - i \zeta}} \right], \nonumber \\
& \int_0^\infty r^2  dr  \int_{-1}^{1} dx e^{i p r x - \eta r} \left[ -(1-x) \zeta ~_1F_1(1 + i \zeta, 2, i p r (1 - x)) + i x  ~_1F_1(i \zeta, 1, i p r (1 - x)) \right] \nonumber \\
& = 2 \frac{d}{dp} \left[ \frac{(\eta - i p)^{-2 i \zeta}}{(p^2 + \eta^2)^{1 - i \zeta}} \right], 
\label{identsm}
 \end{align}
 The second line of eq.~\ref{identsm} can be rewritten as:
\begin{align} I_1 & = \int_0^\infty r^2  dr  \int_{-1}^{1} dx e^{i p r x - \eta r} (x-1) ~_1F_1(1 + i \zeta, 2, i p r (1 - x)) \nonumber \\
  & = \frac{1}{\zeta} \left\{ 2 \frac{d}{dp} \left[ \frac{(\eta - i p)^{-2 i \zeta}}{(p^2 + \eta^2)^{1 - i \zeta}} \right] + 4 p (1 - i \zeta) \frac{(\eta - i p)^{-2 i \zeta}}{(p^2 + \eta^2)^{2 - i \zeta}} \right\} \nonumber \\
& = -4 \frac{(\eta - i p)^{-1 - 2 i \zeta}}{(\eta^2 + p^2)^{1 - i \zeta}}   \end{align}
Differentiating with respect to $\eta$ then gives us:
\begin{align}I_2 & = \int_0^\infty r^2  dr  \int_{-1}^{1} dx e^{i p r x - \eta r} r(x-1) ~_1F_1(1 + i \zeta, 2, i p r (1 - x)) \nonumber \\
& = - 4 (3 \eta + i p - 2 p \zeta) \frac{(\eta - i p)^{-1 - 2 i \zeta}}{(\eta^2 + p^2)^{2 - i \zeta}}  \end{align}
Finally, if $\zeta$ is large, the first line of eq.~\ref{identsm} yields:
 \begin{align} I_3 & =  \int_0^\infty r^2  dr  \int_{-1}^{1} dx e^{i p r x - \eta r} r x (x - 1)  ~_1F_1(1 + i \zeta, 2, i p r (1 - x)) \nonumber \\
 & \approx 4  (\eta^2 - 3p^2 - 2 \eta p \zeta) \frac{(\eta - i p)^{-2 i \zeta}}{(\eta^2 + p^2)^{3 - i \zeta}}.\end{align}

We can further simplify these integrals by taking the limit $p \rightarrow 0$, $p \, \zeta \rightarrow$ constant:
\begin{align} I_1 \rightarrow -4 e^{- 2 p \zeta/\eta}/\eta^3, \quad I_2 \rightarrow - 4 (3 \eta - 2 p \zeta) e^{-2 p \zeta/\eta}/\eta^5, \quad I_3 \rightarrow 4 (\eta - 2 p \zeta)e^{-2 p \zeta/\eta}/\eta^5.
\end{align}

Finally, note also that:
\begin{align}  \int_0^\infty r^2 dr \int_{-1}^{1} dx \left[ r (x^2 - 1) e^{-\eta r} e^{i p r x} ~_1F_1\left(1 + i\zeta, 2, i pr (1 - x)\right) \right] = I_2 + I_3 \rightarrow -8 e^{-2 p \zeta/\eta}/\eta^4. 
\end{align}

In computing the integrals associated with the $\hat{C}_2$ structure, and with the capture rates separated by initial-state partial wave, we have employed the following integrals, which can all be performed by \texttt{Mathematica}:
\begin{align} \int r^2 dr e^{-\eta r} ~_1F_1\left(a, 1, i q r \right) & \approx e^{i a q/\eta} \frac{2\eta^2 + 4 i \eta a q - (a q)^2}{\eta^5}, \nonumber \\
\int r^3 dr e^{-\eta r} ~_1F_1\left(a, 1, i q r \right) & \approx e^{i a q /\eta} \frac{6 \eta^3 + 18 i \eta^2 a q - 9 \eta (a q)^2 - 
   i (a q)^3}{\eta^7}, \nonumber \\
   \int r^3 dr e^{-\eta r} ~_1F_1\left(a, 2, i q r \right)& \approx e^{i a q /\eta} \frac{6 \eta^2 + 6 i \eta a q - (a q)^2}{\eta^6} \nonumber \\
   \int r^3 dr e^{-\eta r} ~_1F_1\left(a, 3, i q r \right)& \approx 2 e^{i a q /\eta} \frac{3 \eta + i a q}{\eta^5}  \end{align}
Here we have evaluated the integrals in the limit $q\rightarrow 0$, but with $a \, q \rightarrow$ constant.

\bibliography{boundstates}
\bibliographystyle{JHEP}

\end{document}